\documentclass[12pt, letterpaper, onecolumn, peerreview]{IEEEtran}
\usepackage{color}
\usepackage{colordvi}
\usepackage{fancybox}
\usepackage{amsmath}
\usepackage{amssymb}
\usepackage{epsfig}
\usepackage{epic}

\newtheorem{theorem}{Theorem}[section]
\newtheorem{lemma}{Lemma}[section]
\newtheorem{corollary}{Corollary}[section]
\newtheorem{proposition}{Proposition}[section]

\newtheorem{property}{Property}[section]
\newtheorem{definition}{Definition}[section]

\begin{document}
\title{\large \bf Source coding and channel requirements for unstable
  processes} 

\author{Anant Sahai\footnote{Department of Electrical Engineering and
Computer Science at the University of California at Berkeley. A few of
these results were presented at ISIT 2004 and a primitive form of
others appeared at ISIT 2000 and in his doctoral dissertation.},
Sanjoy Mitter\footnote{Department of Electrical Engineering and
Computer Science at the Massachusetts Institute of Technology. Support
for S.K.~Mitter was provided by the Army Research Office under the
MURI Grant: Data Fusion in Large Arrays of Microsensors
DAAD19-00-1-0466 and the Department of Defense MURI Grant: Complex
Adaptive Networks for Cooperative Control Subaward \#03-132 and the
National Science Foundation Grant CCR-0325774.} \\ {\small
sahai@eecs.berkeley.edu, mitter@mit.edu}}


\maketitle

\begin{abstract} 
Our understanding of information in systems has been based on the
foundation of memoryless processes. Extensions to stable Markov and
auto-regressive processes are classical. Berger proved a source coding
theorem for the marginally unstable Wiener process, but the
infinite-horizon exponentially unstable case has been open since
Gray's 1970 paper. There were also no theorems showing what is needed
to communicate such processes across noisy channels.

In this work, we give a fixed-rate source-coding theorem for the
infinite-horizon problem of coding an exponentially unstable Markov
process. The encoding naturally results in two distinct bitstreams
that have qualitatively different QoS requirements for communicating
over a noisy medium. The first stream captures the information that is
accumulating within the nonstationary process and requires sufficient
anytime reliability from the channel used to communicate the
process. The second stream captures the historical information that
dissipates within the process and is essentially classical. This
historical information can also be identified with a natural stable
counterpart to the unstable process. A converse demonstrating the
fundamentally layered nature of unstable sources is given by means of
information-embedding ideas.
\end{abstract}

\begin{keywords}
Nonstationary processes, rate-distortion, anytime reliability,
information embedding
\end{keywords}

\IEEEpeerreviewmaketitle

\section{Introduction}
The source and channel models studied in information theory are not
just interesting in their own right, but also provide insights into
the architecture of reliable communication systems. Since Shannon's
work, memoryless sources and channels have always been at the base of
our understanding. They have provided the key insight of separating
source and channel coding with the bit rate alone appearing at the
interface \cite{ShannonOriginalPaper, ShannonLossy}. The basic story
has been extended to many different sources and channels with memory
for point-to-point communication \cite{VembuVerdu}.

However, there are still many issues for which information theoretic
understanding eludes us. Networking in particular has a whole host of
such issues, leading Ephremides and Hajek to entitle their survey
article ``Information Theory and Communication Networks: An
Unconsummated Union!''  \cite{EphremidesHajek}. They comment:
\begin{quotation}
{\small
The interaction of source coding with network-induced delay cuts
across the classical network layers and has to be better
understood. The interplay between the distortion of the source output
and the delay distortion induced on the queue that this source output
feeds into may hold the secret of a deeper connection between
information theory. Again, feedback and delay considerations are
important.}
\end{quotation}

Real communication networks and networked applications are quite
complicated. To move toward a quantitative and qualitative of
understanding of the issues, tractable models that exhibit at least
some of the right qualitative behavior are essential. In
\cite{ControlPartI, ControlPartII}, the problem of stabilization of
unstable plants across a noisy feedback link is considered. There,
delay and feedback considerations become intertwined and the notion of
feedback anytime capacity is introduced. To stabilize an otherwise
unstable plant over a noisy channel, not only is it necessary to have
a channel capable of supporting a certain minimal rate, but the
channel when used with noiseless feedback must also support a high
enough error-exponent (called the anytime reliability) with fixed
delay in a delay-universal fashion.  This turns out to be a sufficient
condition as well, thereby establishing a separation theorem for
stabilization.  In \cite{OurUpperBoundPaper}, upper bounds are given
for the fixed-delay reliability functions of DMCs with and without
feedback, and these bounds are shown to be tight for certain classes
of channels.  Moreover, the fixed-delay reliability functions with
feedback are shown to be fundamentally better than the traditional
fixed-block-length reliability functions.

While the stabilization problem does provide certain important
insights into interactive applications, the separation theorem for
stabilization given in \cite{ControlPartI, ControlPartII} is coarse
--- it only addresses performance as a binary valued entity:
stabilized or not stabilized. All that matters is the tail-behavior of
the closed-loop process. To get a more refined view in terms of
steady-state performance, this paper instead considers the corresponding
open-loop estimation problem. This is the seemingly classical question
of lossy source coding for an {\em unstable} scalar Markov processes
--- mapping the source into bits and then seeing what is required to
communicate such bits using a point-to-point communication system.

\subsection{Communication of Markov processes}
\begin{figure}
\begin{center}
\setlength{\unitlength}{2600sp}
\begingroup\makeatletter\ifx\SetFigFont\undefined%
\gdef\SetFigFont#1#2#3#4#5{%
  \reset@font\fontsize{#1}{#2pt}%
  \fontfamily{#3}\fontseries{#4}\fontshape{#5}%
  \selectfont}%
\fi\endgroup%
\begin{picture}(7516,6270)(202,-5461)
\thinlines
\put(751,-586){\oval(1342,1342)}
\put(6450,-2394){\oval(1342,1342)}
\put(2251,-1186){\framebox(1200,1200){}}
\put(4051,-1186){\framebox(1200,1200){}}
\put(4051,-4486){\framebox(1200,1200){}}
\put(2251,-4486){\framebox(1200,1200){}}
\put(1126,-586){\vector( 1, 0){1125}}
\put(3451,-586){\vector( 1, 0){600}}
\put(4051,-3886){\vector(-1, 0){600}}
\put(2251,-3886){\vector(-1, 0){1200}}
\put(751,539){\line( 0,-1){525}}
\put(751, 14){\vector( 0,-1){225}}
\multiput(3751,539)(0.00000,-119.35484){47}{\line( 0,-1){ 59.677}}
\multiput(5551,539)(0.00000,-119.35484){47}{\line( 0,-1){ 59.677}}
\put(5251,-586){\line( 1, 0){1200}}
\put(6451,-586){\line( 0,-1){1200}}
\put(6451,-1786){\vector( 0,-1){150}}
\multiput(1951,539)(0.00000,-119.35484){47}{\line( 0,-1){ 59.677}}
\put(3451,-436){\vector( 1, 0){600}}
\put(3451,-736){\vector( 1, 0){600}}
\put(3451,-886){\vector( 1, 0){600}}
\put(3451,-286){\vector( 1, 0){600}}
\put(4051,-3736){\vector(-1, 0){600}}
\put(4051,-3586){\vector(-1, 0){600}}
\put(4051,-4036){\vector(-1, 0){600}}
\put(4051,-4186){\vector(-1, 0){600}}
\put(4651,-2161){\makebox(0,0)[b]{\smash{\SetFigFont{10}{14.4}{\rmdefault}{\mddefault}{\updefault}Reliable}}}
\put(751,-886){\makebox(0,0)[b]{\smash{\SetFigFont{10}{14.4}{\rmdefault}{\mddefault}{\updefault}$X_t$}}}
\put(751,-586){\makebox(0,0)[b]{\smash{\SetFigFont{10}{14.4}{\rmdefault}{\mddefault}{\updefault}Source}}}
\put(751,614){\makebox(0,0)[b]{\smash{\SetFigFont{10}{14.4}{\rmdefault}{\mddefault}{\updefault}$W_t$}}}
\put(2851,-2161){\makebox(0,0)[b]{\smash{\SetFigFont{10}{14.4}{\rmdefault}{\mddefault}{\updefault}Source}}}
\put(2851,-2386){\makebox(0,0)[b]{\smash{\SetFigFont{10}{14.4}{\rmdefault}{\mddefault}{\updefault}Coding }}}
\put(2851,-2611){\makebox(0,0)[b]{\smash{\SetFigFont{10}{14.4}{\rmdefault}{\mddefault}{\updefault}Layer}}}
\put(751,-3886){\makebox(0,0)[b]{\smash{\SetFigFont{10}{14.4}{\rmdefault}{\mddefault}{\updefault}$\widehat{X}_t$}}}
\put(751,-4261){\makebox(0,0)[b]{\smash{\SetFigFont{10}{14.4}{\rmdefault}{\mddefault}{\updefault}Estimates}}}
\put(2851,-3886){\makebox(0,0)[b]{\smash{\SetFigFont{10}{14.4}{\rmdefault}{\mddefault}{\updefault}Source}}}
\put(2851,-4111){\makebox(0,0)[b]{\smash{\SetFigFont{10}{14.4}{\rmdefault}{\mddefault}{\updefault}Decoder(s)}}}
\put(3751,-5236){\makebox(0,0)[b]{\smash{\SetFigFont{10}{14.4}{\rmdefault}{\mddefault}{\updefault}``Bitstreams''}}}
\put(3751,-5461){\makebox(0,0)[b]{\smash{\SetFigFont{10}{14.4}{\rmdefault}{\mddefault}{\updefault}Interface}}}
\put(4651,-3886){\makebox(0,0)[b]{\smash{\SetFigFont{10}{14.4}{\rmdefault}{\mddefault}{\updefault}Channel}}}
\put(4651,-4111){\makebox(0,0)[b]{\smash{\SetFigFont{10}{14.4}{\rmdefault}{\mddefault}{\updefault}Decoder}}}
\put(4651,-2611){\makebox(0,0)[b]{\smash{\SetFigFont{10}{14.4}{\rmdefault}{\mddefault}{\updefault}Layer}}}
\put(4651,-2386){\makebox(0,0)[b]{\smash{\SetFigFont{10}{14.4}{\rmdefault}{\mddefault}{\updefault}Transmission}}}
\put(6451,-2536){\makebox(0,0)[b]{\smash{\SetFigFont{10}{14.4}{\rmdefault}{\mddefault}{\updefault}Channel}}}
\put(6451,-2311){\makebox(0,0)[b]{\smash{\SetFigFont{10}{14.4}{\rmdefault}{\mddefault}{\updefault}Noisy}}}
\put(4651,-586){\makebox(0,0)[b]{\smash{\SetFigFont{10}{14.4}{\rmdefault}{\mddefault}{\updefault}Channel}}}
\put(4651,-811){\makebox(0,0)[b]{\smash{\SetFigFont{10}{14.4}{\rmdefault}{\mddefault}{\updefault}Encoder}}}
\put(2851,-811){\makebox(0,0)[b]{\smash{\SetFigFont{10}{14.4}{\rmdefault}{\mddefault}{\updefault}Encoder(s)}}}
\put(2851,-586){\makebox(0,0)[b]{\smash{\SetFigFont{10}{14.4}{\rmdefault}{\mddefault}{\updefault}Source}}}
\put(6451,-3886){\vector(-1, 0){1200}}
\put(6451,-2836){\line( 0,-1){1050}}
\end{picture}
\end{center}
\caption{The point-to-point communication problem considered here. The
  goal is to minimize end-to-end average distortion
  $\rho(X_t,\widehat{X}_t)$. Finite, but possible large, end-to-end delay
  will be permitted. One of the key issues explored is what must be
  made available at the source/channel interface.}
\label{fig:mainproblem}
\end{figure}
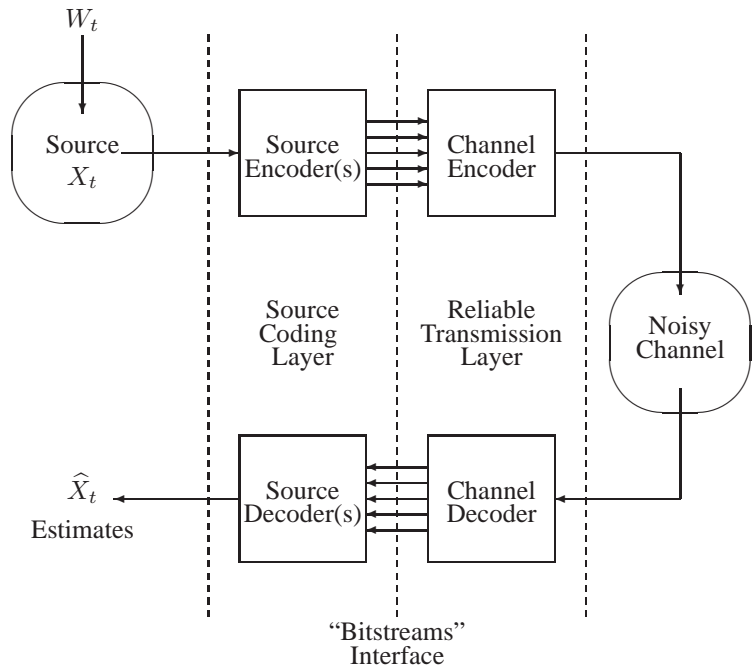

Coding theorems for stable Markov and auto-regressive processes under
mean-squared-error distortion are now well established in the
literature~\cite{Gallager, Gray70}. We consider real-valued Markov
processes, modeled as
\begin{equation} \label{eqn:process}
X_{t+1} = \lambda X_t + W_t
\end{equation}
where $\{W_t\}_{t \geq 0}$ are white and $X_0$ is an independent
initial condition uniformly distributed on
$[-\frac{\Omega_0}{2},+\frac{\Omega_0}{2}]$ where $\Omega_0 > 0$ 
is small. The essence of the problem is depicted in
Fig.~\ref{fig:mainproblem}: to minimize the rate of the encoding while
maintaining an adequate fidelity of reconstruction. Once the source
has been compressed, the resulting bitstreams can presumably be
reliably communicated across a wide variety of noisy channels.

The infinite-horizon source-coding problem is to design a source code
minimizing the rate $R$ used to encode the process while keeping the
reconstruction close to the original source in an average sense
$\lim_{n \rightarrow \infty} \frac{1}{n} \sum_{t=1}^n E[|X_t -
\widehat{X}_t|^\eta]$. {\em The key issue is that any given
  encoder/decoder system must have a bounded delay when used over a
  fixed-rate noiseless channel.} The encoder is not permitted to look
into the entire infinite future before committing to an encoding for
$\widehat{X}_t$.  To allow the laws of large numbers to work, a
finite, but potentially large, end-to-end delay is allowed between
when the encoder observes $X_t$ and when the decoder emits
$\widehat{X}_t$. However, this delay must remain bounded and not grow
with $t$.

For the stable cases $|\lambda| < 1$, standard block-coding arguments
work since long blocks separated by an intervening block look
relatively independent of each other and are in their stationary
distributions.  The ability to encode blocks in an independent way
also tells us that Shannon's classical sense of $\epsilon$-reliability
also suffices for communicating the encoded bits across a noisy
channel. The study of unstable cases $|\lambda| \geq 1$ is
substantially more difficult since they are neither ergodic nor
stationary and furthermore their variance grows unboundedly with
time. As a result, Gray was able to prove only finite horizon results
for such nonstationary processes and the general infinite-horizon
unstable case has remained essentially open since Gray's 1970 paper
\cite{Gray70}. As he put it:
\begin{quotation}
{\small
It should be emphasized that when the source is non-stationary, the
above theorem is not as powerful as one would like. Specifically, it
does not show that one can code a long sequence by breaking it up into
smaller blocks of length $n$ and use the same code to encode each
block. The theorem is strictly a ``one-shot'' theorem unless the
source is stationary, simply because the blocks $[(k-1)n, kn]$ do not
have the same distribution for unequal $k$ when the source is not
stationary. }
\end{quotation}

On the computational side, Hashimoto and Arimoto gave a parametric
form for computing the $R(d)$ function for unstable auto-regressive
Gaussian processes \cite{Hashimoto80} and mean-square distortion. Toby
Berger gave an explicit coding theorem for an important sub-case, the
marginally unstable Wiener process with $\lambda = 1$, by introducing
an ingenious parallel stream methodology. He noticed that although the
Wiener process is nonstationary, it does have stationary and
independent increments \cite{BergerPaper}. However, Berger's
source-coding theorem said nothing about what is required from a noisy
channel. In his own words:\cite{BergerBook}
\begin{quotation}
{\small
It is worth stressing that we have proved only a source coding theorem 
for the Wiener process, not an information transmission theorem. If
uncorrected channel errors were to occur, even in extremely rare
instances, the user would eventually lose track of the Wiener process
completely. It appears (although it has never been proved) that, even
if a {\em noisy} feedback link were provided, it still would not be
possible to achieve a finite [mean squared error] per letter as $t
\rightarrow \infty$. } 
\end{quotation}

In an earlier conference work \cite{SahaiISIT2001} and the first
author's dissertation \cite{SahaiThesis}, we gave a variable rate
coding theorem that showed the $R(d)$ bound is achievable in the
infinite-horizon case if variable-rate codes are allowed. The question
of whether or not fixed-rate and finite-delay codes could be made to
work was left open, and is resolved here along with a full information
transmission theorem.


\subsection{Asymptotic equivalences and direct reductions}

Beyond the technical issue of fixed or variable rate lies a deeper
question regarding the nature of ``information'' in such processes.
\cite{MitterNewtonKalman} contains an analysis of the traditional
Kalman-Bucy filter in which certain entropic expressions are
identified with the accumulation and dissipation of information within
a filter. No explicit source or channel coding is involved, but the
idea of different kinds of information flows is raised through the
interpretation of certain mutual information quantities. In the
stabilization problem of \cite{ControlPartI}, it is hard to see if any
qualitatively distinct kinds of information are present since to an
external observer, the closed-loop process is stable. 

Similarly, the variable-rate code given earlier in
\cite{SahaiISIT2001, SahaiThesis} does not distinguish between
kinds of information since the same high QoS requirements were imposed
on all bits. However, it was clear that all the bits {\em do not
  require} the same treatment since there are examples in which access
to an additional lower reliability medium can be used to improve
end-to-end performance \cite{SahaiAllerton00, SahaiThesis}. The true
nature of the information within the unstable process was left open
and while exponentially unstable processes certainly appeared to be
accumulating information, there was no way to make this interpretation
precise and quantify the amount of accumulation.

In order to understand the nature of information, this paper builds
upon the ``asymptotic communication problem equivalence'' perspective
introduced at the end of \cite{ControlPartI}. This approach associates
communication problems (e.g.~communicating bits reliably at rate $R$
or communicating iid Gaussian random variables to average distortion
$\leq d$) with the set of channels that are good enough to solve that
problem (e.g.~noisy channels with capacity $C > R$).  This parallels
the ``asymptotic computational problem equivalence'' perspective in
computational complexity theory \cite{Hopcraft} except that the
critical resource shifts from computational operations to noisy
channel uses. The heart of the approach is the use of ``reductions''
that show that a system made to solve one communication problem can be
used as a black box to solve another communication problem. Two
problems are asymptotically equivalent if they can be reduced to each
other.

The equivalence perspective is closely related to the traditional
source/channel separation theorems. The main difference is that
traditional separation theorems give a privileged position to one
communication problem --- reliable bit-transport in the Shannon sense
--- and use reductions in only one direction: from the source to bits.
The ``converse'' direction is usually proved using properties of
mutual information. In \cite{MukulAllerton06, OurDirectConverse}, we
give a direct proof of the ``converse'' for classical problems by
showing the existence of randomized codes that embed iid message bits
into iid seeming source symbols at rate $R$. The embedding is done so
that the bits can be recovered with high probability from 
distorted reconstructions as long as the average distortion on long
blocks stays below the distortion-rate function $D(R)$. Similar
results are obtained for the conditional distortion-rate
function. This equivalence approach to separation theorems considers
the privileged position of reliable bit-transport to be purely a
pedagogical matter.

This paper uses the results from \cite{MukulAllerton06,
  OurDirectConverse} to extend the results of \cite{ControlPartI} from
the control context to the estimation context. We demonstrate that the
problem of communicating an unstable Markov process to within average
distortion $d$ is asymptotically equivalent to a pair of communication
problems: classical reliable bit-transport at a rate $\approx R(d) -
\log_2 |\lambda|$ and anytime-reliable bit-transport at a rate $\approx \log_2
|\lambda|$. This gives a precise interpretation to the nature of information
flows in such processes.

\subsection{Outline}

Section~\ref{sec:mainresults} states the main results of this paper. A
brief numerical example for the Gaussian case is given to illustrate
the behavior of such unstable processes. The proofs follow in the
subsequent sections.

Section~\ref{sec:source_coding} considers lossy source coding for
unstable Markov processes with the driving disturbance $W_t$
constrained to have bounded support. A fixed-rate code at a rate
arbitrarily close to $R(d)$ is constructed by encoding process into
two simultaneous fixed-rate message streams.  The first stream has a
bit-rate arbitrarily close to $\log_2 |\lambda|$ and encodes what is
needed from the past to understand the future. It captures the
information that is accumulating within the unstable process. The
other stream encodes those aspects of the past that are not relevant
to the future and so captures the purely historical aspects of the
unstable process in a way that meets the average distortion
constraint. This second stream can be made to have a rate arbitrarily
close to $R(d) - \log_2 |\lambda|$.

Section~\ref{sec:time_reverse} then examines this historical
information more carefully by looking at the process formally going
backward in time. The $R(d)$ curve for the unstable process is shown
to have a shape that is the stable historical part translated by
$\log_2 |\lambda|$ to account for the unstable accumulation of
information. 

Section~\ref{sec:qos.sufficiency} first reviews the fact that random
codes exist achieving anytime reliability over noisy channels even
without any feedback. Then, for $\eta$-difference distortion measures,
an anytime reliability $> \eta \log_2 |\lambda|$ is shown to be
sufficient to encode the first bitstream of the code of
Section~\ref{sec:source_coding} across a noisy channel. The second
bitstream is shown to only require classical Shannon
$\epsilon$-reliability. This completes the reduction of the
lossy-estimation problem to a two-tiered reliable bit-transportation
problem and resolves the conjecture posed by Berger regarding an
information transmission theorem for unstable processes.

Section~\ref{sec:qos.necessity} tackles the other direction. The
problem of anytime-reliable bit-transport is directly reduced to the
problem of lossy-estimation for a decimated version of the unstable
process. This is done using the ideas in \cite{ControlPartI},
reinterpreted as information-embedding and shows that the higher QoS
requirements for the first stream are unavoidable for these processes.
A second stream of messages is then embedded into the historical
segments of the unstable process and this stream is recovered in the
classical Shannon $\epsilon$-reliable sense. {\em Exponentially
  unstable Markov processes are thus the first nontrivial examples of
  stochastic processes that naturally generate two qualitatively
  distinct kinds of information.}

In Section~\ref{sec:gaussians}, the results are then extended to cover
the Gauss-Markov case with the usual squared-error
distortion. Although the proofs are given in terms of Gaussian
processes and squared error, the results actually generalize to any
$\eta$-distortion as well as driving noise distributions $W$ that have
at least an exponentially decaying tail.

This paper focuses throughout on scalar Markov processes for
clarity. It is possible to extend all the arguments to cover the
general autoregressive moving average (ARMA) case. The techniques used
to cover the ARMA case are discussed in the control context in
\cite{ControlPartII} where a state-space formulation is used. A brief
discussion of how to apply those techniques is present here in
Section~\ref{sec:vector}.

\section{Main results} \label{sec:mainresults}

\subsection{Performance bound in the limit of large delays} \label{sec:performance_bounds}

To define $R(d)$ for unstable Markov processes, the infinite-horizon
problem is viewed as the limit of a sequence of finite-horizon
problems: 
\begin{definition}
Given the scalar Markov source given by (\ref{eqn:process}), the {\em
finite $n$-horizon } version of the source is defined to be the
random variables $X_0^{n-1} = (X_0,X_1,\ldots,X_{n-1})$.
\end{definition}
\begin{definition}
The {\em $\eta-$distortion} measure is $\rho(X_i,\widehat{X}_i) = |X_i -
\widehat{X}|^\eta$. It is an additive distortion measure when applied
to blocks. 
\end{definition}

\vspace{0.1in}

The standard information-theoretic rate-distortion function for the
finite-horizon problem using $\eta$-difference distortion is: 
\begin{equation} \label{eqn:finitehorizonratedistortion}
 R_n^X(d) = \inf_{\{{\cal P}(Y_0^{n-1} | X_0^{n-1}) : \frac{1}{n}\sum_{i=0}^{n-1}
 E\left[|X_i - Y_i|^\eta \right] \leq d\}} \frac{1}{n} I(X_0^{n-1};Y_0^{n-1})
\end{equation}

We can consider the block $X_1^n$ as a single vector-valued random
variable $\vec{X}$. The $R_n^X(d)$ defined by
(\ref{eqn:finitehorizonratedistortion}) is related to
$R_1^{\vec{X}}(d)$ by $R_n^X(d) = \frac{1}{n} R_1^{\vec{X}}(nd)$ with
the distortion measure on $\vec{X}$ given by
$\rho(\vec{X},\widehat{\vec{X}}) = \sum_{i=0}^{n-1} |X_i -
\widehat{X}|^\eta$.

The infinite-horizon case is then defined as a limit: 
\begin{equation} \label{eqn:infinitehorizonratedistortion}
 R_{\infty}^X(d) = \liminf_{n \rightarrow \infty} R_n^X(d)
\end{equation}

The distortion-rate function $D_{\infty}^X(R)$ is also defined in
the same manner, except that the mutual-information is fixed and the
distortion is what is infimized.

\subsection{The stable counterpart to the unstable process}
It is insightful to consider what the stable counterpart to this
unstable process would be. There is a natural choice, just formally
turn the recurrence relationship around and flip the order of
time. This gives the ``backwards in time process'' governed by the
recursion 
\begin{equation} \label{eqn:reverse_evolution}
\overleftarrow{X}_{t} = \lambda^{-1} \overleftarrow{X}_{t+1} -
\lambda^{-1}W_{t}.
\end{equation}

This is purely a formal reversal. In place of an initial condition
$X_0$, it is natural to consider a $\overleftarrow{X}_n$ for
some time $n$ and then consider time going backwards from there. Since
$|\lambda^{-1}| < 1,$ this is a stable Markov process and falls under
the classical theorems of \cite{Gray70}. 

\subsection{Encoders and decoders} 

For notational convenience, time is synchronized between the source
and the channel. Thus both delay and rate can be measured against
either source symbols or channel uses.

\begin{definition}
A {\em discrete time channel} is a probabilistic system with an
input. At every time step $t$, it takes an input $a_t \in {\cal A}$
and produces an output $c_t \in {\cal C}$ with
probability $p(C_t|a_1^t,c_1^{t-1})$ where the notation $a_1^t$ is
shorthand for the sequence $(a_1, a_2, \ldots, a_t)$. In general, the
current channel output is allowed to depend on all inputs so far as
well as on past outputs.

The channel is {\em memoryless} if conditioned on $a_t$, the random
variable $C_t$ is independent of any other random variable in the
system that occurs at time $t$ or earlier.  So all that needs to be
specified is $p_t(C_t|a_t)$. The channel is memoryless and stationary if
$p_t(C_t|a_t) = p(C_t|a_t)$ for all times $t$.  
\end{definition}
\vspace{0.1in}

\begin{definition} \label{def:encoder} A {\em rate $R$ source-encoder
    ${\cal E}_s$} is a sequence of maps $\{{\cal E}_{s,i}\}$. The
  range of each map is a single bit $b_i \in \{0,1\}$ if it is a pure
  source encoder and is from the channel input alphabet ${\cal A}$ if
  it is a joint source-channel encoder. The $i$-th map takes as input
  the available source symbols $X_1^{\lfloor \frac{i}{R} \rfloor}$.

  Similarly, a {\em rate $R$ channel-encoder ${\cal E}_c$ without
    feedback} is a sequence of maps $\{{\cal E}_{c,t}\}$. The range of
  each map is the channel input alphabet $\cal A$. The $t$-th map
  takes as input the available message bits $B_1^{\lfloor Rt
    \rfloor}$.

{\em Randomized encoders} also have access to random variables
denoting the common randomness available in the system. This common
randomness is independent of the source and channel. 
\end{definition}
\vspace{0.1in}

\begin{definition} \label{def:decoder} A {\em delay $\phi$ rate $R$
    source-decoder} is a sequence of maps $\{{\cal D}_{s,t}\}$. The
  range of each map is just an estimate $\widehat{X}_t$ for the $t$-th
  source symbol. For pure source decoders, the $t$-th map takes as
  input the available message bits $B_1^{\lfloor (t+ \phi)R
    \rfloor}$. For joint source-channel decoders, it takes as input
  the available channel outputs $C_1^{t + \phi}$. Either way, it can
  see $\phi$ time units beyond the time when the desired source symbol
  first had a chance to impact its inputs.

Similarly, a {\em delay $\phi$ rate $R$ channel-decoder} is a sequence
of maps $\{{\cal D}_{c,i}\}$. The range of each map is just an
estimate $\widehat{B}_i$ for the $i$-th bit taken from $\{0,1\}$. The
$i$-th map takes as input the available channel outputs $C_1^{\lceil 
\frac{i}{R}\rceil + \phi}$ which means that it can see $\phi$ time
units beyond the time when the desired message bit first had a chance
to impact the channel inputs.

{\em Randomized decoders} also have access to the random variables
denoting common randomness.  
\end{definition}
\vspace{0.1in}

\begin{figure}
\begin{center}
\setlength{\unitlength}{3000sp}%
\begingroup\makeatletter\ifx\SetFigFont\undefined%
\gdef\SetFigFont#1#2#3#4#5{%
  \reset@font\fontsize{#1}{#2pt}%
  \fontfamily{#3}\fontseries{#4}\fontshape{#5}%
  \selectfont}%
\fi\endgroup%
\begin{picture}(8124,3289)(439,-2594)
\thinlines
{\color[rgb]{0,0,0}\put(451,-661){\vector( 1, 0){8100}}
}%
{\color[rgb]{0,0,0}\put(451,-511){\line( 1,-2){150}}
}%
{\color[rgb]{0,0,0}\put(751,-511){\line( 1,-2){150}}
}%
{\color[rgb]{0,0,0}\put(1051,-511){\line( 1,-2){150}}
}%
{\color[rgb]{0,0,0}\put(1351,-511){\line( 1,-2){150}}
}%
{\color[rgb]{0,0,0}\put(1651,-511){\line( 1,-2){150}}
}%
{\color[rgb]{0,0,0}\put(1951,-511){\line( 1,-2){150}}
}%
{\color[rgb]{0,0,0}\put(2251,-511){\line( 1,-2){150}}
}%
{\color[rgb]{0,0,0}\put(2551,-511){\line( 1,-2){150}}
}%
{\color[rgb]{0,0,0}\put(2851,-511){\line( 1,-2){150}}
}%
{\color[rgb]{0,0,0}\put(3151,-511){\line( 1,-2){150}}
}%
{\color[rgb]{0,0,0}\put(3451,-511){\line( 1,-2){150}}
}%
{\color[rgb]{0,0,0}\put(3751,-511){\line( 1,-2){150}}
}%
{\color[rgb]{0,0,0}\put(4051,-511){\line( 1,-2){150}}
}%
{\color[rgb]{0,0,0}\put(4351,-511){\line( 1,-2){150}}
}%
{\color[rgb]{0,0,0}\put(4651,-511){\line( 1,-2){150}}
}%
{\color[rgb]{0,0,0}\put(4951,-511){\line( 1,-2){150}}
}%
{\color[rgb]{0,0,0}\put(5251,-511){\line( 1,-2){150}}
}%
{\color[rgb]{0,0,0}\put(5551,-511){\line( 1,-2){150}}
}%
{\color[rgb]{0,0,0}\put(5851,-511){\line( 1,-2){150}}
}%
{\color[rgb]{0,0,0}\put(6151,-511){\line( 1,-2){150}}
}%
{\color[rgb]{0,0,0}\put(6451,-511){\line( 1,-2){150}}
}%
{\color[rgb]{0,0,0}\put(6751,-511){\line( 1,-2){150}}
}%
{\color[rgb]{0,0,0}\put(7051,-511){\line( 1,-2){150}}
}%
{\color[rgb]{0,0,0}\put(7351,-511){\line( 1,-2){150}}
}%
{\color[rgb]{0,0,0}\put(7651,-511){\line( 1,-2){150}}
}%
{\color[rgb]{0,0,0}\put(7951,-511){\line( 1,-2){150}}
}%
{\color[rgb]{0,0,0}\put(1951,-2311){\vector( 1, 0){2250}}
}%
{\color[rgb]{0,0,0}\put(3001,-1261){\vector( 0,-1){450}}
}%
{\color[rgb]{0,0,0}\put(3601,-1261){\vector( 0,-1){450}}
}%
{\color[rgb]{0,0,0}\put(4201,-1261){\vector( 0,-1){450}}
}%
{\color[rgb]{0,0,0}\put(4801,-1261){\vector( 0,-1){450}}
}%
{\color[rgb]{0,0,0}\put(5401,-1261){\vector( 0,-1){450}}
}%
{\color[rgb]{0,0,0}\put(6001,-1261){\vector( 0,-1){450}}
}%
{\color[rgb]{0,0,0}\put(6601,-1261){\vector( 0,-1){450}}
}%
{\color[rgb]{0,0,0}\put(7201,-1261){\vector( 0,-1){450}}
}%
{\color[rgb]{0,0,0}\put(7801,-1261){\vector( 0,-1){450}}
}%
{\color[rgb]{0,0,0}\put(6751,464){\vector( 0,-1){450}}
}%
{\color[rgb]{0,0,0}\put(7351,464){\vector( 0,-1){450}}
}%
{\color[rgb]{0,0,0}\put(7951,464){\vector( 0,-1){450}}
}%
{\color[rgb]{0,0,0}\put(751,464){\vector( 0,-1){450}}
}%
{\color[rgb]{0,0,0}\put(1351,464){\vector( 0,-1){450}}
}%
{\color[rgb]{0,0,0}\put(1951,464){\vector( 0,-1){450}}
}%
{\color[rgb]{0,0,0}\put(2551,464){\vector( 0,-1){450}}
}%
{\color[rgb]{0,0,0}\put(3151,464){\vector( 0,-1){450}}
}%
{\color[rgb]{0,0,0}\put(3751,464){\vector( 0,-1){450}}
}%
{\color[rgb]{0,0,0}\put(4351,464){\vector( 0,-1){450}}
}%
{\color[rgb]{0,0,0}\put(4951,464){\vector( 0,-1){450}}
}%
{\color[rgb]{0,0,0}\put(5551,464){\vector( 0,-1){450}}
}%
{\color[rgb]{0,0,0}\put(6151,464){\vector( 0,-1){450}}
}%
\put(451,-436){\makebox(0,0)[b]{\smash{\SetFigFont{7}{6}{\rmdefault}{\mddefault}{\updefault}{\color[rgb]{0,0,0}$A_{1}$}%
}}}
\put(751,-436){\makebox(0,0)[b]{\smash{\SetFigFont{7}{6}{\rmdefault}{\mddefault}{\updefault}{\color[rgb]{0,0,0}$A_{2}$}%
}}}
\put(1051,-436){\makebox(0,0)[b]{\smash{\SetFigFont{7}{6}{\rmdefault}{\mddefault}{\updefault}{\color[rgb]{0,0,0}$A_{3}$}%
}}}
\put(1351,-436){\makebox(0,0)[b]{\smash{\SetFigFont{7}{6}{\rmdefault}{\mddefault}{\updefault}{\color[rgb]{0,0,0}$A_{4}$}%
}}}
\put(1651,-436){\makebox(0,0)[b]{\smash{\SetFigFont{7}{6}{\rmdefault}{\mddefault}{\updefault}{\color[rgb]{0,0,0}$A_{5}$}%
}}}
\put(1951,-436){\makebox(0,0)[b]{\smash{\SetFigFont{7}{6}{\rmdefault}{\mddefault}{\updefault}{\color[rgb]{0,0,0}$A_{6}$}%
}}}
\put(2251,-436){\makebox(0,0)[b]{\smash{\SetFigFont{7}{6}{\rmdefault}{\mddefault}{\updefault}{\color[rgb]{0,0,0}$A_{7}$}%
}}}
\put(2551,-436){\makebox(0,0)[b]{\smash{\SetFigFont{7}{6}{\rmdefault}{\mddefault}{\updefault}{\color[rgb]{0,0,0}$A_{8}$}%
}}}
\put(2851,-436){\makebox(0,0)[b]{\smash{\SetFigFont{7}{6}{\rmdefault}{\mddefault}{\updefault}{\color[rgb]{0,0,0}$A_{9}$}%
}}}
\put(3151,-436){\makebox(0,0)[b]{\smash{\SetFigFont{7}{6}{\rmdefault}{\mddefault}{\updefault}{\color[rgb]{0,0,0}$A_{10}$}%
}}}
\put(3451,-436){\makebox(0,0)[b]{\smash{\SetFigFont{7}{6}{\rmdefault}{\mddefault}{\updefault}{\color[rgb]{0,0,0}$A_{11}$}%
}}}
\put(3751,-436){\makebox(0,0)[b]{\smash{\SetFigFont{7}{6}{\rmdefault}{\mddefault}{\updefault}{\color[rgb]{0,0,0}$A_{12}$}%
}}}
\put(4051,-436){\makebox(0,0)[b]{\smash{\SetFigFont{7}{6}{\rmdefault}{\mddefault}{\updefault}{\color[rgb]{0,0,0}$A_{13}$}%
}}}
\put(4351,-436){\makebox(0,0)[b]{\smash{\SetFigFont{7}{6}{\rmdefault}{\mddefault}{\updefault}{\color[rgb]{0,0,0}$A_{14}$}%
}}}
\put(4651,-436){\makebox(0,0)[b]{\smash{\SetFigFont{7}{6}{\rmdefault}{\mddefault}{\updefault}{\color[rgb]{0,0,0}$A_{15}$}%
}}}
\put(4951,-436){\makebox(0,0)[b]{\smash{\SetFigFont{7}{6}{\rmdefault}{\mddefault}{\updefault}{\color[rgb]{0,0,0}$A_{16}$}%
}}}
\put(5251,-436){\makebox(0,0)[b]{\smash{\SetFigFont{7}{6}{\rmdefault}{\mddefault}{\updefault}{\color[rgb]{0,0,0}$A_{17}$}%
}}}
\put(5551,-436){\makebox(0,0)[b]{\smash{\SetFigFont{7}{6}{\rmdefault}{\mddefault}{\updefault}{\color[rgb]{0,0,0}$A_{18}$}%
}}}
\put(5851,-436){\makebox(0,0)[b]{\smash{\SetFigFont{7}{6}{\rmdefault}{\mddefault}{\updefault}{\color[rgb]{0,0,0}$A_{19}$}%
}}}
\put(6151,-436){\makebox(0,0)[b]{\smash{\SetFigFont{7}{6}{\rmdefault}{\mddefault}{\updefault}{\color[rgb]{0,0,0}$A_{20}$}%
}}}
\put(6451,-436){\makebox(0,0)[b]{\smash{\SetFigFont{7}{6}{\rmdefault}{\mddefault}{\updefault}{\color[rgb]{0,0,0}$A_{21}$}%
}}}
\put(6751,-436){\makebox(0,0)[b]{\smash{\SetFigFont{7}{6}{\rmdefault}{\mddefault}{\updefault}{\color[rgb]{0,0,0}$A_{22}$}%
}}}
\put(7051,-436){\makebox(0,0)[b]{\smash{\SetFigFont{7}{6}{\rmdefault}{\mddefault}{\updefault}{\color[rgb]{0,0,0}$A_{23}$}%
}}}
\put(7351,-436){\makebox(0,0)[b]{\smash{\SetFigFont{7}{6}{\rmdefault}{\mddefault}{\updefault}{\color[rgb]{0,0,0}$A_{24}$}%
}}}
\put(7651,-436){\makebox(0,0)[b]{\smash{\SetFigFont{7}{6}{\rmdefault}{\mddefault}{\updefault}{\color[rgb]{0,0,0}$A_{25}$}%
}}}
\put(7951,-436){\makebox(0,0)[b]{\smash{\SetFigFont{7}{6}{\rmdefault}{\mddefault}{\updefault}{\color[rgb]{0,0,0}$A_{26}$}%
}}}
\put(601,-1111){\makebox(0,0)[b]{\smash{\SetFigFont{7}{6}{\rmdefault}{\mddefault}{\updefault}{\color[rgb]{0,0,0}$C_{1}$}%
}}}
\put(901,-1111){\makebox(0,0)[b]{\smash{\SetFigFont{7}{6}{\rmdefault}{\mddefault}{\updefault}{\color[rgb]{0,0,0}$C_{2}$}%
}}}
\put(1201,-1111){\makebox(0,0)[b]{\smash{\SetFigFont{7}{6}{\rmdefault}{\mddefault}{\updefault}{\color[rgb]{0,0,0}$C_{3}$}%
}}}
\put(1501,-1111){\makebox(0,0)[b]{\smash{\SetFigFont{7}{6}{\rmdefault}{\mddefault}{\updefault}{\color[rgb]{0,0,0}$C_{4}$}%
}}}
\put(1801,-1111){\makebox(0,0)[b]{\smash{\SetFigFont{7}{6}{\rmdefault}{\mddefault}{\updefault}{\color[rgb]{0,0,0}$C_{5}$}%
}}}
\put(2101,-1111){\makebox(0,0)[b]{\smash{\SetFigFont{7}{6}{\rmdefault}{\mddefault}{\updefault}{\color[rgb]{0,0,0}$C_{6}$}%
}}}
\put(2401,-1111){\makebox(0,0)[b]{\smash{\SetFigFont{7}{6}{\rmdefault}{\mddefault}{\updefault}{\color[rgb]{0,0,0}$C_{7}$}%
}}}
\put(2701,-1111){\makebox(0,0)[b]{\smash{\SetFigFont{7}{6}{\rmdefault}{\mddefault}{\updefault}{\color[rgb]{0,0,0}$C_{8}$}%
}}}
\put(3001,-1111){\makebox(0,0)[b]{\smash{\SetFigFont{7}{6}{\rmdefault}{\mddefault}{\updefault}{\color[rgb]{0,0,0}$C_{9}$}%
}}}
\put(3301,-1111){\makebox(0,0)[b]{\smash{\SetFigFont{7}{6}{\rmdefault}{\mddefault}{\updefault}{\color[rgb]{0,0,0}$C_{10}$}%
}}}
\put(3601,-1111){\makebox(0,0)[b]{\smash{\SetFigFont{7}{6}{\rmdefault}{\mddefault}{\updefault}{\color[rgb]{0,0,0}$C_{11}$}%
}}}
\put(3901,-1111){\makebox(0,0)[b]{\smash{\SetFigFont{7}{6}{\rmdefault}{\mddefault}{\updefault}{\color[rgb]{0,0,0}$C_{12}$}%
}}}
\put(4201,-1111){\makebox(0,0)[b]{\smash{\SetFigFont{7}{6}{\rmdefault}{\mddefault}{\updefault}{\color[rgb]{0,0,0}$C_{13}$}%
}}}
\put(4501,-1111){\makebox(0,0)[b]{\smash{\SetFigFont{7}{6}{\rmdefault}{\mddefault}{\updefault}{\color[rgb]{0,0,0}$C_{14}$}%
}}}
\put(4801,-1111){\makebox(0,0)[b]{\smash{\SetFigFont{7}{6}{\rmdefault}{\mddefault}{\updefault}{\color[rgb]{0,0,0}$C_{15}$}%
}}}
\put(5101,-1111){\makebox(0,0)[b]{\smash{\SetFigFont{7}{6}{\rmdefault}{\mddefault}{\updefault}{\color[rgb]{0,0,0}$C_{16}$}%
}}}
\put(5401,-1111){\makebox(0,0)[b]{\smash{\SetFigFont{7}{6}{\rmdefault}{\mddefault}{\updefault}{\color[rgb]{0,0,0}$C_{17}$}%
}}}
\put(5701,-1111){\makebox(0,0)[b]{\smash{\SetFigFont{7}{6}{\rmdefault}{\mddefault}{\updefault}{\color[rgb]{0,0,0}$C_{18}$}%
}}}
\put(6001,-1111){\makebox(0,0)[b]{\smash{\SetFigFont{7}{6}{\rmdefault}{\mddefault}{\updefault}{\color[rgb]{0,0,0}$C_{19}$}%
}}}
\put(6301,-1111){\makebox(0,0)[b]{\smash{\SetFigFont{7}{6}{\rmdefault}{\mddefault}{\updefault}{\color[rgb]{0,0,0}$C_{20}$}%
}}}
\put(6601,-1111){\makebox(0,0)[b]{\smash{\SetFigFont{7}{6}{\rmdefault}{\mddefault}{\updefault}{\color[rgb]{0,0,0}$C_{21}$}%
}}}
\put(6901,-1111){\makebox(0,0)[b]{\smash{\SetFigFont{7}{6}{\rmdefault}{\mddefault}{\updefault}{\color[rgb]{0,0,0}$C_{22}$}%
}}}
\put(7201,-1111){\makebox(0,0)[b]{\smash{\SetFigFont{7}{6}{\rmdefault}{\mddefault}{\updefault}{\color[rgb]{0,0,0}$C_{23}$}%
}}}
\put(7501,-1111){\makebox(0,0)[b]{\smash{\SetFigFont{7}{6}{\rmdefault}{\mddefault}{\updefault}{\color[rgb]{0,0,0}$C_{24}$}%
}}}
\put(7801,-1111){\makebox(0,0)[b]{\smash{\SetFigFont{7}{6}{\rmdefault}{\mddefault}{\updefault}{\color[rgb]{0,0,0}$C_{25}$}%
}}}
\put(8101,-1111){\makebox(0,0)[b]{\smash{\SetFigFont{7}{6}{\rmdefault}{\mddefault}{\updefault}{\color[rgb]{0,0,0}$C_{26}$}%
}}}
\put(3076,-2536){\makebox(0,0)[b]{\smash{\SetFigFont{7}{6}{\rmdefault}{\mddefault}{\updefault}{\color[rgb]{0,0,0}fixed delay $d=7$}%
}}}
\put(6001,-2011){\makebox(0,0)[b]{\smash{\SetFigFont{8}{6}{\rmdefault}{\mddefault}{\updefault}{\color[rgb]{0,0,0}$\widehat{B}_6$}%
}}}
\put(7201,-2011){\makebox(0,0)[b]{\smash{\SetFigFont{8}{6}{\rmdefault}{\mddefault}{\updefault}{\color[rgb]{0,0,0}$\widehat{B}_8$}%
}}}
\put(3001,-2011){\makebox(0,0)[b]{\smash{\SetFigFont{8}{6}{\rmdefault}{\mddefault}{\updefault}{\color[rgb]{0,0,0}$\widehat{B}_1$}%
}}}
\put(3601,-2011){\makebox(0,0)[b]{\smash{\SetFigFont{8}{6}{\rmdefault}{\mddefault}{\updefault}{\color[rgb]{0,0,0}$\widehat{B}_2$}%
}}}
\put(4201,-2011){\makebox(0,0)[b]{\smash{\SetFigFont{8}{6}{\rmdefault}{\mddefault}{\updefault}{\color[rgb]{0,0,0}$\widehat{B}_3$}%
}}}
\put(4801,-2011){\makebox(0,0)[b]{\smash{\SetFigFont{8}{6}{\rmdefault}{\mddefault}{\updefault}{\color[rgb]{0,0,0}$\widehat{B}_4$}%
}}}
\put(5401,-2011){\makebox(0,0)[b]{\smash{\SetFigFont{8}{6}{\rmdefault}{\mddefault}{\updefault}{\color[rgb]{0,0,0}$\widehat{B}_5$}%
}}}
\put(6601,-2011){\makebox(0,0)[b]{\smash{\SetFigFont{8}{6}{\rmdefault}{\mddefault}{\updefault}{\color[rgb]{0,0,0}$\widehat{B}_7$}%
}}}
\put(7801,-2011){\makebox(0,0)[b]{\smash{\SetFigFont{8}{6}{\rmdefault}{\mddefault}{\updefault}{\color[rgb]{0,0,0}$\widehat{B}_9$}%
}}}
\put(751,539){\makebox(0,0)[b]{\smash{\SetFigFont{8}{6}{\rmdefault}{\mddefault}{\updefault}{\color[rgb]{0,0,0}$B_1$}%
}}}
\put(1351,539){\makebox(0,0)[b]{\smash{\SetFigFont{8}{6}{\rmdefault}{\mddefault}{\updefault}{\color[rgb]{0,0,0}$B_2$}%
}}}
\put(1951,539){\makebox(0,0)[b]{\smash{\SetFigFont{8}{6}{\rmdefault}{\mddefault}{\updefault}{\color[rgb]{0,0,0}$B_3$}%
}}}
\put(2551,539){\makebox(0,0)[b]{\smash{\SetFigFont{8}{6}{\rmdefault}{\mddefault}{\updefault}{\color[rgb]{0,0,0}$B_4$}%
}}}
\put(3151,539){\makebox(0,0)[b]{\smash{\SetFigFont{8}{6}{\rmdefault}{\mddefault}{\updefault}{\color[rgb]{0,0,0}$B_5$}%
}}}
\put(3751,539){\makebox(0,0)[b]{\smash{\SetFigFont{8}{6}{\rmdefault}{\mddefault}{\updefault}{\color[rgb]{0,0,0}$B_6$}%
}}}
\put(4351,539){\makebox(0,0)[b]{\smash{\SetFigFont{8}{6}{\rmdefault}{\mddefault}{\updefault}{\color[rgb]{0,0,0}$B_7$}%
}}}
\put(4951,539){\makebox(0,0)[b]{\smash{\SetFigFont{8}{6}{\rmdefault}{\mddefault}{\updefault}{\color[rgb]{0,0,0}$B_8$}%
}}}
\put(5551,539){\makebox(0,0)[b]{\smash{\SetFigFont{8}{6}{\rmdefault}{\mddefault}{\updefault}{\color[rgb]{0,0,0}$B_9$}%
}}}
\put(6151,539){\makebox(0,0)[b]{\smash{\SetFigFont{8}{6}{\rmdefault}{\mddefault}{\updefault}{\color[rgb]{0,0,0}$B_{10}$}%
}}}
\put(6751,539){\makebox(0,0)[b]{\smash{\SetFigFont{8}{6}{\rmdefault}{\mddefault}{\updefault}{\color[rgb]{0,0,0}$B_{11}$}%
}}}
\put(7351,539){\makebox(0,0)[b]{\smash{\SetFigFont{8}{6}{\rmdefault}{\mddefault}{\updefault}{\color[rgb]{0,0,0}$B_{12}$}%
}}}
\put(7951,539){\makebox(0,0)[b]{\smash{\SetFigFont{8}{6}{\rmdefault}{\mddefault}{\updefault}{\color[rgb]{0,0,0}$B_{13}$}%
}}}
\end{picture}
\end{center}
\caption{The timeline in a rate $\frac{1}{2}$ delay $7$ channel code. Both the
encoder and decoder must be causal so $A_i$ and $\widehat{B}_i$ are
functions only of quantities to the left of them on the timeline. If
noiseless feedback is available, the $A_i$ can also have an explicit
functional dependence on the $C_1^{i-1}$ that lie to the left on the
timeline.}
\label{fig:timeline}
\end{figure}
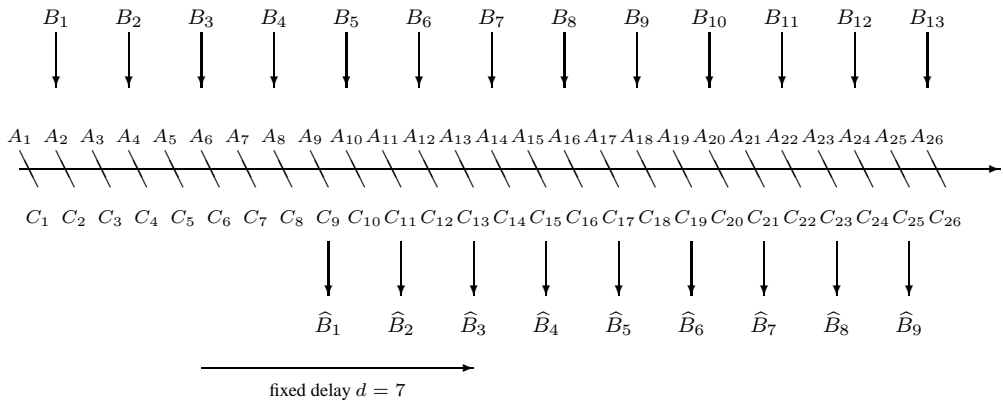

The timeline is illustrated in Fig.~\ref{fig:timeline} for channel
coding and a similar timeline holds for either pure source coding or
joint source-channel coding.

For a specific channel, the maximum rate achievable for a given sense
of reliable communication is called the associated capacity. Shannon's
classical $\epsilon$-reliability requires that for a suitably large
end-to-end delay $\phi$, the probability of error on each bit is
below a specified $\epsilon$.

\begin{definition} \label{def:anytime}
A rate $R$ {\em anytime communication system} over a noisy channel is
a single channel encoder ${\cal E}_c$ and decoder ${\cal
  D}_{c}^{\phi}$ {\em family} for all end-to-end delays $\phi$.

A rate $R$ communication system achieves {\em anytime reliability}
$\alpha$ if there exists a constant $K$ such that:
\begin{equation} \label{eqn:anytime_req}
 {\cal P}(\widehat{B}_1^i(t) \neq B_1^i) \leq K 2^{-\alpha (t-\frac{i}{R})}
\end{equation}
holds for every $i$. The probability is taken over the channel noise,
the message bits $B$, and all of the common randomness available in the
system. If (\ref{eqn:anytime_req}) holds for every possible
realization of the message bits $B$, then we say that the system achieves
{\em uniform anytime reliability} $\alpha$.

Communication systems that achieve {\em anytime reliability} are
called {\em anytime codes} and similarly for {\em uniform anytime
codes}. 
\end{definition}

The important thing to understand about anytime reliability is that it
is not considered to be a proxy used to study encoder/decoder
complexity as traditional reliability functions often are
\cite{Gallager}. Instead, the anytime reliability parameter $\alpha$
indexes a sense of reliable transmission for a bitstream in which the
probability of bit error tends to zero exponentially as time goes on.

\subsection{Main results}

\begin{figure}
\begin{center}
\setlength{\unitlength}{2500sp}%
\begingroup\makeatletter\ifx\SetFigFont\undefined%
\gdef\SetFigFont#1#2#3#4#5{%
  \reset@font\fontsize{#1}{#2pt}%
  \fontfamily{#3}\fontseries{#4}\fontshape{#5}%
  \selectfont}%
\fi\endgroup%
\begin{picture}(7292,1224)(374,-823)
\thinlines
\put(4801,-811){\framebox(1800,1200){}}
\put(5701,-136){\makebox(0,0)[b]{\smash{\SetFigFont{7}{14.4}{\rmdefault}{\mddefault}{\updefault}Source}}}
\put(5701,-361){\makebox(0,0)[b]{\smash{\SetFigFont{7}{14.4}{\rmdefault}{\mddefault}{\updefault}Decoder}}}
\put(1201,-811){\framebox(1800,1200){}}
\put(601,-211){\vector( 1, 0){600}}
\put(3001,-361){\vector( 1, 0){1800}}
\put(3001,-61){\vector( 1, 0){1800}}
\put(6601,-211){\vector( 1, 0){600}}
\put(751,-136){\makebox(0,0)[b]{\smash{\SetFigFont{7}{14.4}{\rmdefault}{\mddefault}{\updefault}$\{X_t\}$}}}
\put(2101,-136){\makebox(0,0)[b]{\smash{\SetFigFont{7}{14.4}{\rmdefault}{\mddefault}{\updefault}Source}}}
\put(2101,-361){\makebox(0,0)[b]{\smash{\SetFigFont{7}{14.4}{\rmdefault}{\mddefault}{\updefault}Encoder}}}
\put(7051,-136){\makebox(0,0)[b]{\smash{\SetFigFont{7}{14.4}{\rmdefault}{\mddefault}{\updefault}$\{\widehat{X}_t\}$}}}
\put(3901,194){\makebox(0,0)[b]{\smash{\SetFigFont{7}{14.4}{\rmdefault}{\mddefault}{\updefault}Coded Bitstreams}}}
\put(3901,-11){\makebox(0,0)[b]{\smash{\SetFigFont{7}{14.4}{\rmdefault}{\mddefault}{\updefault}$R_1$}}}
\put(3901,-556){\makebox(0,0)[b]{\smash{\SetFigFont{7}{14.4}{\rmdefault}{\mddefault}{\updefault}$R_2$}}}
\end{picture}
\end{center}
\caption{The source-coding problem of translating the source into two
  simultaneous bitstreams of fixed rates $R_1$ and $R_2$. End-to-end 
  delay is permitted but must remain bounded for all time. The goal is
  to get $R_1 \approx \log_2 |\lambda|$ and $R_2 \approx R(d) - \log_2
  |\lambda|$.} 
\label{fig:sourcecodingproblem}
\end{figure}
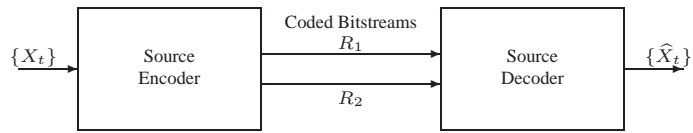

The first result concerns the source coding problem illustrated in
Fig.~\ref{fig:sourcecodingproblem} for unstable Markov
processes with bounded-support driving noise.

\begin{theorem} \label{thm:boundedsupportfixedrate} Assume both the
  source encoder and source decoder can be randomized. Given the unstable
  ($|\lambda| > 1$) scalar Markov process from (\ref{eqn:process})
  driven by independent noise  $\{W_t\}_{t\geq 0}$ with bounded
  support, it is possible to encode the process to average fidelity
  $E[|X_i - \widehat{X}_i|^\eta]$ arbitrarily close to $d$ using two
  fixed-rate bitstreams and a suitably high end-to-end delay $\phi$.

  The first stream (called the {\em checkpoint stream}) can be made to
  have rate $R_1$ arbitrarily close to $\log_2 |\lambda|$ while the second
  (called the {\em historical stream}) can have rate $R_2$ arbitrarily
  close to $R_\infty^X(d) - \log_2 |\lambda|$. 
\end{theorem}
{\em Proof:} See Section~\ref{sec:source_coding}.
\vspace{0.1in}

In a very real sense, the first stream in
Theorem~\ref{thm:boundedsupportfixedrate} represents an initial
description of the process to some fidelity, while the second
represents a refinement of the description \cite{Rimoldi}. These two
descriptions turn out to be qualitatively different when it comes to
communicating them across a noisy channel.

\begin{theorem} \label{thm:anytime_sufficient}

  Suppose that a communication system provides uniform anytime
  reliability $\alpha > \eta \log_2 |\lambda|$ for the checkpoint message
  stream at bit-rate $R_1$. Then given sufficient end-to-end delay
  $\phi$, it is possible to reconstruct the checkpoints to arbitrarily
  high fidelity in the $\eta$-distortion sense.
\end{theorem}
{\em Proof:} See Section~\ref{sec:checkpointqossufficient}.
\vspace{0.1in}
\begin{theorem} \label{thm:history_sufficient}
  Suppose that a communication system can reliably communicate
  message bits meeting any bit-error probability $\epsilon$ given a  
  long enough delay. Then, that communication system can be used to
  reliably communicate the historical information message stream
  generated by the fixed-rate source code of
  Theorem~\ref{thm:boundedsupportfixedrate} in that the expected
  end-to-end distortion can be made arbitrarily close to the
  distortion achieved by the code over a noiseless channel.
\end{theorem}
{\em Proof:} See Section~\ref{sec:historyQoS}.
\vspace{0.1in}

The Gauss-Markov case with mean squared error is covered by corollaries:
\begin{corollary} \label{cor:gaussianfixedrate} Assume both the
  encoder and decoder are randomized and the finite end-to-end delay
  $\phi$ can be chosen to be arbitrarily large. Given an unstable ($|\lambda|
  > 1$) scalar Markov process (\ref{eqn:process}) driven by iid
  Gaussian noise $\{W_t\}_{t\geq 0}$ with zero mean and variance
  $\sigma^2$, it is possible to encode the process to average fidelity
  $E[|X_i - \widehat{X}_i|^2]$ arbitrarily close to $d$ using two
  fixed-rate bitstreams.

  The checkpoint stream can be made to have rate $R_1$ arbitrarily
  close to $\log_2 |\lambda|$ while the historical stream can have rate $R_2$
  arbitrarily close to $R_\infty^X(d) - \log_2 |\lambda|$. 
\end{corollary}
{\em Proof:} See Section~\ref{sec:gaussianfixedrate}.
\vspace{0.1in}
\begin{corollary} \label{cor:sufficient_gaussian}
Suppose that a communication system provides us with the ability to
carry two message streams. One at rate $R_1 > \log_2 |\lambda|$ with
uniform anytime reliability $\alpha > 2 \log_2 |\lambda|$, and another
with classical Shannon reliability at rate $R_2 > R_\infty^X(d) -
\log_2 |\lambda|$ where $R_\infty^X(d)$ is the rate-distortion
function for an unstable Gauss-Markov process with unstable gain
$|\lambda| \geq 1$ and squared-error distortion. 

Then it is possible to successfully transport the two-stream code of
Corollary~\ref{cor:gaussianfixedrate} using this communication system
by picking a sufficiently large end-to-end delay $\phi$. The mean
squared error of the resulting system will be as close to $d$ as
desired.
\end{corollary}
{\em Proof:} See Section~\ref{sec:gaussianchannelsufficiency}.
\vspace{0.1in}

Theorems \ref{thm:anytime_sufficient} and \ref{thm:history_sufficient}
together with the source code of
Theorem~\ref{thm:boundedsupportfixedrate} combine to establish a
reduction of the $d$-lossy joint source/channel coding problem to the
problem of communicating message bits at rate $R(d)$ over the same
channel, wherein a substream of message bits 
at rate $\approx \log_2 |\lambda|$ is given an anytime reliability of at least
$\eta \log_2 |\lambda|$. This reduction is in the sense of Section VII of
\cite{ControlPartI}: any channel that is good enough to solve the
second pair of problems is good enough to solve the first problem.

The asymptotic relationship between the forward and backward
rate-distortion functions is captured in the following theorem.

\begin{theorem} \label{thm:backwards_bound} Let $X$ be the unstable
  Markov process of (\ref{eqn:process}) with $|\lambda|>1$ and let the
  stable backwards-in-time version from (\ref{eqn:reverse_evolution})
  be denoted $\overleftarrow{X}$. Assume that the iid driving noise
  $W$ has a Riemann-integrable density $f_W$ and there exists a
  constant $K$ so that $E[|\sum_{i=1}^t \lambda^{-i} W_{i}|^\eta] \leq
  K$ for all $t\geq 1$. Furthermore for the purpose of calculating
  the rate-distortion functions below, assume that for the
  backwards-in-time version is initialized with $\overleftarrow{X}_n =
  0$. Let $Q_{\Delta}$ be the uniform quantizer that maps its input to
  the nearest neighbor of the form $k\Delta$ for integer $k$. 

\begin{equation} \label{eqn:backwards_bound}
R_\infty^{\overleftarrow{X}}(d) =_{(a)}
\lim_{\Delta \rightarrow 0} \lim_{n\rightarrow \infty}
R_n^{\overleftarrow{X}|Q_{\Delta}(\overleftarrow{X}_0)}(d)  
=_{(b)} 
\lim_{n\rightarrow \infty}
R_n^{\overleftarrow{X}|\overleftarrow{X}_{0}}(d)
=_{(c)} 
R_\infty^{X}(d) - \log_2 |\lambda|.
\end{equation}
or expressed in terms of distortion-rate functions for $R > \log_2 |\lambda|$: 
$$D_\infty^X(R) = D_\infty^{\overleftarrow{X}}(R - \log_2 |\lambda|).$$

This implies that the process generally undergoes a phase transition
from infinite to bounded average distortion at the critical rate
$\log_2 |\lambda|$. 
\end{theorem}
{\em Proof: }See Section~\ref{sec:thmbackwardsbound}.
\vspace{0.1in}

Notice that there are no explicitly infinite distortions in the
original setup of the problem. Consequently, the appearance of infinite
distortions is interesting as is the abrupt transition from infinite
to finite distortions around the critical rate of $\log_2
|\lambda|$. This abrupt transition gives a further indication that
there is something fundamentally nonclassical about the rate $\log_2
|\lambda|$ information inside the process. 

To make this precise, a converse is needed. Classical rate-distortion
results only point out that the mutual information across the
communication system must be at least $R(d)$ on average. However, as
\cite{VerduHanCapacity} points out, having enough mutual information
is not enough to guarantee a reliable-transport capacity since the
situation here is not stationary and ergodic. The following theorem
gives the converse, but adds an intuitively required additional
condition that the probability of excess average distortion over any
long enough segment can be made as small as desired. 

\begin{theorem} \label{thm:strict_converse}
Consider the unstable process given by (\ref{eqn:process}) with the
iid driving noise $W$ having a Riemann-integrable density
$f_W$ satisfying the conditions of Theorem~\ref{thm:backwards_bound}.

Suppose there exists a family (indexed by window size $n$) of
joint source-channel codes $({\cal E}_s,{\cal D}_s)$) so that the
$n$-th member of the family has reconstructions that satisfy
\begin{equation} \label{eqn:distortion_condition}
E[|X_{kn} - \widehat{X}_{kn}|^\eta] \leq d
\end{equation}
for every positive integer $k$. Furthermore, assume the family
collectively also satisfies 
\begin{equation} \label{eqn:extracondition}
\lim_{n \rightarrow \infty} 
\sup_{\tau \geq 0}
{\cal P}(\frac{1}{n}\sum_{i=\tau}^{\tau + n-1} |\widehat{X}_i -
X_i|^\eta > d) = 0
\end{equation}
so that the probability of excess distortion can be made
arbitrarily small on long enough blocks.

Then for any $R_1 < \log_2 |\lambda|, \alpha < \eta \log_2 |\lambda|, R_2 < 
R_\infty^{X}(d) - \log_2 |\lambda|, P_e > 0$, the channel must support the
simultaneous carrying of a bit-rate $R_1$ priority message stream with
anytime reliability $\alpha$ along with a second message stream of
bit-rate $R_2$ with a probability of bit error $\leq P_e$ for some
end-to-end delay $\phi$. 
\end{theorem} 
{\em Proof:} See Section~\ref{sec:necessity}.
\vspace{0.1in}

Note that a Gaussian disturbance $W$ is covered by Theorems
\ref{thm:backwards_bound} and \ref{thm:strict_converse}, even if the
difference distortion measure is not mean squared-error.


\subsection{An example and comparison to the sequential rate
  distortion problem} \label{sec:example} In the case of Gauss-Markov
processes with squared-error distortion, Hashimoto and Arimoto in
\cite{Hashimoto80} give an explicit way of calculating
$R(d)$. Tatikonda in \cite{TatikondaThesis, OurMainLQGPaper} gives a
similar explicit lower bound to the rate required when the
reconstruction $\widehat{X}_t$ is forced to be causal in that it can
only depend on $X_j$ observations for $j \leq t$.

Assuming unit variance for the driving noise $W$ and $\lambda > 1$,
Hashimoto's formula is parametric in terms of the water-filling
parameter $\kappa$ and for the Gauss-Markov case considered here
simplifies to:
\begin{eqnarray}
D(\kappa) & = & \frac{1}{2\pi} \int_{-\pi}^{\pi} \min\left[\kappa,
  \frac{1}{1 - 2\lambda\cos(\omega) + \lambda^2}\right] d\omega,
\nonumber \\
R(\kappa) & = & \log_2 \lambda + 
\frac{1}{2\pi} \int_{-\pi}^{\pi} \max\left[0,
\frac{1}{2} \log_2 \frac{1}{\kappa (1 - 2\lambda\cos(\omega) + \lambda^2)}\right]
  d\omega. \label{eqn:parametricgaussiancalculation}
\end{eqnarray}

The rate-distortion function for the stable counterpart given in
(\ref{eqn:reverse_evolution}) has a water-filling solution that is
identical to \ref{eqn:parametricgaussiancalculation}, except without
the $\log_2 \lambda$ term in the $R(\kappa)$! Thus, in the Gaussian
case with squared error distortion direct calculation
verifies the claim $$R_{\infty}^X(d) = \log_2 \lambda +
R_{\infty}^{\overleftarrow{X}}(d)$$ from Theorem~\ref{thm:backwards_bound}.


For the unstable process, Tatikonda's formula for causal
reconstructions is given by 
\begin{equation}
R_{\mbox{seq}}(d) = \frac{1}{2} \log_2(\lambda^2 + \frac{1}{d}).
\end{equation}

\begin{figure}
\begin{center}
\mbox{\epsfxsize=4in \epsfysize=3in \epsfbox{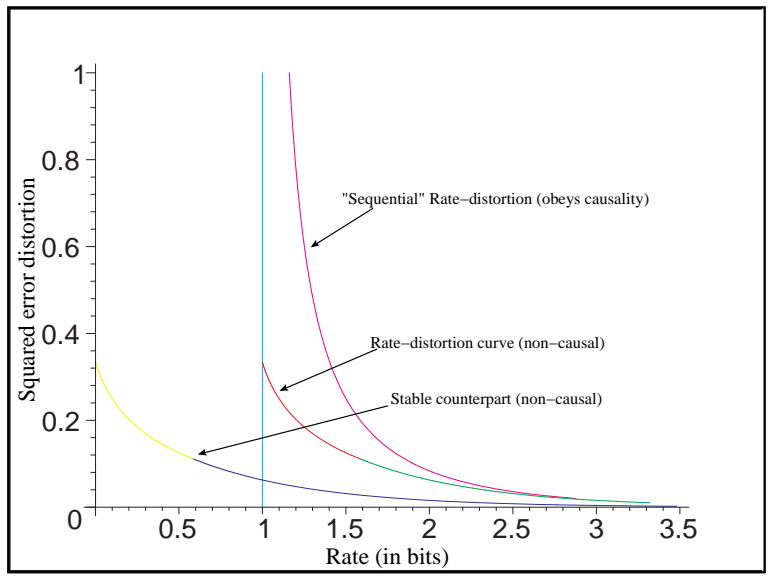}}
\end{center}
\caption{The distortion-rate curves for an unstable Gauss-Markov
process with $\lambda=2$ and its stable backwards-version. The stable and
unstable $D(R)$ curves are related by a simple translation by $1$ bit
per symbol.}
\label{fig:curves}
\end{figure}

Fig.~\ref{fig:curves} shows the distortion-rate frontier for both the
original unstable process and backwards stable process. It is easy to
see that the forward and backward process curves are translations of
each other. In addition, the sequential rate-distortion curve for the
forward process is qualitatively distinct. $D_{\mbox{seq}}(R)$ goes to
infinity as $R \downarrow \log_2 \lambda$ while $D(R)$ approaches a finite
limit.

The results in this paper show that the lower curve for the regular
distortion-rate frontier can be approached arbitrarily closely by
increasing the acceptable (finite) end-to-end delay. This suggests
that it takes some time for the randomness entering the unstable
process through $W$ to sort itself into the two categories of
fundamental accumulation and transient history. The difference in the
resulting distortion is not that significant at high rates, but
becomes unboundedly large as the rate approaches $\log_2 \lambda$.  It
is open whether similar information-embedding theorems similar to
Theorem~\ref{thm:strict_converse} exist that give an operational
meaning to the gap between $R_{seq}(d)$ and $R(d)$. If a communication
system can be used to satisfy distortion $d$ in a causal way, does
that mean the underlying communication resources also must be able to
support messages at this higher rate $R_{seq}(d)$?

\section{Two stream source encoding: approaching $R(d)$} \label{sec:source_coding}

This section proves Theorem~\ref{thm:boundedsupportfixedrate}. 


\subsection{Proof strategy}

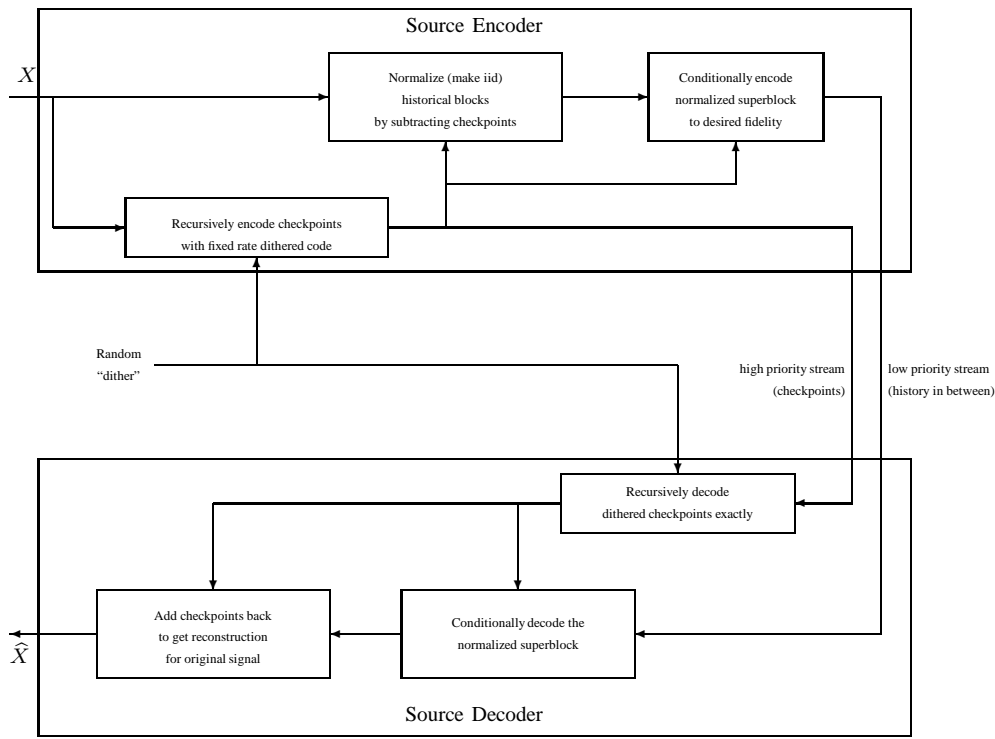
\begin{figure}
\begin{center}
\setlength{\unitlength}{2400sp}%
\begingroup\makeatletter\ifx\SetFigFont\undefined%
\gdef\SetFigFont#1#2#3#4#5{%
  \reset@font\fontsize{#1}{#2pt}%
  \fontfamily{#3}\fontseries{#4}\fontshape{#5}%
  \selectfont}%
\fi\endgroup%
\begin{picture}(9324,7524)(-11,-6823)
\thinlines
{\color[rgb]{0,0,0}\put(5701,-4711){\framebox(2400,600){}}
}%
\put(6901,-4336){\makebox(0,0)[b]{\smash{\SetFigFont{5}{14.4}{\rmdefault}{\mddefault}{\updefault}{\color[rgb]{0,0,0}Recursively decode}%
}}}
\put(6901,-4561){\makebox(0,0)[b]{\smash{\SetFigFont{5}{14.4}{\rmdefault}{\mddefault}{\updefault}{\color[rgb]{0,0,0}dithered checkpoints exactly}%
}}}
{\color[rgb]{0,0,0}\put(  1,-211){\vector( 1, 0){3300}}
}%
{\color[rgb]{0,0,0}\put(4501,-1561){\vector( 0, 1){900}}
}%
{\color[rgb]{0,0,0}\put(3901,-1561){\line( 1, 0){4800}}
\put(8701,-1561){\line( 0,-1){2850}}
\put(8701,-4411){\vector(-1, 0){600}}
}%
{\color[rgb]{0,0,0}\put(301,-2011){\framebox(9000,2700){}}
}%
{\color[rgb]{0,0,0}\put(6601,-661){\framebox(1800,900){}}
}%
{\color[rgb]{0,0,0}\put(5701,-4411){\line(-1, 0){3600}}
\put(2101,-4411){\vector( 0,-1){900}}
}%
{\color[rgb]{0,0,0}\put(901,-5761){\vector(-1, 0){900}}
}%
{\color[rgb]{0,0,0}\put(301,-6811){\framebox(9000,2850){}}
}%
{\color[rgb]{0,0,0}\put(451,-211){\line( 0,-1){1350}}
\put(451,-1561){\vector( 1, 0){750}}
}%
{\color[rgb]{0,0,0}\put(3301,-661){\framebox(2400,900){}}
}%
{\color[rgb]{0,0,0}\put(2551,-2986){\vector( 0, 1){1125}}
}%
{\color[rgb]{0,0,0}\put(1501,-2986){\line( 1, 0){5400}}
\put(6901,-2986){\vector( 0,-1){1125}}
}%
{\color[rgb]{0,0,0}\put(5701,-211){\vector( 1, 0){900}}
}%
{\color[rgb]{0,0,0}\put(7501,-1111){\vector( 0, 1){450}}
\put(7501,-1111){\line(-1, 0){1350}}
\put(6151,-1111){\line(-1, 0){1650}}
}%
{\color[rgb]{0,0,0}\put(1201,-1861){\framebox(2700,600){}}
}%
{\color[rgb]{0,0,0}\put(4051,-5761){\vector(-1, 0){750}}
}%
{\color[rgb]{0,0,0}\put(901,-6211){\framebox(2400,900){}}
}%
{\color[rgb]{0,0,0}\put(5251,-4411){\vector( 0,-1){900}}
}%
{\color[rgb]{0,0,0}\put(4051,-6211){\framebox(2400,900){}}
}%
{\color[rgb]{0,0,0}\put(8401,-211){\line( 1, 0){600}}
\put(9001,-211){\line( 0,-1){5550}}
\put(9001,-5761){\vector(-1, 0){2550}}
}%
\put(3901,-361){\makebox(0,0)[b]{\smash{\SetFigFont{5}{14.4}{\rmdefault}{\mddefault}{\updefault}{\color[rgb]{0,0,0} }%
}}}
\put(4801,464){\makebox(0,0)[b]{\smash{\SetFigFont{8}{14.4}{\rmdefault}{\mddefault}{\updefault}{\color[rgb]{0,0,0}Source Encoder}%
}}}
\put(4801,-6661){\makebox(0,0)[b]{\smash{\SetFigFont{8}{14.4}{\rmdefault}{\mddefault}{\updefault}{\color[rgb]{0,0,0}Source Decoder}%
}}}
\put(  1,-6061){\makebox(0,0)[lb]{\smash{\SetFigFont{8}{14.4}{\rmdefault}{\mddefault}{\updefault}{\color[rgb]{0,0,0}$\widehat{X}$}%
}}}
\put( 76,-61){\makebox(0,0)[lb]{\smash{\SetFigFont{8}{14.4}{\rmdefault}{\mddefault}{\updefault}{\color[rgb]{0,0,0}$X$}%
}}}
\put(8626,-3061){\makebox(0,0)[rb]{\smash{\SetFigFont{5}{14.4}{\rmdefault}{\mddefault}{\updefault}{\color[rgb]{0,0,0}high priority stream}%
}}}
\put(7501,-61){\makebox(0,0)[b]{\smash{\SetFigFont{5}{14.4}{\rmdefault}{\mddefault}{\updefault}{\color[rgb]{0,0,0}Conditionally encode}%
}}}
\put(8626,-3286){\makebox(0,0)[rb]{\smash{\SetFigFont{5}{14.4}{\rmdefault}{\mddefault}{\updefault}{\color[rgb]{0,0,0}(checkpoints)}%
}}}
\put(4501,-511){\makebox(0,0)[b]{\smash{\SetFigFont{5}{14.4}{\rmdefault}{\mddefault}{\updefault}{\color[rgb]{0,0,0}by subtracting checkpoints}%
}}}
\put(1126,-2911){\makebox(0,0)[b]{\smash{\SetFigFont{5}{14.4}{\rmdefault}{\mddefault}{\updefault}{\color[rgb]{0,0,0}Random}%
}}}
\put(1126,-3136){\makebox(0,0)[b]{\smash{\SetFigFont{5}{14.4}{\rmdefault}{\mddefault}{\updefault}{\color[rgb]{0,0,0}``dither''}%
}}}
\put(2551,-1561){\makebox(0,0)[b]{\smash{\SetFigFont{5}{14.4}{\rmdefault}{\mddefault}{\updefault}{\color[rgb]{0,0,0}Recursively encode checkpoints}%
}}}
\put(2551,-1786){\makebox(0,0)[b]{\smash{\SetFigFont{5}{14.4}{\rmdefault}{\mddefault}{\updefault}{\color[rgb]{0,0,0}with fixed rate dithered code}%
}}}
\put(2101,-5611){\makebox(0,0)[b]{\smash{\SetFigFont{5}{14.4}{\rmdefault}{\mddefault}{\updefault}{\color[rgb]{0,0,0}Add checkpoints back}%
}}}
\put(5251,-5686){\makebox(0,0)[b]{\smash{\SetFigFont{5}{14.4}{\rmdefault}{\mddefault}{\updefault}{\color[rgb]{0,0,0}Conditionally decode the}%
}}}
\put(5251,-5911){\makebox(0,0)[b]{\smash{\SetFigFont{5}{14.4}{\rmdefault}{\mddefault}{\updefault}{\color[rgb]{0,0,0}normalized superblock}%
}}}
\put(4501,-61){\makebox(0,0)[b]{\smash{\SetFigFont{5}{14.4}{\rmdefault}{\mddefault}{\updefault}{\color[rgb]{0,0,0}Normalize (make iid)}%
}}}
\put(4501,-286){\makebox(0,0)[b]{\smash{\SetFigFont{5}{14.4}{\rmdefault}{\mddefault}{\updefault}{\color[rgb]{0,0,0}historical blocks}%
}}}
\put(7501,-286){\makebox(0,0)[b]{\smash{\SetFigFont{5}{14.4}{\rmdefault}{\mddefault}{\updefault}{\color[rgb]{0,0,0}normalized superblock}%
}}}
\put(7501,-511){\makebox(0,0)[b]{\smash{\SetFigFont{5}{14.4}{\rmdefault}{\mddefault}{\updefault}{\color[rgb]{0,0,0}to desired fidelity}%
}}}
\put(2101,-5836){\makebox(0,0)[b]{\smash{\SetFigFont{5}{14.4}{\rmdefault}{\mddefault}{\updefault}{\color[rgb]{0,0,0}to get reconstruction}%
}}}
\put(2101,-6061){\makebox(0,0)[b]{\smash{\SetFigFont{5}{14.4}{\rmdefault}{\mddefault}{\updefault}{\color[rgb]{0,0,0}for original signal}%
}}}
\put(9076,-3061){\makebox(0,0)[lb]{\smash{\SetFigFont{5}{14.4}{\rmdefault}{\mddefault}{\updefault}{\color[rgb]{0,0,0}low priority stream}%
}}}
\put(9076,-3286){\makebox(0,0)[lb]{\smash{\SetFigFont{5}{14.4}{\rmdefault}{\mddefault}{\updefault}{\color[rgb]{0,0,0}(history in between)}%
}}}
\end{picture}

\caption{A flowchart showing how to do fixed-rate source coding for
  Markov sources using two streams and how the streams are
  decoded.}\label{fig:twostreams}
\end{center}
\end{figure}

The code for proving Theorem~\ref{thm:boundedsupportfixedrate} is
illustrated in Fig.~\ref{fig:twostreams}. Without loss of generality,
assume $\lambda = |\lambda|  > 1$ to avoid the notational
complication of keeping track of the sign.
\begin{itemize}
\item Look at time in blocks of size $n$ and encode the values of
  endpoints $(X_{kn-1},X_{kn})$ recursively to very high precision
  using rate $n(\log_2 \lambda + \epsilon_1)$ per pair. Each block
  $X_{kn},X_{kn+1},\ldots,X_{(k+1)n-1}$ will have encoded checkpoints
  $(\check{X}_{kn},\check{X}_{kn+n-1})$ at both ends.

 \item Use the encoded checkpoints $\{\check{X}_{kn}\}$ at the start
   of the blocks to transform the process segments in between (the
   history) so that they look like an iid~sequence of finite horizon
   problems $\vec{X}$. 

 \item Use the checkpoints $\{\check{X}_{kn + n -1}\}$ at the end of
   the blocks as side-information to encode the history to fidelity
   $d$ at a rate of $n(R_\infty^X(d) - \log_2 \lambda + \epsilon_2 + o(1))$
   per block. 

 \item ``Stationarize'' the encoding by choosing a random starting
       offset so that no times $t$ are {\em a priori} more vulnerable
       to distortion.
\end{itemize}

The source decoding proceeds in the same manner and first recovers the
checkpoints, and then uses them as known side-information to decode
the history. The two are then recombined to give a reconstruction of
the original source to the desired fidelity. 

The above strategy follows the spirit of Berger's
encoding\cite{BergerPaper}. In Berger's code for the Wiener process,
the first stream's rate is negligible relative to that of the second
stream.  In our case, the first stream's rate is significant and
cannot be averaged away by using large blocks $n$.


The detailed constructions and proof for this theorem are in the next
few subsections, with some technical aspects relegated to the appendices.

\subsection{Recursively encoding checkpoints} \label{sec:checkpointcode}

This section relies on the assumption of bounded
support for the driving noise $|W_t| \leq \frac{\Omega}{2}$, but does
not care about any other property of the $\{W_t\}_{t\geq 0}$ like
independence or stationarity. The details of the distortion measure are
also not important for this section.

\begin{proposition} \label{prop:checkpointfidelity} Given the unstable
  ($\lambda > 1$) scalar Markov process of (\ref{eqn:process})
  driven by noise $\{W_t\}_{t\geq 0}$ with bounded support, and any
  $\Delta > 0$, it is possible to causally and recursively encode
  checkpoints spaced by $n$ so that $|\check{X}_{kn} - X_{kn}| \leq
  \frac{\Delta}{2}$. For any $R_1 > \log_2 \lambda$, this can be done with
  rate $nR_1$ bits per checkpoint by choosing $n$ large
  enough. Furthermore, if an iid~sequence of independent pairs of
  continuous uniform random variables $\{\Theta_i,\Theta_i'\}_{i \geq
    0}$ is available to both the encoder and decoder for dithering,
  the errors $(\check{X}_{kn-1} - X_{kn-1}, \check{X}_{kn} - X_{kn})$
  can be made an iid~sequence of pairs of independent uniform random
  variables on $[-\frac{\Delta}{2},+\frac{\Delta}{2}]$. 
\end{proposition}
\vspace{0.1in}
{\em Proof:} First, consider the initial condition at $X_0$. It can
be quantized to be within an interval of size $\Delta$ by using
$\log_2 \lceil \frac{\Omega_0}{\Delta} \rceil$ bits. 

With a block length of $n$, the successive endpoints are related by:
\begin{equation}\label{eqn:checkpointupdate}
X_{(k+1)n} = \lambda^n X_{kn} + [\lambda^{n-1} \sum_{i=0}^{n-1} \lambda^{-i} W_{kn+i}]
\end{equation}
The second term $[\cdots]$ on the left of (\ref{eqn:checkpointupdate})
can be denoted $\widetilde{W}_k$ and bounded using

\begin{equation} \label{eqn:checkpointjump}
|\widetilde{W}_k| 
= 
|\lambda^{n-1} \sum_{i=0}^{n-1} \lambda^{-i} W_{kn+i} |
\leq
|\lambda^{n-1}| \sum_{i=0}^{n-1} \lambda^{-i} \frac{\Omega}{2} 
<
\lambda^{n} \frac{\Omega}{2(\lambda-1)}.
\end{equation}

Proceed by induction. Assume that $\check{X}_{kn}$ satisfies
$|X_{kn} - \check{X}_{kn}| \leq \frac{\Delta}{2}$. This clearly holds
for $k=0$. Without any further information, it is known that
$X_{(k+1)n}$ must lie within an interval of size $\lambda^n \Delta + \lambda^n
\frac{\Omega}{\lambda-1}$. By using $nR_1'$ bits (where $R_1'$ is chosen to
guarantee an integer $nR_1'$) to encode where the true value lies,
the uncertainty is cut by a factor of $2^{nR_1'}$. To have the
resulting interval of size $\Delta$ or smaller, we must have:
$$\Delta \geq 2^{-nR_1'} \lambda^n (\Delta + \frac{\Omega}{(\lambda-1)}).$$ 

Dividing through by $\Delta 2^{-nR_1'} \lambda^n$ and taking logarithms gives
 $$n(R_1' - \log_2 \lambda) \geq \log_2 (1 + \frac{\Omega}{\Delta (\lambda-1)}).$$

Encoding $\check{X}_{kn - 1}$ given $\check{X}_{kn}$ requires very little
additional rate since $|X_{kn - 1} - \lambda^{-1} \check{X}_{kn}| <
\Omega + \Delta$ and so $\log_2 \lceil \frac{\Omega}{\Delta} + 1
\rceil < \log_2(2 + \frac{\Omega}{\Delta})$ additional bits are good
enough to encode both checkpoints. Putting everything together in
terms of the original $R_1$ gives
\begin{equation} \label{eqn:checkpointRconstraint}
R_1 \geq \max \left(
    \log_2 \lambda + \frac{\log_2 (1 + \frac{\Omega}{\Delta (\lambda-1)}) + \log_2(2+\frac{\Omega}{\Delta})}{n},
    \frac{\log_2 \lceil \frac{\Omega_0}{\Delta} \rceil}{n} \right).
\end{equation}

It is clear from (\ref{eqn:checkpointRconstraint}) that no matter how
small a $\Delta$ we choose, by picking an $n$ large enough the rate
$R_1$ can get as close to $\log_2 \lambda$ as desired. In particular,
picking $n = K (\log_2 \frac{1}{\Delta})^2$ works with large $K$ and
small $\Delta$. 

To get the uniform nature of the final error $\check{X}_{kn} -
X_{kn}$, subtractive dithering can be used \cite{Zamir}. This is
accomplished by adding a small iid~random variable $\Theta_k$,
uniform on $[-\frac{\Delta}{2},+\frac{\Delta}{2}]$, to the $X_{kn}$,
and only then quantizing $(X_{kn}+\Theta_k)$ to resolution $\Delta$.
At the decoder,  $\Theta_k$ is subtracted from the result to get
$\check{X}_{kn}$. Similarly for $\check{X}_{kn-1}$. This results in the
checkpoint error sequence $(X_{kn-1} - \check{X}_{kn-1},X_{kn} -
\check{X}_{kn})$ being iid~uniform pairs over
$[-\frac{\Delta}{2},+\frac{\Delta}{2}]$. These pairs are also
independent of all the $W_t$ and initial condition $X_0$. \hfill
$\Box$ \vspace{0.1in} 

In what follows, we always assume that $\Delta$ is chosen to be of
high fidelity relative to the target distortion $d$ (e.g.~For
squared-error distortion, this means that $\Delta^2 \ll d$.) as well
as small relative to the the initial condition so $\Delta \ll
\Omega_0$.

\subsection{Transforming and encoding the history}

Having dealt with the endpoints, focus attention on the historical
information between them. Here the bounded support assumption is not
needed for the $\{W_t\}$, but the iid~assumption is important. First,
the encoded checkpoints are used to transform the historical
information so that each historical segment looks iid.  Then, it is
shown that these segments can be encoded to the appropriate fidelity
and rate when the decoder has access to the encoded checkpoints as
side information.

\subsubsection{Forward transformation}

The simplest transformation is to effectively restart the process at
every checkpoint and view time going forward. This can be considered
normalizing each of the historical segments $X_{kn}^{(k+1)n-1}$ to
$(\widetilde{X}_{(k,i)}, 0\leq i \leq n-1)$ for $k=0,1,2,\ldots$.

\begin{equation} \label{eqn:forwardtransformation}
\widetilde{X}_{(k,i)} = X_{kn + i} - \lambda^i \check{X}_{kn}
\end{equation}

For each $k$, the block $\widetilde{X}_k = \{\widetilde{X}_{(k,i)}\}_{0 \leq i
  \leq n-1}$ satisfies $\widetilde{X}_{(k,i+1)} = \lambda \widetilde{X}_{(k,i)} +
W_{(k,i)}$. By dithered quantization, the initial condition ($i=0$) of each
block is a uniform random variable of support $\Delta$ that is
independent of all the other random variables in the system. The
initial conditions are iid~across the different $k$. Thus, except for
the initial condition, the blocks $\widetilde{X}_k$ are identically 
distributed to the finite horizon versions of the problem.

Since $\Delta < \Omega_0$, each $\widetilde{X}_k$ block starts with a
tighter initial condition than the original $X$ process did.  Since
the initial condition is uniform, this can be viewed as a genie-aided
version of the original problem where a genie reveals a few bits of
information about the initial condition. Since the initial condition
enters the process dynamics in a linear way and the distortion measure 
$\rho$ depends only on the difference, this implies that the new
process with the smaller initial condition requires no more bits per
symbol to achieve a distortion $d$ than did the original
process. Thus: 
$$R_n^{X}(d) - \frac{1 + \log_2\frac{\Omega_0}{\Delta}}{n} \leq R_n^{\widetilde{X}}(d) \leq R_n^{X}(d)$$
for all $n$ and $d$. So in the limit of large $n$
\begin{equation} \label{eqn:historyislikefinite}
R_\infty^{\widetilde{X}}(d) =
R_\infty^{X}(d).
\end{equation}

In simple terms, the normalized history behaves like the finite
horizon version of the problem when $n$ is large.


\subsubsection{Conditional encoding}
The idea is to encode the normalized history between two checkpoints
conditioned on the ending checkpoint. The decoder has access to the
exact values of these checkpoints through the first bitstream.


For a given $k$, shift the encoded ending checkpoint
$\check{X}_{(k+1)n - 1}$ to 
\begin{equation} \label{eqn:Zdef}
Z_k^q = \check{X}_{(k+1)n - 1} - \lambda^{n-1} \check{X}_{kn}.
\end{equation}
$Z_k^q$ is clearly available at both the encoder and the decoder since
it only depends on the encoded checkpoints. Furthermore, it is clear
that
$$\widetilde{X}_{(k,n-1)} - Z_k^q  = (X_{(k+1)n-1} - \lambda^{n-1}
\check{X}_{kn}) - (\check{X}_{(k+1)n-1}  - \lambda^{n-1}
\check{X}_{kn}) = X_{(k+1)n-1} - \check{X}_{(k+1)n-1}$$
which is a uniform random variable on
$[-\frac{\Delta}{2},+\frac{\Delta}{2}]$. Thus $Z^q_k$ is just a
dithered quantization to $\Delta$ precision of the endpoint
$\widetilde{X}_{(k,n-1)}$.

Define the conditional rate-distortion function
$R_\infty^{X|Z^q,\Theta}(d)$ for the limit of long historical blocks
$\widetilde{X}_{k,0}^{n-1}$ conditioned on their quantized endpoint as
\begin{equation} \label{eqn:conditionaldefinition}
R_\infty^{X|Z^q,\Theta}(d) = \liminf_{n \rightarrow \infty}
\frac{1}{n} \inf_{\{{\cal P}(Y_0^{n-1} | \widetilde{X}_0^{n-1}, Z^q, \Theta) : \frac{1}{n}\sum_{i=0}^{n-1}
 E\left[|\widetilde{X}_i - Y_i|^\eta \right] \leq d\}} \frac{1}{n}
I(\widetilde{X}_0^{n-1};Y_0^{n-1}|Z^q,\Theta).
\end{equation}

\begin{proposition} \label{prop:conditionaldistortion}
 Given an unstable ($\lambda > 1$) scalar Markov process
 $\{\widetilde{X}_{k,t}\}$ obeying (\ref{eqn:process}) and whose
 driving noise satisfies $E[|\sum_{i=1}^t \lambda^{-i} W_{i}|^\eta]
 \leq K$ for all $t\geq 1$ for some constant $K$, together with
 its encoded endpoint $Z_k^q$ obtained by $\Theta$-dithered
 quantization to within a uniform random variable with small support
 $\Delta$, the limiting conditional rate-distortion function 
\begin{equation} \label{eqn:conditionalshift}
R_\infty^{X|Z^q,\Theta}(d) = R_\infty^X(d) - \log_2 \lambda.
\end{equation}
\end{proposition}
\vspace{0.1in}
{\em Proof:} See Appendix~\ref{app:propconditionaldistortion}.
\vspace{0.1in}

The case of driving noise with bounded support clearly satisfies the
conditions of this proposition since geometric sums converge. The
conditional rate-distortion function in
Proposition~\ref{prop:conditionaldistortion} has a corresponding
coding theorem: 

\begin{proposition} \label{prop:historyencoding} 
  Given an unstable ($\lambda > 1$) scalar Markov process $\{X_t\}$
  given by (\ref{eqn:process}) together with its $n$-spaced pairs of
  encoded checkpoints $\{\check{X}\}$ obtained by dithered
  quantization to within iid~uniform random variables with small
  support $\Delta$, for every $\epsilon_4 > 0$ there exists an $M$
  large enough so that a conditional source-code exists that maps a
  length $M$ superblock of the historical information
  $\{\widetilde{X}_k\}_{0 \leq k < M}$ into a superblock $\{T_k\}_{0
    \leq k < M}$ satisfying
\begin{equation} \label{eqn:superblocksatisfies}
\frac{1}{M} \sum_{k=0}^{M-1} \frac{1}{n} \sum_{j=1}^{n}
   E[\rho(\widetilde{X}_{(k,j)}, T_{(k,j)})] \leq d + \epsilon_4.
 \end{equation}
 
 By choosing $n$ large enough, the rate of the superblock code can be
 made as close as desired to  $R_\infty^X(d) - \log_2 \lambda$ if the
 decoder is also assumed to have access to the encoded checkpoints
 $\check{X}_{kn}$. 
\end{proposition}
\vspace{0.1in}
{\em Proof:} $M$ of the $\widetilde{X}_k$ blocks
are encoded together using conditioning on the encoded
checkpoints at the end of each block. The pair $(\widetilde{X}_{k},
Z_k^q)$ have a joint distribution, but are iid~across $k$ by 
the independence properties of the subtractive dither and the driving
noise $W_{(k,i)}$. Furthermore, the $\widetilde{X}_{(k,i)}$ are
bounded and as a result, the all zero reconstruction results in a
bounded distortion on the $\widetilde{X}$ vector that depends on
$n$. Even without the bounded support assumption,
Theorem~\ref{thm:backwards_bound} reveals that there is a 
reconstruction based on the $Z^q_k$ alone that has bounded average
distortion where the bound does not even depend on $n$. 

Since the side information $Z^q_k$ is available at both encoder and
decoder, the classical conditional rate-distortion coding theorems of
\cite{LeinerGray} tell us that there exists a block-length $M(n)$ so
that codes exist satisfying (\ref{eqn:superblocksatisfies}). The rate
can be made arbitrarily close to $R_n^{X|Z^q,\Theta}(d)$. By letting
$n$ get large, Proposition~\ref{prop:conditionaldistortion} reveals
that this rate can be made as close as desired to $R_\infty^X(d) -
\log_2 \lambda$. \hfill $\Box$  

\vspace{0.1in}


\subsection{Putting history together with checkpoints}
The next step is to show how the decoder can combine the two
streams to get the desired rate/distortion performance.

The rate side is immediately obvious since there is $\log_2 \lambda$
from Proposition~\ref{prop:checkpointfidelity} and $R_\infty^X(d) -
\log_2 \lambda$ from Proposition~\ref{prop:historyencoding}. The sum
is as close to $R_\infty^X(d)$ as desired. On the distortion side, the
decoder runs (\ref{eqn:forwardtransformation}) in reverse to get
reconstructions. Suppose that $T_{(k,i)}$ are the encoded transformed
source symbols from the code in
Proposition~\ref{prop:historyencoding}. Then $\widehat{X}_{kn + i} =
T_{(k,i)} + \lambda^i \check{X}_{kn}$ and so $X_{kn + i} -
\widehat{X}_{kn + i} = \widetilde{X}_{(k,i)} - T_{(k,i)}$. Since the
differences are the same, so is the distortion.

\subsection{``Stationarizing'' the code}
The underlying $X_t$ process is non-stationary so there is no hope to
make the encoding truly stationary. However, as it stands, only the
average distortion across each of the $Mn$ length superblocks is close
to $d$ in expectation giving the resulting code a potentially
``cyclostationary'' character. Nothing guarantees that source symbols
at every time will have the same level of expected fidelity. To fix
this, a standard trick can be applied by making the encoding have two
phases:
\begin{itemize}
\item An initialization phase that lasts for a random $T$
  time-steps. $T$ is a random integer chosen uniformly from $0, 1,
  \ldots Mn-1$ based on common randomness available to the encoder and
  decoder. During the first phase, all source symbols are encoded to
  fidelity $\Delta$ recursively using the code of
  Proposition~\ref{prop:checkpointfidelity} with $n=1$.

\item A main phase that applies the two-part code described above
   but starts at time $T+1$. 
\end{itemize}

The extra rate required in the first phase is negligible on average
since it is a one-time cost. This takes a finite amount of time to
drain out through the rate $R_1$ message stream. This time can be
considered an additional delay that must be suffered for everything in
the second phase. Thus it adds to the delay of $n$ required by the
causal recursive code for the checkpoints. The rest of the end-to-end
delay is determined by the total length $Mn$ of the superblock chosen
inside Proposition~\ref{prop:historyencoding}.

Let $d_i$ be such that the original super-block code gives expected
distortion $d_i$ at position $i$ ranging from $0$ to $Mn-1$. It is
known from Proposition~\ref{prop:historyencoding} that $\frac{1}{Mn}
\sum_{i=0}^{Mn-1} d_i \leq d + \epsilon_4$. Because the first phase is
guaranteed to be high fidelity and all other time positions are
randomly and uniformly assigned positions within the superblock of
size $Mn$, the expected distortion $E[|X_i - \widehat{X}_i|^\eta]
\leq d + \epsilon_4$ for every bit position $i$.

The code actually does better than that since the probability of
excess average distortion over a long block is also guaranteed to go
to zero. This property is inherited from the repeated use of
independent conditional rate-distortion codes in the second stream
\cite{LeinerGray}.

This completes the proof of Theorem~\ref{thm:boundedsupportfixedrate}. 

\section{Time-reversal and the essential phase transition} \label{sec:time_reverse}

It is interesting to note that the distortion of the code in the
previous section turns out to be entirely based on the conditional
rate-distortion performance for the historical segments. The
checkpoints merely contribute a $\log_2 \lambda$ term in the rate. 


The nature of historical information in the unstable Markov process
described by (\ref{eqn:process}) can be explored more fully by
transforming the historical blocks going locally backward in time.
The informational distinction between the process going forward and
the purely historical information parallels the concepts of
information production and dissipation explored in the context of the
Kalman Filter \cite{MitterNewtonKalman}. 

First, the original problem is formally decomposed into forward and
backward parts. Then, Theorem~\ref{thm:backwards_bound} is proved.


\subsection{Endpoints and history} \label{sec:twoanalog}

It is useful to think of the original problem as being broken down
into two analog sub-problems:  

\subsubsection{The $n$-endpoint problem} This is the communication of
the process $\{X_{kn}\}$ where each sample arrives every $n$ time
steps and the   samples are related to each other through
(\ref{eqn:checkpointupdate}) with $\widetilde{W}_k$ being iid~and
having the same distribution as $\lambda^{n-1} \sum_{i=0}^{n-1}
\lambda^{-i} W_{i}$. 

This process must be communicated so that $E[|X_{kn} -
  \widehat{X}_{kn}|^\eta] \leq K$ for some performance $K$. This is
  essentially a decimated version of the original problem. 

\subsubsection{The conditional history problem} 
The stable $\overleftarrow{X}$ process defined in
(\ref{eqn:reverse_evolution}) can be viewed in blocks of length
$n$. The conditional history problem is thus the problem of
communicating an iid~sequence of $n$-vectors $\vec{X}_k^{-}  
= (\overleftarrow{X}_{k,1},\ldots,\overleftarrow{X}_{k,n-1})$
conditioned on iid~$Z_k$ that are known perfectly at the encoder and
decoder. The joint distribution of $\vec{X}^{-},Z$ are given by: 
\begin{eqnarray*}
 Z   & = & -\sum_{i=0}^{n-1} \lambda^{-i} W_{i} \\
 \overleftarrow{X}_{n-1} & = & -\lambda^{-1}W_{n-1} \\
 \overleftarrow{X}_{t} & = & \lambda^{-1}\overleftarrow{X}_{t+1} - \lambda^{-1} W_{t} \\
\end{eqnarray*}
where the underlying $\{W_t\}$ are iid. Unrolling the recursion gives 
$\overleftarrow{X}_{t} = -\sum_{i=0}^{n-1-t} \lambda^{-i-1} W_{t+i}$. The $Z$ is
thus effectively the endpoint $Z = \overleftarrow{X}_0$.  The vectors
$\vec{X}^{-}_k$ are made available to the encoder every $n$ time
units along with their corresponding side-information $Z_k$. The goal 
is to communicate these to a receiver that has access to the
side-information $Z_k$ so that $\frac{1}{n}\sum_{i=1}^{n-1} 
E[\rho(\overleftarrow{X}_{k,i},\widehat{X}^{-}_{k,i})] \leq d$ for all
$k$. 


The relevant rate distortion function for the above problem is the
conditional rate-distortion function $R_n^{\overleftarrow{X}|Z}(d)$.
The proof of Theorem~\ref{thm:boundedsupportfixedrate} in the previous
section involves a slightly modified version of the above where the
side-information $Z$ is known only to some quantization precision
$\Delta$. The quantized side-information is $Z^q =
Q_{(\Delta,\Theta)}(Z)$. The relevant conditional rate-distortion
function is $R_n^{\overleftarrow{X}|Z^q,\Theta}(d)$.

\subsubsection{Reductions back to the original problem} It is obvious
how to put these two problems together to construct an unstable $\{X_t\}$
stream: the endpoints problem provides the skeleton and the
conditional history interpolates in between. To reduce the endpoints
problem to the original unstable source communication problem, just use
randomness at the transmitter to sample from the interpolating
distribution and fill in the history.  

To reduce the conditional history problem to the original unstable
source communication problem, just use the iid $Z_k$ to simulate the
endpoints problem and use the interpolating $\vec{X}$ history to fill
out $\{X_t\}$. Because the distortion measure is a difference
distortion measure, the perfectly known endpoint process allows us to
translate everything so that the same average distortion is
attained.

\subsection{Rate-distortion relationships
  proved} \label{sec:thmbackwardsbound} 
Theorem~\ref{thm:backwards_bound} tells us that the unstable
$|\lambda|>1$ Markov processes are nonclassical only as they evolve
into the future.  The historical information is a stable Markov
process that fleshes out the unstable skeleton of the nonstationary
process. This fact also allows a simplification in the code depicted
in Fig.~\ref{fig:twostreams}. Since the side-information does not
impact the rate-distortion curve for the stable historical process,
the encoding of the historical information can be done unconditionally
and on a block-by-block basis. There is no need for superblocks. 

The remainder of this section proves
Theorem~\ref{thm:backwards_bound}. 

{\em Proof: }
\subsubsection{(a)}
It is easy to see that $R_\infty^{\overleftarrow{X}}(d) =
\lim_{n\rightarrow \infty}
R_n^{\overleftarrow{X}|Q_{\Delta}(\overleftarrow{X}_0)}(d)$ since the 
endpoint $\overleftarrow{X}_0$ is distributed like $-\sum_{i=1}^t
\lambda^{-i} W_{i}$ and has a finite $\eta$-th moment by
assumption. By Lemma~\ref{lem:momentboundentropyquantized} (in the
Appendix), the entropy of $Q_{\Delta}(\overleftarrow{X}_0)$ is bounded
below a constant that depends only on the precision $\Delta$. This
finite number is then amortized away as $n \rightarrow \infty$. 

\subsubsection{(b)} Next, we show
\begin{equation}
\lim_{\Delta \rightarrow 0}
R_\infty^{\overleftarrow{X}|Q_{\Delta}(\overleftarrow{X}_0)}(d) =
\lim_{n\rightarrow \infty} R_n^{\overleftarrow{X}|\overleftarrow{X}_0}(d).
\end{equation}

For notational convenience, let $Z^q =
Q_{\Delta}(\overleftarrow{X}_0)$. First,
$R_n^{\overleftarrow{X}|\overleftarrow{X}_0}(d)$ is 
immediately bounded above by $R_n^{\overleftarrow{X}|Z^q}(d)$
since knowledge of $\overleftarrow{X}_0$ exactly is better than 
knowledge of only the quantized $Z^q$. To get a lower bound, imagine a
hypothetical problem that is one time-step longer and consider the
choice between knowing $\overleftarrow{X}_0$ to fine precision
$\Delta$ or knowing $\overleftarrow{X}_{-1}$ exactly.  

\begin{eqnarray*}
R^{{\overleftarrow{X}}_0^{n-1}|\overleftarrow{X}_{-1}}(d) 
&\geq_{(i)}&
R^{{\overleftarrow{X}}_0^{n-1}|\overleftarrow{X}_{-1},Z^q}(d) \\
&\geq_{(ii)}&
R^{{\overleftarrow{X}}_0^{n-1}|\overleftarrow{X}_{-1},Z^q,C_\gamma,G_\delta,W_\delta''}(d) 
\end{eqnarray*}
where (i) and (ii)
above hold since added conditioning can only reduce the conditional
rate-distortion function, and $C_\gamma, G_\delta, W_\delta''$ are
from the following lemma applied to the hypothesized $W_{-1}$ driving
noise.

\begin{lemma} \label{lem:riemannmixture}
Given a random variable $W$ with density $f_W$, arbitrary $1 > \gamma >
0$, there exists a $\delta > 0$ so that it is possible to realize $W$
as
\begin{equation} \label{eqn:noisedecomposition}
W = (1-C_\gamma)(G_\delta+U_\delta) + C_\gamma W_{\delta}''
\end{equation}
where 
\begin{itemize}
 \item $C_\gamma$ is a Bernoulli random variable with probability
   $\gamma$ of being 1.

 \item $U_\delta$ is a continuous uniform random variable on
   $[-\frac{\delta}{2}, +\frac{\delta}{2}]$.

 \item $G_\delta$ and $W_{\delta}''$ are some random variables whose
   distributions depend on $f_W, \delta, \gamma$.
 
 \item $C_\gamma, U_\delta, G_\delta, W_\delta''$ are all independent
   of each other.
 \end{itemize}
\end{lemma}
{\em Proof: }See Appendix~\ref{app:riemannmixtures}.
\vspace{0.1in}

Pick $\gamma$ small and then choose $\Delta \ll \delta$. Notice that
$\overleftarrow{X}_{-1} = \lambda^{-1}\overleftarrow{X}_0 - 
\lambda^{-1}(1-C_\gamma)(G_\delta+U_\delta) + \lambda^{-1}C_\gamma W_\delta''$
where $C_\gamma, U_\delta, G_\delta, W_\delta''$ are independent of
each other as well as the entire vector
${\overleftarrow{X}}_0^{n-1}$. Because the $\{\overleftarrow{X}_t\}$
process is Markov, the impact of the observations
$\overleftarrow{X}_{-1},Z^q,C_\gamma,G_\delta,W_\delta''$ on
the conditional rate-distortion function is factored entirely through
the posterior distribution for $\overleftarrow{X}_{0}$.

There are two cases:
\begin{itemize}
 \item $C_\gamma=1$ The value for $\overleftarrow{X}_{0}$ is entirely
   revealed by the observations. The posterior is a Dirac delta.

 \item $C_\gamma=0$ There are two independent measurements of
   $\overleftarrow{X}_{0}$. The first is the quantization 
   $Z^q$. The second is $\lambda\overleftarrow{X}_{-1} + G_\delta =
   \overleftarrow{X}_{0} - U_\delta$. This is just
   $\overleftarrow{X}_{0}$ blurred by uniform noise.

   It is useful to view them as coming one after the other. After
   seeing $Z^q = Q_{\Delta}(\overleftarrow{X}_{0}) = z_1$,
   the posterior distribution ${\cal P}(\overleftarrow{X}_{0}|Z^q =
   z_1)$ has support only within $[z_1 - \frac{\Delta}{2}, z_1 +
   \frac{\Delta}{2}]$. 

   The distribution ${\cal P}(Z_2 | Z^q = z_1)$ for the second
   observation $Z_2 = \overleftarrow{X}_{0} - U_\delta$ conditioned on
   the first observation has a pair of interesting properties. First,
   it has support only on $[z_1 - \frac{\delta + \Delta}{2}, z_1 +
   \frac{\delta + \Delta}{2}]$. Second, the distribution is uniform
   over the interval $(z_1 - \frac{\delta - \Delta}{2}, z_1 +
   \frac{\delta - \Delta}{2})$ since the ${\cal
     P}(\overleftarrow{X}_{0}|Z^q = z_1)$ has support with total span
   $\Delta \ll \delta$. 

   Consider the posterior ${\cal P}(\overleftarrow{X}_{0}|Z^q =
   z_1,Z_2 = z_2)$ for $z_2 \in (z_1 - \frac{\delta - \Delta}{2}, z_1
   + \frac{\delta - \Delta}{2})$ and apply Bayes rule:
\begin{eqnarray*}
{\cal P}(\overleftarrow{X}_{0} \leq x|Z^q = z_1,Z_2 = z_2)
& = & 
\frac{{\cal P}(\overleftarrow{X}_{0} \leq x, Z_2 = z_2| Z^q = z_1)}{{\cal P}(Z_2 =
  z_2| Z^q = z_1)} \\
& = & 
\delta {\cal P}(\overleftarrow{X}_{0} \leq x, Z_2 = z_2| Z^q = z_1) \\
& = & 
\left(\delta {\cal P}(Z_2 = z_2| Z^q = z_1, \overleftarrow{X}_{0} \leq
  x) \right)  {\cal P}(\overleftarrow{X}_{0}
\leq x | Z^q = z_1) \\
& = &  {\cal P}(\overleftarrow{X}_{0} \leq x | Z^q = z_1).
\end{eqnarray*}
So if it lands in this region, the second observation is
useless. Notice that $U_\delta \in (\frac{\delta - 2\Delta}{2},
\frac{\delta - 2\Delta}{2})$ forces the second observation to be
inside this region. Thus the second observation is useless with
probability at least $(1-\gamma)\frac{\delta - 2\Delta}{\delta}$
regardless of what the actual ${\overleftarrow{X}}_0^{n-1}$ are.
\end{itemize}

Define a new hypothetical observation $Z'$ that with probability
$(1-\gamma)\frac{\delta - 2\Delta}{\delta}$ is just equal to $Z^q$ and
is equal to $\overleftarrow{X}_{0}$ otherwise. The above tells us that
this is a more powerful observation than than the original 
$(\overleftarrow{X}_{-1},Z^q,C_\gamma,G_\delta,W_\delta'')$. Thus
\begin{eqnarray*}
R^{{\overleftarrow{X}}_0^{n-1}|\overleftarrow{X}_{-1}}(d) 
& \geq &
R^{{\overleftarrow{X}}_0^{n-1}|Z'}(d) \\
& = & 
(1-\gamma)\frac{\delta - 2\Delta}{\delta}
R^{{\overleftarrow{X}}_0^{n-1}|Z^q}(d) + 
\left(1-(1-\gamma)\frac{\delta - 2\Delta}{\delta} \right)
R^{{\overleftarrow{X}}_0^{n-1}|\overleftarrow{X}_{0}}(d) \\
& \geq & 
(1-\gamma)\frac{\delta - 2\Delta}{\delta}
R^{{\overleftarrow{X}}_0^{n-1}|Z^q}(d).
\end{eqnarray*}
Simple algebra then reveals that
\begin{eqnarray*}
R_{n}^{\overleftarrow{X}|Z^q}
& = &
\frac{1}{n} R^{{\overleftarrow{X}}_0^{n-1}|Z^q}(d) \\
& \leq &
\frac{\delta R^{{\overleftarrow{X}}_0^{n-1}|\overleftarrow{X}_{-1}}(d)}
{n (1-\gamma) {\delta - 2\Delta}} \\
& = & 
\frac{\delta (n+1)}{n (1-\gamma) {\delta - 2\Delta}} 
R_{n+1}^{\overleftarrow{X}|Z}(d).
\end{eqnarray*}
Taking the limits of $n \rightarrow \infty, \frac{\Delta}{\delta}
\rightarrow 0, \delta \rightarrow 0, \gamma \rightarrow 0$ establishes
the desired result. 

Notice that an identical argument works to show that 
$$\lim_{\Delta \rightarrow 0} \lim_{n\rightarrow \infty} R_n^{X|Z^q}(d) 
= 
\lim_{n\rightarrow \infty}
R_n^{X|X_{n}}(d)$$
for the forward unstable process. It does not matter if it is
conditioned on the exact endpoint or a finely quantized version of
it. Notice also that the argument is unchanged if the quantization was
dithered rather than undithered.

\subsubsection{(c)} This follows almost immediately from
(\ref{eqn:conditionalshift}) from
Proposition~\ref{prop:conditionaldistortion}. The only remaining task
is to show that $$R_\infty^{\overleftarrow{X}|Z}(d) = R_\infty^{X|Z}(d).$$

It is clear that the iid $\{Z_k\}$ in the ``conditional
history'' problem are just scaled-down (by a factor of $\lambda^{-(n-1)}$)
versions of the $\{\widetilde{W}_k\}$ from the ``endpoints'' problem. The
forward $\vec{X}_k = (X_{k,1},\ldots,X_{k,n-1})$ can be recovered
using a simple translation of $\vec{X}^{-}_k$ by the 
vector $(Z_k,\lambda Z_k,\ldots,\lambda^{n-1}Z_k)$ since
\begin{eqnarray*}
X_t 
& = & \sum_{i=0}^{t-1} \lambda^{t-i-1}W_i \\
& = & \sum_{i=0}^{n-1} \lambda^{t-i-1}W_i  
- \sum_{i=t}^{n-1}  \lambda^{t-i-1}W_i \\
& = & \lambda^{t-1}\sum_{i=0}^{n-1} \lambda^{-i}W_i  
- \sum_{i=0}^{n-1-t} \lambda^{-i-1}W_{t+i} \\
& = & \lambda^{t-1}Z + \overleftarrow{X}_t.
\end{eqnarray*}

Similarly, the conditional history problem can be recovered from the
forward one by another simple translation of $\vec{X}_k$ by the 
vector $(-\lambda^{-(n-1)}Z_k,\ldots,-\lambda^{-1}Z_k,-Z_k)$.

Thus, the problem of encoding the conditional history to distortion
$d$ conditioned on its endpoints is the same whether we are
considering the unstable forward or stable backwards processes. 

\subsubsection{Phase transition}
At rates strictly less than $\log_2 \lambda$, the distortion for the
original $X$ process is necessarily infinite. This is shown in 
Lemma~\ref{lem:anytime_necessity} where finite distortion implies  
the ability to carry $\approx \log_2 \lambda$ bits through the communication
medium. \hfill $\Box$

\section{Quality of service requirements for communicating unstable
  processes: sufficiency} \label{sec:qos.sufficiency}

In Section~\ref{sec:anytime}, the sense of anytime reliability is
reviewed from \cite{ControlPartI} and related to classical results on
sequential coding for noisy channels. Then in
Section~\ref{sec:checkpointqossufficient}, anytime reliable
communication is shown to be sufficient for protecting the encoding of
the checkpoint process, thereby proving
Theorem~\ref{thm:anytime_sufficient}. Finally in
Section~\ref{sec:historyQoS}, it is shown that it is sufficient to
communicate the historical information using traditional Shannon 
$\epsilon$-reliability, thereby proving Theorem~\ref{thm:history_sufficient}.

\subsection{Anytime reliability} \label{sec:anytime} It should be
clear that the encoding given for the checkpoint process in
Section~\ref{sec:checkpointcode} is very sensitive to bit errors since
it is decoded recursively in a way that propagates errors in an
unbounded fashion. To block this propagation of errors, the channel
code must guarantee not only that every bit eventually is received
correctly, but that this happens fast enough. This is what motivates
the definition of anytime reliability given in
Definition~\ref{def:anytime}. The relationship of anytime reliability
to classical concepts of error exponents as well as bounds are given
in \cite{OurUpperBoundPaper, ControlPartI}.

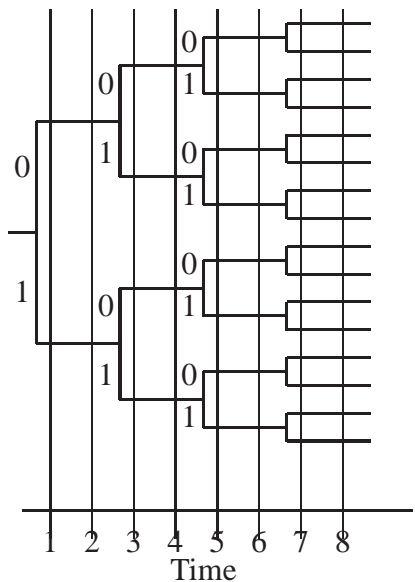
\begin{figure}
\begin{center}
\setlength{\unitlength}{2300sp}%
\begingroup\makeatletter\ifx\SetFigFont\undefined%
\gdef\SetFigFont#1#2#3#4#5{%
  \reset@font\fontsize{#1}{#2pt}%
  \fontfamily{#3}\fontseries{#4}\fontshape{#5}%
  \selectfont}%
\fi\endgroup%
\begin{picture}(4384,6162)(279,-5311)
\thinlines
\put(451,-4561){\line( 1, 0){4200}}
\put(751,839){\line( 0,-1){5700}}
\put(1201,839){\line( 0,-1){5700}}
\put(1651,839){\line( 0,-1){5700}}
\put(2101,839){\line( 0,-1){5700}}
\put(2551,839){\line( 0,-1){5700}}
\put(3001,839){\line( 0,-1){5700}}
\put(3451,839){\line( 0,-1){5700}}
\put(3901,839){\line( 0,-1){5700}}
\thicklines
\put(301,-1561){\line( 1, 0){300}}
\put(601,-1561){\line( 0, 1){1200}}
\put(601,-361){\line( 1, 0){900}}
\put(601,-1561){\line( 0,-1){1200}}
\put(601,-2761){\line( 1, 0){900}}
\put(1501,-2761){\line( 0,-1){600}}
\put(1501,-3361){\line( 1, 0){900}}
\put(1501,-2761){\line( 0, 1){600}}
\put(1501,-2161){\line( 1, 0){900}}
\put(2401,-2161){\line( 0,-1){300}}
\put(2401,-2461){\line( 1, 0){900}}
\put(2401,-2161){\line( 0, 1){300}}
\put(2401,-1861){\line( 1, 0){900}}
\put(2401,-3361){\line( 0,-1){300}}
\put(2401,-3661){\line( 1, 0){900}}
\put(2401,-3361){\line( 0, 1){300}}
\put(2401,-3061){\line( 1, 0){900}}
\put(2401,-961){\line( 0,-1){300}}
\put(2401,-1261){\line( 1, 0){900}}
\put(2401,-961){\line( 0, 1){300}}
\put(2401,-661){\line( 1, 0){900}}
\put(2401,239){\line( 0,-1){300}}
\put(2401,-61){\line( 1, 0){900}}
\put(2401,239){\line( 0, 1){300}}
\put(2401,539){\line( 1, 0){900}}
\put(1501,-361){\line( 0, 1){600}}
\put(1501,239){\line( 1, 0){900}}
\put(1501,-361){\line( 0,-1){600}}
\put(1501,-961){\line( 1, 0){900}}
\put(3301,539){\line( 0, 1){150}}
\put(3301,689){\line( 1, 0){900}}
\put(3301,539){\line( 0,-1){150}}
\put(3301,389){\line( 1, 0){900}}
\put(3301,-61){\line( 0, 1){150}}
\put(3301, 89){\line( 1, 0){900}}
\put(3301,-61){\line( 0,-1){150}}
\put(3301,-211){\line( 1, 0){900}}
\put(3301,-661){\line( 0, 1){150}}
\put(3301,-511){\line( 1, 0){900}}
\put(3301,-661){\line( 0,-1){150}}
\put(3301,-811){\line( 1, 0){900}}
\put(3301,-1261){\line( 0, 1){150}}
\put(3301,-1111){\line( 1, 0){900}}
\put(3301,-1261){\line( 0,-1){150}}
\put(3301,-1411){\line( 1, 0){900}}
\put(3301,-1861){\line( 0, 1){150}}
\put(3301,-1711){\line( 1, 0){900}}
\put(3301,-1861){\line( 0,-1){150}}
\put(3301,-2011){\line( 1, 0){900}}
\put(3301,-2461){\line( 0, 1){150}}
\put(3301,-2311){\line( 1, 0){900}}
\put(3301,-2461){\line( 0,-1){150}}
\put(3301,-2611){\line( 1, 0){900}}
\put(3301,-3061){\line( 0, 1){150}}
\put(3301,-2911){\line( 1, 0){900}}
\put(3301,-3061){\line( 0,-1){150}}
\put(3301,-3211){\line( 1, 0){900}}
\put(3301,-3661){\line( 0, 1){150}}
\put(3301,-3511){\line( 1, 0){900}}
\put(3301,-3661){\line( 0,-1){150}}
\put(3301,-3811){\line( 1, 0){900}}
\put(451,-961){\makebox(0,0)[b]{\smash{\SetFigFont{12}{7.2}{\rmdefault}{\mddefault}{\updefault}0}}}
\put(451,-2311){\makebox(0,0)[b]{\smash{\SetFigFont{12}{7.2}{\rmdefault}{\mddefault}{\updefault}1}}}
\put(1351,-61){\makebox(0,0)[b]{\smash{\SetFigFont{12}{7.2}{\rmdefault}{\mddefault}{\updefault}0}}}
\put(1351,-811){\makebox(0,0)[b]{\smash{\SetFigFont{12}{7.2}{\rmdefault}{\mddefault}{\updefault}1}}}
\put(1351,-2461){\makebox(0,0)[b]{\smash{\SetFigFont{12}{7.2}{\rmdefault}{\mddefault}{\updefault}0}}}
\put(1351,-3211){\makebox(0,0)[b]{\smash{\SetFigFont{12}{7.2}{\rmdefault}{\mddefault}{\updefault}1}}}
\put(2251,389){\makebox(0,0)[b]{\smash{\SetFigFont{12}{7.2}{\rmdefault}{\mddefault}{\updefault}0}}}
\put(2251,-61){\makebox(0,0)[b]{\smash{\SetFigFont{12}{7.2}{\rmdefault}{\mddefault}{\updefault}1}}}
\put(2251,-811){\makebox(0,0)[b]{\smash{\SetFigFont{12}{7.2}{\rmdefault}{\mddefault}{\updefault}0}}}
\put(2251,-1261){\makebox(0,0)[b]{\smash{\SetFigFont{12}{7.2}{\rmdefault}{\mddefault}{\updefault}1}}}
\put(2251,-2011){\makebox(0,0)[b]{\smash{\SetFigFont{12}{7.2}{\rmdefault}{\mddefault}{\updefault}0}}}
\put(2251,-2461){\makebox(0,0)[b]{\smash{\SetFigFont{12}{7.2}{\rmdefault}{\mddefault}{\updefault}1}}}
\put(2251,-3211){\makebox(0,0)[b]{\smash{\SetFigFont{12}{7.2}{\rmdefault}{\mddefault}{\updefault}0}}}
\put(2251,-3661){\makebox(0,0)[b]{\smash{\SetFigFont{12}{7.2}{\rmdefault}{\mddefault}{\updefault}1}}}
\put(1201,-5011){\makebox(0,0)[b]{\smash{\SetFigFont{12}{7.2}{\rmdefault}{\mddefault}{\updefault}2}}}
\put(1651,-5011){\makebox(0,0)[b]{\smash{\SetFigFont{12}{7.2}{\rmdefault}{\mddefault}{\updefault}3}}}
\put(2101,-5011){\makebox(0,0)[b]{\smash{\SetFigFont{12}{7.2}{\rmdefault}{\mddefault}{\updefault}4}}}
\put(2551,-5011){\makebox(0,0)[b]{\smash{\SetFigFont{12}{7.2}{\rmdefault}{\mddefault}{\updefault}5}}}
\put(3001,-5011){\makebox(0,0)[b]{\smash{\SetFigFont{12}{7.2}{\rmdefault}{\mddefault}{\updefault}6}}}
\put(3451,-5011){\makebox(0,0)[b]{\smash{\SetFigFont{12}{7.2}{\rmdefault}{\mddefault}{\updefault}7}}}
\put(3901,-5011){\makebox(0,0)[b]{\smash{\SetFigFont{12}{7.2}{\rmdefault}{\mddefault}{\updefault}8}}}
\put(2401,-5311){\makebox(0,0)[b]{\smash{\SetFigFont{12}{7.2}{\rmdefault}{\mddefault}{\updefault}Time}}}
\put(751,-5011){\makebox(0,0)[b]{\smash{\SetFigFont{12}{7.2}{\rmdefault}{\mddefault}{\updefault}1}}}
\end{picture}
\end{center}
\caption{A channel encoder viewed as a tree. At every integer time,
  each path of the tree has a channel input symbol. The path taken
  down the tree is determined by the message bits to be sent. Infinite
  trees have no intrinsic target delay and bit/path estimates can get
  better as time goes on.}
\label{fig:encodertree}
\end{figure}

Here, the focus is on the case where there is no explicit feedback
of channel outputs. Consider maximum-likelihood decoding \cite{ForneyML} or
sequential-decoding \cite{JelinekSequential} as applied to
an infinite tree code like the one illustrated in
Fig.~\ref{fig:encodertree}. The estimates $\widehat{B}_i(t)$ describe
the current estimate for the most likely path through the tree based
on the channel outputs received so far.  Because of the possibility of
``backing up,'' in principle the estimate for $\widehat{B}_i$ could
change at any point in time. The theory of both ML and sequential
decoding tells us that generically, the probability of bit error on
bit $i$ approaches zero exponentially with increasing delay. 

%
%
%

%
 
In traditional analysis, random ensembles of infinite tree codes are
viewed as idealizations used to study the asymptotic behavior of
finite sequential encoding schemes such as convolutional codes. We can
instead interpret the traditional analysis as telling us that random
infinite tree codes achieve anytime reliability. In particular, we
know from the analysis of \cite{ForneyML} that at rate $R$ bits per
channel use, we can achieve anytime reliability $\alpha$ equal to the
block random coding error exponent. Pinsker's argument in
\cite{PinskerNoFeedback} as generalized in \cite{OurUpperBoundPaper}
tells us also that we cannot hope to do any better, at least in the
high-rate regime for symmetric channels. We summarize this
interpretation in the following theorem:

\begin{theorem} Random anytime codes exist for all
  DMCs\label{thm:anytime_codes_exist}  
For a stationary discrete memoryless channel (DMC) with capacity $C$,
randomized anytime codes exist without feedback at all rates $R < C$
and have anytime reliability $\alpha = E_r(R)$ where $E_r(R)$ is the
random coding error exponent as calculated in base $2$.
\end{theorem}
\vspace{0.1in} {\em Proof:} See
Appendix~\ref{app:randomanytimecodesexist}. \vspace{0.1in} 



\subsection{Sufficiency for the checkpoint process}
\label{sec:checkpointqossufficient}
The effect of any bit error in the checkpoint encoding of
Section~\ref{sec:checkpointcode} will be to throw us into a wrong bin
of size $\Delta$. This bin can be at most $\lambda^n
\frac{\Omega}{\lambda-1}$ away from the true bin. The error will then
propagate and grow by a factor $\lambda^n$ as we move from checkpoint
to checkpoint.  

If we are interested in the $\eta-$difference distortion, then the
distortion is growing by a factor of $\lambda^{n \eta}$ per checkpoint, or a
factor of $\lambda^\eta$ per unit of time. As long as the probability of
error on the message bits goes down faster than that, the expected
distortion will be small. This parallels Theorem~4.1 in
\cite{ControlPartI} and results in this proof for
Theorem~\ref{thm:anytime_sufficient}. 

{\em Proof:} Let $\check{X}'_{kn}(\phi)$ be the best estimate of the checkpoint
$\check{X}_{kn}$ at time $kn + \phi$. By the anytime
reliability property, grouping the message bits into groups of $nR_1$ at
a time, and the nature of exponentials, it is easy to see
that there exists a constant $K'$ so that:
\begin{eqnarray*}
 E[|\check{X}'_{kn}(\phi) - \check{X}_{kn}|^\eta] 
& \leq & 
 \sum_{j=0}^k 
 K' 2^{-\alpha(\phi + nj)} \lambda^{jn\eta} \frac{\Omega}{\lambda-1} \\
& = & K'' 2^{-\alpha \phi} \sum_{j=0}^k 2^{-jn(\alpha - \eta \log_2 \lambda)} \\
& \leq & K'' 2^{-\alpha \phi} \sum_{j=0}^\infty 2^{-jn(\alpha - \eta
   \log_2 \lambda)} \\
& = & K''' 2^{-\alpha \phi}
\end{eqnarray*}
where $K'''$ is a constant that depends on the anytime code, rate
$R_1$, support $\Omega$, and unstable $\lambda$. Thus by making sure
$\alpha > \eta \log_2 \lambda$ and choosing $\phi$ large enough,
$2^{-\alpha \phi}$ will become small enough so that $K''' 2^{-\alpha
  \phi}$ is as small as we like and the checkpoints will be
reconstructed to arbitrarily high fidelity. \hfill $\Box$
\vspace{0.1in}

Theorem~\ref{thm:anytime_sufficient} applies even in the case that
$\lambda=1$ and hence answers the question posed by Berger in
\cite{BergerBook} regarding the ability to track an unstable
process over a noisy channel without perfect
feedback. Theorem~\ref{thm:anytime_codes_exist} tells us that it is in
principle possible to get anytime reliability without any feedback at
all, and thus also with only noisy feedback.

This idea of tracking an unstable process using an anytime code is
useful beyond the source-coding context. In \cite{ITWpaper,
  DraperISIT06Paper, DraperECCPaper}, anytime codes are used over a
noisy feedback link to study the reliability functions for
communication using ARQ schemes and expected delay. The sequence
numbers of blocks are considered to be an unstable process that needs
to be tracked at the encoder. The random requests for retransmissions
make it behave like a random walk with a forward drift, but that can
stop and wait from time to time.

\subsection{Sufficiency for the history process} \label{sec:historyQoS}
It is easy to see that the history information for the two stream code
does not propagate errors from superblock to superblock and so does
not require any special QoS beyond what one would need for an
iid~or stationary-ergodic process. This is the basis for proving
Theorem~\ref{thm:history_sufficient}. 

{\em Proof:} Since the impact of a bit error is felt only within the
superblock, no propagation of errors needs to be considered. Theorem 
\ref{thm:backwards_bound} tells us that there is a maximum possible
distortion on the historical component. Thus the standard
achievability argument \cite{Gallager} for $D(R)$ tells us that as
long as the probability of block error can be made arbitrarily small
$\epsilon$ with increasing block-length, then the additional expected
distortion induced by decoding errors will also be arbitrarily
small. The desired probability of bit error can then be set to be
$\epsilon$ divided by the superblock length. \hfill $\Box$ \vspace{0.1in}

The curious fact here is that the QoS requirements of the second
stream of messages only need to hold on a superblock-by-superblock
basis. To achieve a small ensemble average distortion, there is no
need to have a secondary bitstream available with error probability
that gets arbitrarily small with increased delay! The secondary
channel could be nonergodic and go into outage for the entire
semi-infinite length of time as long as that outage event occurs
sufficiently rarely so that the average on each superblock is kept
small. Thus the second stream of messages is compatible with the
approach put forth in \cite{GoldsmithDistortion}.

\section{Quality of service requirements for communicating unstable
  processes: necessity} \label{sec:qos.necessity}

The goal is to prove Theorem~\ref{thm:strict_converse} by showing that
unstable Markov processes require communication channels capable of
supporting two-tiered service: a high priority core of rate $\log_2
\lambda$ with anytime-reliability of at least $\eta \log_2 \lambda$,
and the rest with Shannon reliable bit-transport. To do this, this
section proceeds in stages and follows the asymptotic equivalence
approach of \cite{ControlPartI}.

This section builds on Section~\ref{sec:twoanalog} where the pair of
communication problems (the endpoint communication problem and conditional
history communication problem) were introduced. In
Section~\ref{sec:necessity}, it is shown that the anytime-reliable
bit-transport problem reduces to the first problem (endpoint
communication) in the pair. Then Section~\ref{sec:finishnecessity}
finishes the necessity argument by showing how traditional
Shannon-reliable bit-transport reduces to the second problem and that
the two of them can be put together. This reduces a pair of
data-communication problems --- anytime-reliable bit transport and
Shannon-reliable bit-transport --- to the original problem of
communicating a single unstable process to the desired fidelity.

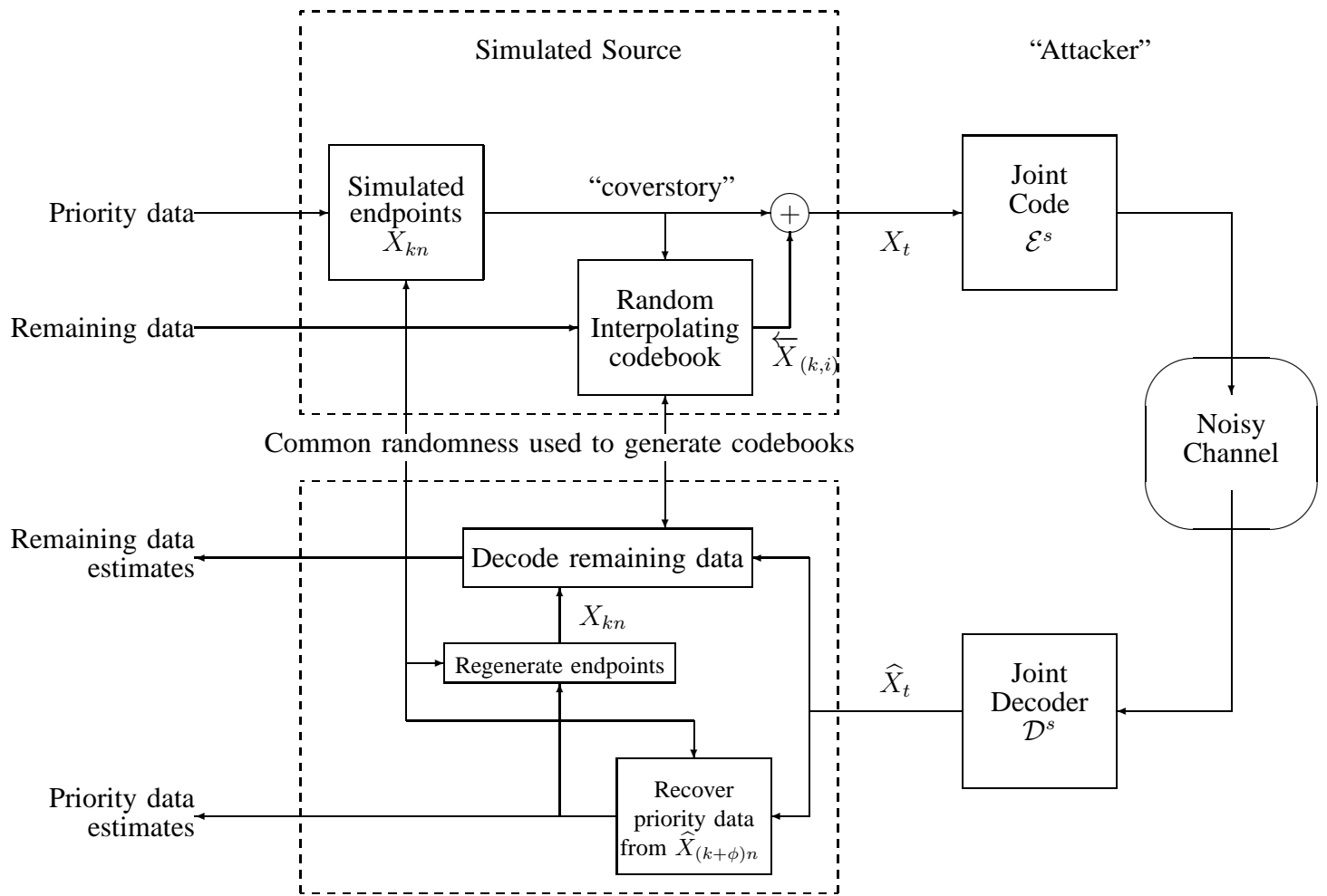
\begin{figure}
\begin{center}
\setlength{\unitlength}{3500sp}%
\begingroup\makeatletter\ifx\SetFigFont\undefined%
\gdef\SetFigFont#1#2#3#4#5{%
  \reset@font\fontsize{#1}{#2pt}%
  \fontfamily{#3}\fontseries{#4}\fontshape{#5}%
  \selectfont}%
\fi\endgroup%
\begin{picture}(8790,6924)(339,-7573)
{\color[rgb]{0,0,0}\thinlines
\put(9150,-4044){\oval(1342,1342)}
}%
{\color[rgb]{0,0,0}\put(5851,-2236){\vector( 1, 0){1200}}
}%
{\color[rgb]{0,0,0}\put(3301,-2236){\vector( 1, 0){2250}}
}%
{\color[rgb]{0,0,0}\put(7051,-2836){\framebox(1200,1200){}}
}%
{\color[rgb]{0,0,0}\put(8251,-2236){\line( 1, 0){900}}
\put(9151,-2236){\vector( 0,-1){1425}}
}%
{\color[rgb]{0,0,0}\put(5401,-3136){\line( 1, 0){300}}
\put(5701,-3136){\vector( 0, 1){750}}
}%
{\color[rgb]{0,0,0}\put(1876,-3811){\dashbox{60}(4200,3150){}}
}%
{\color[rgb]{0,0,0}\put(1876,-7561){\dashbox{60}(4200,3225){}}
}%
{\color[rgb]{0,0,0}\put(7051,-6136){\line(-1, 0){1200}}
\put(5851,-6136){\line( 0,-1){825}}
\put(5851,-6961){\vector(-1, 0){300}}
}%
{\color[rgb]{0,0,0}\put(7051,-6736){\framebox(1200,1200){}}
}%
{\color[rgb]{0,0,0}\put(8251,-6136){\makebox(1.6667,11.6667){\SetFigFont{5}{6}{\rmdefault}{\mddefault}{\updefault}.}}
}%
\put(4701,-2101){\makebox(0,0)[b]{\smash{\SetFigFont{12}{14.4}{\rmdefault}{\mddefault}{\updefault}{\color[rgb]{0,0,0}``coverstory''}%
}}}
\put(2701,-2101){\makebox(0,0)[b]{\smash{\SetFigFont{12}{14.4}{\rmdefault}{\mddefault}{\updefault}{\color[rgb]{0,0,0}Simulated}%
}}}
\put(6526,-2536){\makebox(0,0)[b]{\smash{\SetFigFont{12}{14.4}{\rmdefault}{\mddefault}{\updefault}{\color[rgb]{0,0,0}$X_t$}%
}}}
\put(5701,-2311){\makebox(0,0)[b]{\smash{\SetFigFont{12}{14.4}{\rmdefault}{\mddefault}{\updefault}{\color[rgb]{0,0,0}$+$}%
}}}
\put(7651,-2011){\makebox(0,0)[b]{\smash{\SetFigFont{12}{14.4}{\rmdefault}{\mddefault}{\updefault}{\color[rgb]{0,0,0}Joint}%
}}}
\put(7651,-2236){\makebox(0,0)[b]{\smash{\SetFigFont{12}{14.4}{\rmdefault}{\mddefault}{\updefault}{\color[rgb]{0,0,0}Code}%
}}}
\put(2701,-2311){\makebox(0,0)[b]{\smash{\SetFigFont{12}{14.4}{\rmdefault}{\mddefault}{\updefault}{\color[rgb]{0,0,0}endpoints}%
}}}
\put(4726,-2986){\makebox(0,0)[b]{\smash{\SetFigFont{12}{14.4}{\rmdefault}{\mddefault}{\updefault}{\color[rgb]{0,0,0}Random}%
}}}
\put(4726,-3436){\makebox(0,0)[b]{\smash{\SetFigFont{12}{14.4}{\rmdefault}{\mddefault}{\updefault}{\color[rgb]{0,0,0}codebook}%
}}}
\put(4051,-1036){\makebox(0,0)[b]{\smash{\SetFigFont{12}{14.4}{\rmdefault}{\mddefault}{\updefault}{\color[rgb]{0,0,0}Simulated Source}%
}}}
\put(8051,-1036){\makebox(0,0)[b]{\smash{\SetFigFont{12}{14.4}{\rmdefault}{\mddefault}{\updefault}{\color[rgb]{0,0,0}``Attacker''}%
}}}
\put(9151,-4186){\makebox(0,0)[b]{\smash{\SetFigFont{12}{14.4}{\rmdefault}{\mddefault}{\updefault}{\color[rgb]{0,0,0}Channel}%
}}}
\put(9151,-3961){\makebox(0,0)[b]{\smash{\SetFigFont{12}{14.4}{\rmdefault}{\mddefault}{\updefault}{\color[rgb]{0,0,0}Noisy}%
}}}
\put(7651,-5911){\makebox(0,0)[b]{\smash{\SetFigFont{12}{14.4}{\rmdefault}{\mddefault}{\updefault}{\color[rgb]{0,0,0}Joint}%
}}}
\put(7651,-6136){\makebox(0,0)[b]{\smash{\SetFigFont{12}{14.4}{\rmdefault}{\mddefault}{\updefault}{\color[rgb]{0,0,0}Decoder}%
}}}
{\color[rgb]{0,0,0}\put(5701,-2236){\circle{336}}
}%
{\color[rgb]{0,0,0}\put(4726,-2236){\vector( 0,-1){375}}
}%
{\color[rgb]{0,0,0}\put(4051,-3661){\framebox(1350,1050){}}
}%
{\color[rgb]{0,0,0}\put(9151,-4411){\line( 0,-1){1725}}
\put(9151,-6136){\vector(-1, 0){900}}
}%
{\color[rgb]{0,0,0}\put(2701,-3961){\vector( 0, 1){1200}}
}%
{\color[rgb]{0,0,0}\put(4726,-3961){\vector( 0, 1){300}}
}%
{\color[rgb]{0,0,0}\put(1051,-3136){\vector( 1, 0){3000}}
}%
{\color[rgb]{0,0,0}\put(1051,-2236){\vector( 1, 0){1050}}
}%
{\color[rgb]{0,0,0}\put(2101,-2761){\framebox(1200,1050){}}
}%
{\color[rgb]{0,0,0}\put(2701,-4186){\line( 0,-1){2025}}
\put(2701,-6211){\line( 1, 0){2250}}
\put(4951,-6211){\vector( 0,-1){300}}
}%
{\color[rgb]{0,0,0}\put(4351,-7411){\framebox(1200,900){}}
}%
{\color[rgb]{0,0,0}\put(4351,-6961){\vector(-1, 0){3300}}
}%
{\color[rgb]{0,0,0}\put(3901,-6961){\vector( 0, 1){1050}}
}%
{\color[rgb]{0,0,0}\put(3001,-5911){\framebox(1800,300){}}
}%
{\color[rgb]{0,0,0}\put(2701,-5761){\vector( 1, 0){300}}
}%
{\color[rgb]{0,0,0}\put(5851,-6136){\line( 0, 1){1200}}
\put(5851,-4936){\vector(-1, 0){450}}
}%
{\color[rgb]{0,0,0}\put(4726,-4186){\vector( 0,-1){525}}
}%
{\color[rgb]{0,0,0}\put(3901,-5611){\vector( 0, 1){450}}
}%
{\color[rgb]{0,0,0}\put(3151,-5161){\framebox(2250,450){}}
}%
{\color[rgb]{0,0,0}\put(3151,-4936){\vector(-1, 0){2100}}
}%
\put(4726,-3211){\makebox(0,0)[b]{\smash{\SetFigFont{12}{14.4}{\rmdefault}{\mddefault}{\updefault}{\color[rgb]{0,0,0}Interpolating}%
}}}
\put(3901,-4111){\makebox(0,0)[b]{\smash{\SetFigFont{12}{14.4}{\rmdefault}{\mddefault}{\updefault}{\color[rgb]{0,0,0}Common randomness used to generate codebooks}%
}}}
\put(1051,-3211){\makebox(0,0)[rb]{\smash{\SetFigFont{12}{14.4}{\rmdefault}{\mddefault}{\updefault}{\color[rgb]{0,0,0}Remaining data}%
}}}
\put(1051,-2311){\makebox(0,0)[rb]{\smash{\SetFigFont{12}{14.4}{\rmdefault}{\mddefault}{\updefault}{\color[rgb]{0,0,0}Priority data}%
}}}
\put(4951,-6811){\makebox(0,0)[b]{\smash{\SetFigFont{10}{14.4}{\rmdefault}{\mddefault}{\updefault}{\color[rgb]{0,0,0}Recover}%
}}}
\put(4951,-7036){\makebox(0,0)[b]{\smash{\SetFigFont{10}{14.4}{\rmdefault}{\mddefault}{\updefault}{\color[rgb]{0,0,0}priority data}%
}}}
\put(3901,-5836){\makebox(0,0)[b]{\smash{\SetFigFont{10}{14.4}{\rmdefault}{\mddefault}{\updefault}{\color[rgb]{0,0,0}Regenerate endpoints}%
}}}
\put(4276,-5011){\makebox(0,0)[b]{\smash{\SetFigFont{12}{14.4}{\rmdefault}{\mddefault}{\updefault}{\color[rgb]{0,0,0}Decode remaining data}%
}}}
\put(1051,-4861){\makebox(0,0)[rb]{\smash{\SetFigFont{12}{14.4}{\rmdefault}{\mddefault}{\updefault}{\color[rgb]{0,0,0}Remaining data}%
}}}
\put(1051,-6886){\makebox(0,0)[rb]{\smash{\SetFigFont{12}{14.4}{\rmdefault}{\mddefault}{\updefault}{\color[rgb]{0,0,0}Priority data}%
}}}
\put(1051,-5086){\makebox(0,0)[rb]{\smash{\SetFigFont{12}{14.4}{\rmdefault}{\mddefault}{\updefault}{\color[rgb]{0,0,0}estimates}%
}}}
\put(1051,-7111){\makebox(0,0)[rb]{\smash{\SetFigFont{12}{14.4}{\rmdefault}{\mddefault}{\updefault}{\color[rgb]{0,0,0}estimates}%
}}}
\put(4051,-5461){\makebox(0,0)[lb]{\smash{\SetFigFont{12}{14.4}{\rmdefault}{\mddefault}{\updefault}{\color[rgb]{0,0,0}$X_{kn}$}%
}}}
\put(5551,-3436){\makebox(0,0)[lb]{\smash{\SetFigFont{12}{14.4}{\rmdefault}{\mddefault}{\updefault}{\color[rgb]{0,0,0}$\overleftarrow{X}_{(k,i)}$}%
}}}
\put(2701,-2536){\makebox(0,0)[b]{\smash{\SetFigFont{12}{14.4}{\rmdefault}{\mddefault}{\updefault}{\color[rgb]{0,0,0}$X_{kn}$}%
}}}
\put(7651,-2536){\makebox(0,0)[b]{\smash{\SetFigFont{12}{14.4}{\rmdefault}{\mddefault}{\updefault}{\color[rgb]{0,0,0}${\cal E}^s$}%
}}}
\put(6526,-5986){\makebox(0,0)[b]{\smash{\SetFigFont{12}{14.4}{\rmdefault}{\mddefault}{\updefault}{\color[rgb]{0,0,0}$\widehat{X}_t$}%
}}}
\put(4951,-7261){\makebox(0,0)[b]{\smash{\SetFigFont{10}{14.4}{\rmdefault}{\mddefault}{\updefault}{\color[rgb]{0,0,0}from $\widehat{X}_{(k+\phi)n}$}
}}}
\put(7651,-6361){\makebox(0,0)[b]{\smash{\SetFigFont{12}{14.4}{\rmdefault}{\mddefault}{\updefault}{\color[rgb]{0,0,0}${\cal D}^s$}%
}}}
\end{picture}
\caption{Turning a joint-source-channel code into a two-stream code
  using information embedding. The good joint-source-channel code is
  like an attacker that will not impose too much distortion. Our goal
  is to simulate a source that carries our messages so that they can
  be recovered from the attacker's output. }   
\label{fig:simulator}
\end{center}
\end{figure}

The proof construction is illustrated in Fig.~\ref{fig:simulator}. Two
message streams need to be embedded --- a priority stream that
requires anytime reliability and a remaining stream for which
Shannon-reliability is good enough. The priority stream is used to
generate the endpoints while the the history part is filled in with
the appropriate conditional distribution. This simulated process is
then run through the joint source-channel encoder ${\cal E}_s$ to
generate channel inputs. The channel outputs are given to the joint
source-channel decoder ${\cal D}_s$ which produces, after some delay
$\phi$, a fidelity $d$ reconstruction of the simulated unstable
process. By looking at the reconstructions corresponding to the
endpoints, it is possible to recover the priority message bits in an
anytime reliable fashion. With these in hand, the remaining stream can
also be extracted from the historical reconstructions.

%


\subsection{Necessity of anytime reliability} \label{sec:necessity}

We follow the spirit of information embedding\cite{Moulin} except that
we have no {\em a-priori} covertext. Instead we use a simulated unstable
process that uses common randomness and without loss of generality,
message bits assumed to be from iid~coin tosses. If the message bits were not
fair coin tosses to begin with, XOR them with a one-time pad using
common randomness before embedding them. This section parallels the
necessity story in \cite{ControlPartI}, except that in this 
context, there is the additional complication of having a specified
distribution for the $\{W_t\}$, not just a bound on the allowed $|W_t|$.

The result is proved in stages. First, we assume that the density of
$W$ is a continuous uniform random variable plus something
independent. After that, this assumption is relaxed to having a
Riemann-integrable density $f_W$.

\subsubsection{Uniform driving noise}
\begin{lemma} \label{lem:anytime_necessity_uniform}
Assume the driving noise $W = G + U_{\delta}$ where $G,U_\delta$ are
independent random variables with $U_\delta$ being a uniform random
variable on the interval $[-\frac{\delta}{2},+\frac{\delta}{2}]$ for
some $\delta > 0$.

If a joint source-channel encoder/decoder pair exists for the endpoint
process given by (\ref{eqn:checkpointupdate}) that achieves
(\ref{eqn:distortion_condition}) for every position $kn$, then for
every rational rate $R = \frac{nR}{n} < \log_2 \lambda$, there exists a
randomized anytime code for the channel that achieves an anytime
reliability of $\alpha = \eta \log_2 \lambda$.  
\end{lemma}
\vspace{0.1in}
{\em Proof:} The goal is to simulate the the endpoint process using
the message bits and then to recover the message bits from the
reconstructions of the endpoints. Pick the initial condition $X_0$
using common randomness so it can be ignored in what follows. 

At the encoder, the goal is to simulate 
\begin{eqnarray*}
\widetilde{W}_k 
& = & \lambda^{n-1} \sum_{i=0}^{n-1} \lambda^{-i} W_{k,i} \\
& = & \lambda^{n-1}W_{k,0} + \lambda^{n-1}\sum_{i=0}^{n-1} \lambda^{-i} W_{k,i} \\
& = & \lambda^{n-1}U_{\delta,k} + \lambda^{n-1}(G_k + \sum_{i=0}^{n-1} \lambda^{-i}
W_{k,i}) \\
& = & U_{\lambda^{n-1}\delta,k} + [\lambda^{n-1}(G_k +
\sum_{i=0}^{n-1} \lambda^{-i} W_{k,i})]
\end{eqnarray*}
The $[\lambda^{n-1}(G_k + \sum_{i=0}^{n-1} \lambda^{-i} W_{k,i})]$
term is simulated entirely using common randomness and is hence known
to both the transmitter and receiver. The $U_{\lambda^{n-1}\delta,k}$
term is a uniform random variable on
$[-\frac{\lambda^{n-1}\delta}{2},+\frac{\lambda^{n-1}\delta}{2}]$ and
is simulated using a combination of common randomness and the fair
coin tosses coming from the message bits.

Since a uniform random variable has a binary expansion that is fair
coin tosses, we can write $U_{\lambda^{n-1}\delta,k} =
\frac{\lambda^{n-1}\delta}{2} \sum_{\ell=1}^\infty (\frac{1}{2})^\ell 
S_{k,\ell}$ where the $S_{k,\ell}$ are iid~random variables taking on
values $\pm 1$ each with probability $\frac{1}{2}$.

The idea is to embed the iid~$nR$ message bits into positions $\ell =
1, 2, \ldots , nR$ while letting the rest
--- a uniform random variable $U'_{\delta 2^{nR},k}$ representing the
semi-infinite sequence of bits $(S_{k,nR + 1}, S_{k, nR + 2}, \ldots)$
  --- be chosen using common randomness. The result is: 
\begin{equation} \label{eqn:embeddingrule}
 \widetilde{W}_k = \lambda^{n-1}\frac{\delta}{2} M_{k} + [\lambda^{n-1}(
  U'_{\delta 2^{nR},k}
  G_k + \sum_{i=0}^{n-1} \lambda^{-i} W_{k,i})]
\end{equation}
where $M_{k}$ is the $nR$ bits of the message as represented by $2^{nR}$
equally likely points in the interval $[-1,+1]$ spaced apart by
$2^{1-nR}$, and the rest of the terms $[\cdots]$ are chosen using
common randomness known at both the transmitter and receiver side. 

\begin{figure}
\begin{center}
\mbox{\epsfxsize=3.7in \epsfysize=0.6in \epsfbox{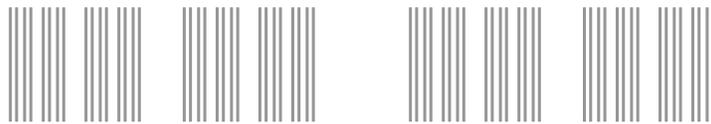}}
\end{center}
\caption{The priority message bits are used to refine a point on a Cantor
  set. The natural tree structure of the Cantor set construction
  allows us to encode bits sequentially. The Cantor set also has
  finite gaps between all points corresponding to bit sequences that
  first differ in a particular bit position. These gaps allow us to
  reliably extract bit values from noisy observations of the Cantor
  set point regardless of which point it is.}
\label{fig:cantorset}
\end{figure}

Since the simulated endpoints process is a linear function of the
$\{\widetilde{W}_k\}$ and the distortion measure is a difference
distortion, it suffices to just consider the $\{X'_{kn}\}$ 
process representing the response to the discrete messages $\{M_k\}$ 
alone. This has a zero initial condition and evolves like
\begin{equation} \label{eqn:Xprime} 
X'_{(k+1)n} = \lambda^n X'_{kn} + \beta M_{k}
\end{equation}
where $\beta = \lambda^{n-1}\frac{\delta}{2}$. Expanding this
recursion out as a sum gives
\begin{equation} \label{eqn:XprimeSum} 
X'_{(k+1)n} = (\lambda^n)^k \beta \sum_{i=0}^k  \lambda^{-ni} M_{k-i}.
\end{equation}
This looks like a generalized binary expansion in base $\lambda^n$ and
therefore implies that the $X'$ process takes values on a growing
Cantor set (illustrated in Fig.~\ref{fig:cantorset} for $nR = 1$) 

The key property is that there are gaps in the Cantor set:

\begin{property} \label{gap_property}
If the rate $R < \log_2 \lambda + \frac{\log_2(1-\lambda^{-n})}{n}$ and the
message-streams $M$ and $\bar{M}$ first differ at position $j$ 
(message $M_j \neq \bar{M}_j$), then at time $k > j$, the encoded $X'_{kn}$ and
$\bar{X}'_{kn}$ corresponding to $M_1^{k -1}$ 
and $\bar{M}_1^{k -1}$ respectively differ by at least:
\begin{equation} \label{eqn:gapbound}
|X'_{kn} - \bar{X}'_{kn}| \leq K \lambda^{n(k-j)}
\end{equation}
for some constant $K>0$ that does not depend on the values of the message
bits, $k$, or $j$.
\end{property}
{\em Proof:} See Appendix~\ref{app:propertygapbound}.
\vspace{0.1in}

In coding theory terms, Property~\ref{gap_property} can be interpreted 
as an infinite Euclidean free-distance for the code with the added
information that the distance increases exponentially as
$\lambda^{n(k-j)}$. Thus, a bit error can only happen if the received
``codeword'' is more than half the minimum distance away.

At the decoder, the common randomness means that the estimation error 
$X_{kn} - \widehat{X}_{kn}$ is the error in estimating $X'_{kn}$. By
applying Markov's inequality to this using
(\ref{eqn:distortion_condition}), we immediately get a bound on the
probability of an error on the prefix $M_0^i$ for $i < k$:
\begin{eqnarray*}
{\cal P}(\widehat{M}_1^i(kn) \neq M_1^i) 
& \leq &
{\cal P}(|\widehat{X}'_{kn} - X'_{kn}| > \frac{K}{2} \lambda^{n(k-i)} ) \\
& = & 
{\cal P}(|\widehat{X}_{kn} - X_{kn}| > \frac{K}{2} \lambda^{n(k-i)} ) \\
& = & 
{\cal P}(|\widehat{X}_{kn} - X_{kn}|^\eta > (\frac{K}{2})^\eta (\lambda^{n(k-i)})^\eta \\
& \leq & 
d (\frac{K}{2})^{-\eta} (\lambda^{n(k-i)})^\eta \\
& = & 
K' 2^{-(\eta \log_2 \lambda) n(k-i)}.
\end{eqnarray*}
But $n(k-i)$ is the delay that is experienced at the $nR$-bit message
level. If bits have to be buffered-up to form messages, then
  the delay at the bit level includes another constant $n$. This only
  increases the constant $K'$ further but does not change the exponent
  with large delays. Thus, the desired anytime reliability is
obtained. \hfill $\Box$ \vspace{0.1in}

\subsubsection{General driving noise} 
Lemma~\ref{lem:anytime_necessity_uniform} can have the technical
smoothness condition weakened to simply requiring a Riemann-integrable
density for the white $W$ driving process. 
\begin{lemma} \label{lem:anytime_necessity}
Assume the driving noise $W$ has a Riemann-integrable density
$f_W$. If there exists a family of joint source-channel
encoder/decoder pairs for a sequence of increasing $n$-endpoint
problems given by (\ref{eqn:checkpointupdate}) that achieve 
(\ref{eqn:distortion_condition}) for every position $kn$, then for
every rate $R < \log_2 \lambda$ and anytime
reliability $\alpha < \eta \log_2 \lambda$, there exists a randomized
anytime code for the underlying channel.

\end{lemma}
\vspace{0.1in}
{\em Proof:} Since the density is Riemann-integrable,
Lemma~\ref{lem:riemannmixture} applies. Choose $\delta$ such that $\gamma < 
\lambda^{-2\eta n}$. When simulating $W_{k,0}$ in the endpoint
process, use common randomness for $C_\gamma$ and $W''_\delta$, and
follow the procedure from the proof of
Lemma~\ref{lem:anytime_necessity_uniform} for $G_\delta$ and
$U_\delta$. 

We can thus interpret a ``heads'' for $C_\gamma$ as an ``erasure''
with probability $\gamma$ since no message can be encoded in that time  
period. From the point of view of
Lemma~\ref{lem:anytime_necessity_uniform}, this can be considered a
known null message. 

Since the outcome of these coin tosses come from common
randomness, the position of these erasures are known to both the
transmitter and the receiver. In this way, it behaves like a packet
erasure channel with feedback. This problem
is studied in Theorem~3.3 of \cite{OurUpperBoundPaper}, and the
delay-optimal coding strategy relative to the erasure channel is to
place incoming packets into a FIFO queue awaiting a non-erased
opportunity for transmission. The following lemma summarizes
the results needed from \cite{OurUpperBoundPaper}.

\begin{lemma} \label{lem:packet-erasure}
Suppose packets arrive deterministically at a rate of $R$ packets per
unit time and enter a FIFO queue drained at constant rate $1$ per unit 
time. 
\begin{itemize}
\item Suppose $\gamma < \frac{1}{16}$. If each packet has a size
  distribution that is bounded below a geometric$(1-\gamma)$
  (i.e. ${\cal P}(\mbox{Size} > s) \leq \gamma^s$ for all non-negative 
  integers $s$), then the random delay $\phi$ experienced by any
  individual packet from arrival to departure from the queue satisfies
  ${\cal P}(\phi > s) \leq K 2^{-\alpha s}$ for all non-negative 
  $s$ and some constant $K$ that does not depend on $s$. Furthermore,
  if $R < \frac{1}{1+2r}$ for some $r > 0$, then $\alpha \geq -\log_2
  \gamma - 2\gamma^r$.  

 \item Assume the rate $R = \frac{1}{n}$ and each packet has a size
  distribution that is bounded by: ${\cal P}(\mbox{Size} > n(1-\epsilon) + s)
  \leq \gamma^s$ for all non-negative integers $s$. Then the
  delay $\phi$ experienced by any individual packet has a tail
  distribution bounded in the same way as for $R' =
  \frac{1}{n\epsilon}$ and packets with geometric$(1-\gamma)$
  size. That is ${\cal P}(\phi > s) \leq K 2^{-\alpha s}$ where
  $\alpha \geq -\log_2 \gamma - 2\gamma^{\frac{n\epsilon - 1}{2}}$. 
\end{itemize}
\end{lemma}
{\em Proof:} See Theorem~3.3 and Corollary~6.1 of \cite{OurUpperBoundPaper}.
\vspace{0.1in}

For our problem, the message bits are arriving
deterministically at bit-rate $R < \log_2 \lambda$ per unit time to the
transmitter. Pick $r>0$ small enough so that $R' = (1+3r)R <
\log_2 \lambda$. Group message bits into packets of size $nR'$. These
packets arrive deterministically at rate $\frac{1}{1 + 3r} <
\frac{1}{1 + 2r}$ packets per $n$ time units. Thus,
Lemma~\ref{lem:packet-erasure} applies and the delay (in $n$ units)
experienced by a packet in the queue has a delay error exponent
$\alpha$ of least 
\begin{eqnarray*} 
-\log_2 \gamma - 2\gamma^r 
& \geq & 
-\log_2 \lambda^{-2\eta n}  - 2 \lambda^{-2\eta n r} \\
& = & n 2 \eta \log_2 \lambda  - 2 \lambda^{-2\eta n r}
\end{eqnarray*}
per $n$ time steps or $2 \eta \log_2 \lambda - \frac{2 \lambda^{-2 \eta n r}}{n}$
per unit time step. When $n$ is large, this exponent is much faster
than the delay exponent of $\eta \log_2 \lambda$ obtained in the proof of
Lemma~\ref{lem:anytime_necessity_uniform}. The two delays
experienced by a bit are independent by construction. Thus, the dominant
delay-exponent remains $\eta \log_2 \lambda$ as desired. \hfill $\Box$ \vspace{0.1in}

Notice that the simulated endpoint process depends only on common
randomness and the message packets. Since the common randomness is
known perfectly at the receiver by assumption and the message packets
are known with a probability that tends to $1$ with delay, the 
endpoint process is also known with zero distortion with a probability
tending to $1$ as the delay increases. 


\subsection{Embedding classical bits} \label{sec:finishnecessity}
All that remains is to embed the classical message bits into the
historical process. The overall construction is described in
Fig.~\ref{fig:simulator}. First, $n$ is chosen to be large enough so
that the $R_1$ stream can be successfully embedded in the endpoint
process by Lemma~\ref{lem:anytime_necessity}. 

Now, $n$ is further increased so that $R_2 <
R_n^{\overleftarrow{X}|\overleftarrow{X}_{0}}(d)$ the conditional  
rate-distortion function for the history given the endpoint. This can
be done since $\lim_{n\rightarrow \infty}
R_n^{\overleftarrow{X}|\overleftarrow{X}_{0}}(d) =  R_\infty^{X}(d) -
\log_2 \lambda$ by Theorem~\ref{thm:backwards_bound}. 

By choosing an appropriate additional delay,
Lemma~\ref{lem:anytime_necessity} assures us that the receiver will 
know all the past high-priority messages and hence simulated endpoints
correctly with an arbitrarily small probability of error
$\epsilon$. As described in Section~\ref{sec:twoanalog}, this means we
now have a family of systems (indexed by $m$) that solve the
conditional history problem. The condition (\ref{eqn:extracondition})
translates into
\begin{equation} \label{eqn:translatedcondition}
\lim_{m \rightarrow \infty} 
\sup_{\tau \geq 0}
{\cal P}(\frac{1}{m} \sum_{k=\tau}^{\tau + m-1} \frac{1}{n}\sum_{i=1}^{n-1} |\overleftarrow{X}_{(k,i)} - \widehat{X}^{-}_{(k,i)}|^\eta > d)
= 0.
\end{equation}
It tells us that by picking $m$ large enough, the probability of
having excess distortion can be made as small as desired. 


The simulated $\{Z_k\}$ containing the high-priority messages are
interpreted as the ``coverstory'' that must be respected when
embedding messages into the $\{\vec{X}_k^{-}\}$ process. The $\{Z_k\}$
are iid by construction and hence Theorem~3 from
\cite{MukulAllerton06} (full proofs in \cite{OurDirectConverse})
applies and tells us that a length $m' > m$ random code with
$\vec{X}^{-}_k$ drawn independently of each other, but conditional on
the iid $Z_k$, can be used to embed information at any rate $n R_2 <
nR_n^{\overleftarrow{X}|Z}(d+\epsilon) = nR_n^{X|X_{n}}(d+\epsilon)$
per vector symbol with arbitrarily low probability of error. \hfill
$\Box$ \vspace{0.1in} 

The ``weak law of large numbers''-like condition
(\ref{eqn:extracondition}), or something like it, is required for the
theorem to hold since there are joint source-channel codes for which
mutual information cannot be turned into the reliable communication of
bits at arbitrarily low probabilities of error. Consider the following
contrived example. Suppose there are two different joint
source-channel codes available: one has a target distortion of $d_1$
and the other has a target distortion of $d_2 = 10 d_1$. The actual
joint code, which is presumed to have access to common randomness,
could decide with probability $\frac{1}{1000}$ to use the second code
rather than the first. In such a case, the ensemble average mutual
information is close to $R(d_1) - \log_2 \lambda$ bits, but with
non-vanishing probability $\frac{1}{1000}$ we might not be able
sustain such a rate over the virtual channel.

We conjecture that for DMCs, if any joint source-channel code exists
that hits the target distortion on average, then one should also exist
that meets (\ref{eqn:extracondition}) and it should be possible to
simultaneously communicate two streams of messages reliably with 
anytime reliability on the first stream and enough residual rate on
the second.

\section{Unstable Gauss-Markov processes with squared-error distortion}\label{sec:gaussians}

\subsection{Source-coding for Gaussian processes} \label{sec:gaussianfixedrate}

The goal here is to prove Corollary~\ref{cor:gaussianfixedrate}.  The
strategy is essentially as before. One simplification is that we can
make full use of Theorem~\ref{thm:backwards_bound} and rely on
$R_\infty^{\overleftarrow{X}}(d) = R_\infty^X(d) - \log_2
\lambda$. There is thus no rate loss in encoding the historical
segments on a block-by-block basis rather than using superblocks and
conditional encodings. The only issue that remains is  dealing with
the unbounded support when encoding the checkpoints.

The overall approach is: (key differences {\em italicized})
\begin{itemize}
 \item[(a)] Look at time in blocks of size $n$ and encode the values 
       of checkpoints $X_{kn}$ recursively to very high precision 
       using a prefix-free {\em variable-length} code with rate
       $n(\log_2 \lambda + \epsilon_1) +  L_k$ bits per value, where
       the $L_k$ are iid random variables with appropriately nice
       properties.  

 \item[(b)] {\em Smooth out the variable-length code by running it through
       a FIFO queue drained at constant rate $R_1 = \log_2 \lambda +
       \epsilon_1 + \epsilon_q$. Make sure that the delay exponent in
       the queue is high enough. }

   \item[(c)] Use the {\em exact value} for the ending checkpoint
     $X_{(k+1)n}$ ({\em instead of the quantized $\check{X}$}) to
     transform the segment immediately before it so that it looks
     exactly like a stable backwards Gaussian process of length $n$
     with initial condition $0$. Encode each block of the backwards
     history process to average-fidelity $d$ using a fixed-rate
     rate-distortion code for the backwards process that operates at
     rate $R^{\overleftarrow{X}}_\infty(d) + \epsilon_s$.

 \item[(d)] At the decoder, wait $\phi$ time units and attempt to
       decode the checkpoints to high fidelity. {\em If the FIFO queue is
       running too far behind, then extrapolate a reconstruction
       based on the last fully decoded checkpoint.}

 \item[(e)] Decode the history process to average-fidelity $d$ and
       combine it with the recursively quantized checkpoints to get
       the reconstruction. 
\end{itemize}

\paragraph{Encoding the checkpoints} 
(\ref{eqn:checkpointupdate}) remains valid, but the term $\widetilde{W}_k
= \lambda^{n-1} \sum_{i=0}^{n-1} \lambda^{-i} W_{kn+i}$ is not bounded since the
$W_i$ are iid Gaussians. The $\widetilde{W}_k$ are instead Gaussian with
variance
\begin{eqnarray*}
\widetilde{\sigma}^2 
& = & 
\lambda^{2(n-1)}\sum_{i=0}^{n-1} \lambda^{-2i} \sigma^2 \\
& \leq & 
\lambda^{2(n-1)} \sigma^2 \sum_{i=0}^{\infty} \lambda^{-2i}  \\
& = & 
\lambda^{2n} \frac{\sigma^2}{\lambda^2 - 1}.
\end{eqnarray*}
The standard deviation $\widetilde{\sigma}$ is therefore
$\lambda^{n}\frac{\sigma}{\sqrt{\lambda^2 -1}}$. Pick $l =
2^{\frac{\epsilon_1}{3}n}$ and essentially pretend that this random
variable $\widetilde{W}_k$ has bounded support of $l\widetilde{\sigma}$ during
the encoding process. By comparing (\ref{eqn:checkpointjump}) to the
above, the effective $\Omega$ is simply $l \sigma
\frac{2(\lambda-1)}{\sqrt{\lambda^2 -1}} =
2^{\frac{\epsilon_1}{3}n}\sigma\sqrt{\frac{\lambda-1}{\lambda+1}}$. Define
$\widetilde{\Omega} = \sigma\sqrt{\frac{\lambda-1}{\lambda+1}}$ so that the effective
$\Omega = 2^{\frac{\epsilon_1}{3}n} \widetilde{\Omega}$.

\begin{figure}
\begin{center}
\begin{tabular}{|c|c|}
\hline
Offset & Codeword \\
0 & 100 \\
\hline
\hline
+1 & 1110 \\
\hline
-1 & 1100 \\
\hline
\hline
+2 & 11110 \\
\hline
-2 & 11010 \\
\hline
\hline
+3 & 111110 \\
\hline
-3 & 110110 \\
\hline
\hline
+4 & 1111110 \\
\hline
-4 & 1101110 \\
\hline
\hline
$\vdots$ & $\vdots$ \\
\hline
\end{tabular}
\end{center}
\caption{Unary encoding of integer offsets to deal with the unbounded
support. The first bit denotes start while the nest two bits reflect
the sign. The length of the rest reflects the magnitude of the
offset with a zero termination. The encoding is prefix-free and
hence uniquely decodable. The length of the encoding of integer $S$ is
bounded by $3 + |S|$} 
\label{fig:unarytable}
\end{figure}

Encode the checkpoint increments recursively as before, only add an
additional variable-length code for the value of $\lfloor
\frac{\widetilde{W}}{l\widetilde{\sigma}} + \frac{1}{2}\rfloor$ while treating
the remainder using the fixed-rate code as before. The variable length
code is a unary encoding that counts how many $l\widetilde{\sigma}$
away from the center the $\widetilde{W}_k$ actually
is. (Fig.~\ref{fig:unarytable} illustrates the unary code.) Let $L_k$
be the length of the $k$-th unary codeword. This is bounded above by
\begin{eqnarray*}
P(L_k \geq 3 + j) 
& = & P(|\widetilde{W}| >  jl\widetilde{\sigma}).
\end{eqnarray*}
Let $N$ be a standard Gaussian random variable and rewrite this
as
\begin{equation} \label{eqn:GaussianTailBound}
P(L_k \geq 3 + j) = P(|N| > j 2^{\frac{\epsilon_1}{3}n}) \leq
\exp(-\frac{1}{2} j^2 2^{\frac{2\epsilon_1}{3}n})
\end{equation}
and so $L_k$ is very likely indeed to be small and certainly has a
finite expectation $\bar{L} < 4$ if $n$ is large. 

The fixed-rate part of the checkpoint encoding has a rate that is
the same as that given by (\ref{eqn:checkpointRconstraint}),
except that $\Omega$ is now mildly a function of $n$. Plugging in 
$2^{\frac{\epsilon_1}{3}n} \widetilde{\Omega}$ for $\Omega$ 
in (\ref{eqn:checkpointRconstraint}) gives
\begin{eqnarray*}
R_{1,f} & \geq & \max \left(
    \log_2 \lambda + \frac{\log_2 (1 + \frac{\Omega}{\Delta (\lambda-1)}) + \log_2(2+\frac{\Omega}{\Delta})}{n},
    \frac{\log_2 \lceil \frac{\Omega_0}{\Delta} \rceil}{n} \right) \\
& = & \max \left(
    \log_2 \lambda + \frac{\log_2 (1 + \frac{2^{\frac{\epsilon_1}{3}n} \widetilde{\Omega}}{\Delta (\lambda-1)}) + \log_2(2+\frac{2^{\frac{\epsilon_1}{3}n} \widetilde{\Omega}}{\Delta})}{n},
    \frac{\log_2 \lceil \frac{\Omega_0}{\Delta} \rceil}{n} \right) \\
& = & \max \left(
    \log_2 \lambda + \frac{2}{3}\epsilon_1 + \frac{\log_2 (2^{-\frac{\epsilon_1}{3}n} +
      \frac{\widetilde{\Omega}}{\Delta (\lambda-1)}) + \log_2(2^{1-\frac{\epsilon_1}{3}n}+\frac{\widetilde{\Omega}}{\Delta})}{n},
    \frac{\log_2 \lceil \frac{\Omega_0}{\Delta} \rceil}{n} \right).
\end{eqnarray*}
Essentially, the required rate $R_{1,f}$ for the fixed-rate part
has only increased by a small constant
$\frac{2}{3}\epsilon_1$. Holding $\Delta$ fixed and assuming $n$ is
large enough, we can see that  
\begin{equation} \label{eqn:checkpointGaussianR}
R_{1,f} = \log_2 \lambda + \epsilon_1
\end{equation}
is sufficient.

\paragraph{Smoothing out the flow} 
The code so far is variable-rate and to turn this into a fixed-rate
$R_1 = \log_2 \lambda + \epsilon_1 + \epsilon_q$ bitstream, it is
smoothed by going through a FIFO queue. First, encode the offset using
the variable-length code and then recursively encode the increment as
was done in the finite support case. All such codes will begin with a
$1$ and thus we can use zeros to pad the end of a codeword whenever
the FIFO is empty. When $n$ is large, the average input rate to the
FIFO is smaller than the output rate and hence it will be empty
infinitely often.

\paragraph{Getting history and encoding it} 
Section~\ref{sec:time_reverse} explains why such a transformation is
possible by subtracting off a scaled version of the endpoint. The
result is a stable Gaussian process and so \cite{Gray70} reveals that
it can be encoded arbitrarily close to its rate-distortion bound
$R_\infty^{\overleftarrow{X}}(d) = R_\infty^{X}(d) -
\log_2 \lambda$ if $n$ is large enough. 

\paragraph{Decoding the checkpoints} The decoder can wait long enough
so that the checkpoint we are interested in is very likely to have
made it through the FIFO queue by now. The ideas here are similar to
\cite{OurUpperBoundPaper, OurSourceCodingPaper} in that a FIFO queue
is used to smooth out the rate variation with good large
deviations performance. There is $n\epsilon_q$ slack that has to
accommodate $L_k$ bits. Because $n$ can be made large, the error
exponent with delay here can be made as large as needed.

More precisely, a packet of size $n(\epsilon_1 + \log_2 \lambda) + L_k$ bits
arrives every $n$ time units where the $L_k$ are iid. This is drained
at rate $R_1 = \epsilon_q + \epsilon_1 + \log_2 \lambda$. An alternative
view is therefore that a point packet arrives deterministically every
$n$ time units and it has a random service time $T_k$ given by
$n\frac{\epsilon_1 + \log_2 \lambda}{\epsilon_q + \epsilon_1 + \log_2 \lambda} +
\frac{L_k}{\epsilon_q + \epsilon_1 + \log_2 \lambda}$. Define
$(1-\epsilon'_q)=\frac{\epsilon_1 + \log_2 \lambda}{\epsilon_q + \epsilon_1
  + \log_2 \lambda}$. Then the random service time $T_k = (1-\epsilon'_q)n +
\frac{L_k}{\epsilon_q + \epsilon_1 + \log_2 \lambda}$ when measured in time
units or $T^b_k = (1-\epsilon'_q)nR_1 + L_k$ when measured in
bit-units. 

This can be analyzed using large-deviations techniques or by applying
standard results in queuing. The important thing is a bound on the
length $L_k$ which is provided by (\ref{eqn:GaussianTailBound}). It is
clear that\footnote{While this
  proof is written for the Gaussian case, the arguments here readily
  generalize to any driving distribution $W$ that has at least an
  exponential tail probability. To accommodate $W$ with power-law tail
  distributions would require the use of logarithmic encodings as
  described in \cite{NairPaper1,NairPaper2}. This does not work for
  our case because the unary nature of the encoding is important when
  we consider transporting such bitstreams across a noisy channel.},
\begin{eqnarray*} 
P(L_k \geq 3 + j) 
& \leq & \exp(-\frac{1}{2} j^2 2^{\frac{2\epsilon_1}{3}n}) \\
& \leq & \exp(- 2^{\frac{2\epsilon_1}{3}n - 1})j).
\end{eqnarray*}
Since an exponential eventually dominates all constants, we know that
for any $\beta > 0$, there exists a sufficiently large $n$ so that:
\begin{equation} \label{eqn:variable-service-bound}
P(L_k - 3 > j) \leq 2^{-\beta j}.
\end{equation}

Thus, the delay (in bits) experienced by a block in the queue will
behave no worse than that of point messages arriving every $nR_1$ bits
where each requires at least $nR_1(1-\epsilon'_q) + 3 =
nR_1(1-\epsilon''_q)$ bits plus an iid 
geometric$(1-p)$ number of bits with $p = 2^{-\beta}$. 

Lemma~\ref{lem:packet-erasure} applies to this queuing problem and the
second part of that lemma tells us that the delay performance is
exactly the same as that of a system with point messages arriving
every $n\epsilon''_q$ bits requiring only an iid geometric number of
bits. Since $\frac{1}{n\epsilon''_q}$ is small, the first part of
Lemma~\ref{lem:packet-erasure} applies. Set $r =
\frac{n\epsilon''_q}{3} - 1$, then the bit-delay exponent $\alpha_b$
is at least 
\begin{eqnarray*}
\alpha_b 
& \geq & -\log_2 2^{-\beta} - 2^{-\beta r} \\
& = & \beta - 2^{-\beta (\frac{n\epsilon''_q}{3} - 1)}
\end{eqnarray*}
which is at least $\beta - 1$ when $n\epsilon''_q \geq 3$. Converting
between bit-delay and time-delay is essentially just a factor of
$\log_2 \lambda$ and so the time-delay exponent is at least $\frac{\beta -
  1}{\log_2 \lambda}$. But $\beta$ can be made as large as we want by
choosing $n$ large enough. 

\paragraph{Getting the final reconstruction} The history process
is added to the recovered checkpoint. This differs from the 
original process by only the error in the history plus the impact of
the error in the checkpoint. The checkpoint reconstruction-error's
impact dies exponentially since the history process is stable. So the
target distortion is achieved if the checkpoint has arrived by the time
reconstruction is attempted. By choosing a large enough 
end-to-end delay $\phi$, the probability of this can be made as high
as we like.

However, the goal is not just to meet the target distortion level $d$
with high probability, it is also to hit the target in
expectation. Thus, we must bound the impact of not having the
checkpoint available in time. When this happens, the un-interpretable
history information is ignored and the most recent checkpoint is
simply extrapolated forward to the current time. The expected squared
errors grow as $\lambda^{2 \psi}$ where $\psi$ is the delay in
time-units. The arguments here exactly parallel those of
Theorem~\ref{thm:anytime_sufficient}, where the FIFO queue is acting
like an anytime code. Since the delay-exponent of the queue is as
large as we want, it can be made larger than $2 \log_2 \lambda$. Thus,
the expected distortion coming from such ``overflow'' situations is as
small as desired. This completes the proof of
Corollary~\ref{cor:gaussianfixedrate}. \hfill $\Box$ \vspace{0.1in}

\subsection{Channel sufficiency for communicating Gaussian processes} \label{sec:gaussianchannelsufficiency}

This section shows why Corollary~\ref{cor:sufficient_gaussian} is
true. The story in the Gaussian case is mostly unchanged since the
historical information is as classical as ever. The only issue is with
the checkpoint stream. An error in a bit $\psi$ steps ago can do more
than propagate through the usual pathway. It could also damage the
bits corresponding to the variable-length offset. 

Because of the unary encoding and the $2^{\frac{\epsilon_1}{3}n}$
expansion in the effective $\Omega$, an uncorrected bit-stream error
$\psi$ time-steps ago can only impact the error in the checkpoint
reconstruction by $4\psi (\log_2 \lambda) 2^{\frac{\epsilon_1}{3}n}$ since the worst
error is clearly to flip the sign bit and keep the unary codeword from
terminating thereby making it at most $2 \psi \log_2 \lambda$ bits long. The
current reconstruction is therefore incorrect by an $O(\psi
2^{\frac{\epsilon_1}{3}n} \lambda^{\psi})$ change in its
value. As far as $\eta$-distortion is concerned, the distortion grows
by a factor $O(\psi^\eta 2^{\eta \frac{\epsilon_1}{3}n} \lambda^{\eta
  \psi})$ from what it would be with correct
reconstruction. Asymptotically, the delay $\psi$ is much larger than
the block-length $n$ and so the polynomial term in front is
insignificant relative to the exponential in $\psi$. If the code has
anytime reliability $\alpha > \eta \log_2 \lambda$, then the same argument
as Theorem~\ref{thm:anytime_sufficient} applies and the Corollary
holds. \hfill $\Box$

\section{Extensions to the vector case} \label{sec:vector} With the
scalar case explored, it is natural to consider what happens for
general finite-dimensional linear models where $\lambda$ is replaced
with a matrix $A$ and $X$ is a vector. In the Gaussian process case,
these will correspond to cases with formally rational power-spectral
densities. Though the details are left to the reader, the story is
sketched here.  No fundamentally new phenomena arise in the vector
case, except that different anytime reliabilities can be required on
different streams arising from the same source as is seen in the
control context \cite{ControlPartII}.

The source-coding results here naturally extend to the fully observed
vector case with generic driving noise distributions. Instead of two
message streams, there is one special stream for each unstable
eigenvalue $\lambda_i$ of $A$ and a singe final stream capturing the
residual information across all dimensions. All the sufficiency
results also generalize in a straightforward manner --- each of the
unstable streams requires a corresponding anytime reliability
depending on the distortion function's $\eta$ and the magnitude of the
eigenvalue. The multiple priority-stream necessity results also follow
generically.\footnote{The required condition is that the the driving
  noise distribution $W$ should not have support isolated to an
  invariant subspace of $A$. If that were to happen, there would
  be modes of the process that are never excited.} This is a
straightforward application of a system diagonalization\footnote{The
  case of non-diagonal Jordan blocks is only a challenge for the
  necessity part regarding anytime reliability. It is covered in
  \cite{ControlPartII} in the control context. The same argument holds
  here with a Riemann-integrable joint-density assumption on the
  driving noise.} argument followed by an eigenvalue by eigenvalue
analysis. The necessity result for the residual rate follows the same
proof as here based on inverse-conditional rate-distortion with the
endpoints in all dimensions used as side-information.

The case of partially observed vector Markov processes where the
observations $C_y \vec{X}$ are linear in the system state requires one
more trick. We need to invoke the observability\footnote{The linear
  observation should not be restricted to a single invariant subspace.
  If it were, we could drop the other subspaces from the model as
  irrelevant to the observed process under consideration.} of the
system state. Instead of a single checkpoint pair, use an appropriate
number\footnote{The appropriate number is twice the number of
  observations required before all of the unstable subspaces show up
  in the observation. This number is bounded above by twice the
  dimensionality of the vector state space. The factor of two is to
  allow each block to have its own beginning and end.} of consecutive
values for the observation and encode them together to high fidelity
$\Delta$. This can be done by transforming
coordinates linearly so that the system is diagonal, though driven by
correlated noise, from checkpoint-block to the next
checkpoint-block. The initial condition is governed by the self-noise
that is unavoidable while trying to observe the state. Each unstable
eigenvalue will contribute its own $\log_2 \lambda_i$ term to the
first stream rate and will require the appropriate anytime
reliability. The overhead continues to be sublinear in $n$ and the
residual information continues to be classical in nature by the same
arguments given here. 

The partially observed necessity story is essentially unchanged on the
information embedding side, except that every long block should be
followed by a miniblock of length equal to the dimensionality $k$
during which no message is embedded and only common-randomness is used
to generate the driving noise. This will allow the decoder to easily
use observability to get noisy access to the unstable state
itself.

In \cite{ControlPartII}, these techniques are applied in the context
of control rather than estimation. The interested reader is referred
there for the details. Some simplifications to the general story might
be possible in the case of SISO autoregressive processes, but we have
not explored them in detail.

\section{Conclusions}
We have characterized the nature of information in an unstable Markov
process. On the source coding side, this was done by giving the
fixed-rate coding Theorem~\ref{thm:boundedsupportfixedrate}. This
theorem's code construction naturally produces two streams --- one
that captures the essential unstable nature of the process and
requires a rate of at least $\log_2 \lambda$, and another that
captures the essentially classical nature of the information left
over. The quantitative distortion is dominated by the encoding of the
second stream, while the first stream serves to ensure its finiteness
as time goes on. The essentially stable nature of the second stream's
information is then made precise by Theorem~\ref{thm:backwards_bound}
which relates the forward $D(R)$ curve to the ``backwards'' one 
corresponding to a stable process.

At the intersection of source and channel coding, the notion of
anytime reliability was reviewed and
Theorem~\ref{thm:anytime_codes_exist} shows that it is nonzero for 
DMCs at rates below capacity. Theorem~\ref{thm:anytime_sufficient} and
Lemma~\ref{lem:anytime_necessity} then shows that the first stream
requires a high-enough anytime reliability from a 
communication system rather than merely enough rate. In contrast,
Theorems \ref{thm:history_sufficient} and \ref{thm:strict_converse}
show that the second stream requires only sufficient rate. Together,
all these results establish the relevant separation principle for such
unstable Markov processes.

This work brings exponentially unstable processes firmly into the fold
of information theory. More fundamentally, it shows that reliability
functions are not a matter purely internal to channel coding. In the
case of unstable processes, the demand for appropriate reliability
arises at the source-channel interface. Thus unstable processes have
the potential to be useful models while taking an
information-theoretic look at QoS issues in communication systems. The
success of the ``reductions and equivalences'' paradigm of
\cite{ControlPartI, OurDirectConverse} here suggests that this
approach might also be useful in understanding other situations in
which classical approaches to separation theorems break down.

\appendices 

\section{Riemann-integrable densities as
  mixtures} \label{app:riemannmixtures} 

It is often conceptually useful to think of generic random variables
with Riemann-integrable densities as being mixtures of a blurred
uniform random variable along with something else. This appendix
proves Lemma~\ref{lem:riemannmixture}. 

Since the density is Riemann-integrable,
\begin{eqnarray*}
\int_{-\infty}^{+\infty} f_W(w) dw
& = & \lim_{\delta \rightarrow 0}
\sum_{i=-\infty}^{+\infty} \delta \min_{x \in [i\delta -
  \frac{\delta}{2}, i\delta + \frac{\delta}{2}]} f_W(x)
\end{eqnarray*}

Thus, $f_W$ can be expressed as a non-negative piecewise constant
function $f'_W$ that only changes every $\delta$ units plus a
non-negative function $f''_W$ representing the ``error'' in
Riemann-integration from below. By choosing $\delta$ small enough, the
total mass in $f''_W$ can be made as small as desired since the
Riemann sums above converge. 

Choose $\delta$ such that the total mass in $f''_W$ is $\gamma$. So
\begin{equation} \label{eqn:densitydecomposition}
f_W = (1-\gamma)(\frac{f'_W}{1-\gamma}) + \gamma
(\frac{f''_W}{\gamma})
\end{equation}
and thus $W$ can thus be simulated in the following way:
\begin{enumerate}
\item Flip an independent biased coin $C_\gamma$ with probability of
      heads $\gamma$.
\item If heads, independently draw the value of $W$ from the
      density $\frac{f''_W}{\gamma}$ corresponding to a random
      variable $W''$.
\item If tails, independently draw the value of $W$ from the random
      variable $W''$ with piecewise constant density
      $\frac{f'_W}{1-\gamma}$. This can clearly be done by using a
      discrete random variable $G_\delta$ plus an independent uniform random
      variable $U_\delta$ so that $W'' = G_\delta + U_\delta$ has density
      $\frac{f'_W}{1-\gamma}$. 
\end{enumerate}
This proves the result. \hfill $\Box$

\section{Entropy bound for quantized random variables with bounded
  moments}

\begin{lemma} \label{lem:momentboundentropyquantized} 
Consider a random variable $Z$ that is quantized to precision $\Delta$
so $Z^q = Q_{\Delta}(Z)$. Further suppose that $E[|Z|^\eta] \leq
K$ where $K > \Delta^\eta$. Then 
\begin{equation} \label{eqn:entropySbound}
H(S) < 7 + \frac{\log_2 K}{\eta} + 2\log_2 \frac{\log_2 K}{\eta}
+ \log_2\frac{1}{\Delta} + 2 \log_2 \log_2\frac{1}{\Delta} 
+ \frac{5 + \ln2}{\eta \ln 2}.
\end{equation}
\end{lemma}
{\em Proof: }Let $Z^q = S\Delta$ where $S$ is an integer. Then $|S|
\leq 1 +  \frac{|Z|}{\Delta}$ and so
\begin{eqnarray*}
E[|S|^\eta] & \leq &  E[(1 + \frac{|Z|}{\Delta})^\eta] \\
& \leq &  E[(2\max\left(1,\frac{|Z|}{\Delta}\right))^\eta] \\
& = &  E[2^\eta \max\left(1^\eta,(\frac{|Z|}{\Delta})^\eta\right)] \\
& \leq & E[2^\eta + 2^\eta \frac{|Z|^\eta}{\Delta^\eta}] \\
& = & 2^\eta + \frac{2^\eta}{\Delta^\eta}E[|Z|^\eta] \\
& \leq & 2^\eta + \frac{2^\eta K}{\Delta^\eta} \\
& < & 2^{\eta + 1} \frac{K}{\Delta^\eta}.
\end{eqnarray*}
Applying the Markov inequality gives 
\begin{equation} \label{eqn:markovForS}
{\cal P}(|S| \geq s) \leq \min(1,\frac{2^{\eta + 1} K}{\Delta^\eta} s^{-\eta}).
\end{equation}
The integer $S$ can be encoded into bits using a self-punctuated 
code using less than $4 + \log_2 (|S|) + 2\log_2 (1 + \log_2 (|S|+1))$ bits to
encode $S \neq 0$ \cite{CoverThomas}. First encode the sign of $S$
using a single bit. There are at most $1+\log_2(|S| + 1)$ digits in
the natural binary expansion of $|S|$. This length can be encoded
using at most $2 + 2\log_2 (1 + \log_2 (|S|+1))$ bits by giving its
binary expansion with each digit followed by a $0$ if it is
not the last digit, and a $1$ if it is the last digit. Finally, $|S|$
itself can be encoded using at most $1 + \log_2 |S|$ bits.

Since the entropy must be less than the expected code-length for any
code, 
\begin{eqnarray*}
H(S) & \leq & 4 + E[\log_2 (|S|)] + 2E[\log_2 (1 + \log_2 (|S|+1))] \\
& = & 4 + \int_0^\infty {\cal P}(\log_2(|S|) > l) dl 
+ 2 \int_0^\infty {\cal P}(\log_2 (1 + \log_2 (|S|+1)) > l) dl.
\end{eqnarray*}
First, we deal with the dominant term
\begin{eqnarray*}
& & \int_0^\infty {\cal P}(\log_2(|S|) > l) dl \\
& = & \int_0^\infty {\cal P}(|S| > 2^l) dl \\
& \leq & \int_0^\infty \min(1,\frac{2^{\eta + 1} K}{\Delta^\eta}
2^{-\eta l}) dl \\
& = & \frac{1}{\eta}\log_2(\frac{2^{\eta + 1}K}{\Delta^\eta}) + 
      \int_{0}^\infty  2^{-\eta u} dl \\
& = & 1 + \frac{\log_2 K}{\eta} + \log_2\frac{1}{\Delta} + 
     \frac{1 + \ln2}{\eta \ln2} 
\end{eqnarray*}

Next, consider the smaller term 
\begin{eqnarray*}
& & 2 \int_0^\infty {\cal P}(\log_2 (1 + \log_2 (|S|+1)) > l) dl \\
& = & 
2 \int_0^\infty {\cal P}(\log_2 (|S|+1) > 2^l - 1) dl \\
& = & 
2 \int_0^\infty {\cal P}(|S|+1 > \frac{2^{2^l}}{2}) dl \\
& \leq & 
2 (1+\int_0^\infty {\cal P}(|S| > 2^{2^l}) dl) \\
& \leq & 
2 + 2\int_0^\infty \min(1,\frac{2^{\eta + 1} K}{\Delta^\eta} 2^{-\eta 2^l}) dl \\
& = & 
2 + 2\log_2 \frac{\log_2 \frac{2^{\eta + 1} K}{\Delta^\eta}}{\eta}
+ 2\int_0^\infty 2^{-\eta 2^l} dl \\
& < & 
2 
+ 2\log_2 \frac{\log_2 K}{\eta}
+ 2\log_2 (1 + \frac{1}{\eta})
+ 2\log_2 \log_2 \frac{1}{\Delta}
+ \frac{2}{\eta \ln2} \\
& \leq & 
2 
+ 2\log_2 \frac{\log_2 K}{\eta}
+ 2\log_2 \log_2 \frac{1}{\Delta}
+ \frac{4}{\eta \ln2} \\
\end{eqnarray*}
where the final inequalities come from the concave $\cap$ nature of
$\log_2$ and lower bounding $2^l$ with just $l$. Putting everything
together gives the desired result. \hfill $\Box$ 

\section{Proof of
  Proposition~\ref{prop:conditionaldistortion}} \label{app:propconditionaldistortion}
From (\ref{eqn:infinitehorizonratedistortion}) and
(\ref{eqn:finitehorizonratedistortion}), we know for every $\epsilon_2
> 0$, if $\Delta$ is small enough and $n$ is large enough, that there
exists a random vector $Y_0^{n-1}$ so that $\frac{1}{n}
\sum_{i=0}^{n-1} \rho(\widetilde{X}_i, Y_i) = d+\epsilon_3$ and that
even the best such vector must satisfy 
$$n(R_\infty^X(d)-\epsilon_2) \leq I(\widetilde{X}_0^{n-1};Y_0^{n-1})
\leq n(R_\infty^X(d)+\epsilon_2).$$ 

Decompose the relevant mutual information as
\begin{equation} \label{eqn:conditionalbreakdown}
I(\widetilde{X}_0^{n-1};Y_0^{n-1}|Z^q,\Theta) 
= - I(\widetilde{X}_0^{n-1};Z^q|\Theta) + I(\widetilde{X}_0^{n-1};Y_0^{n-1},Z^q|\Theta).
\end{equation}

To get the desired result of asymptotic equality, this conditional
mutual information has to be both upper and lower bounded. To upper
bound the conditional mutual information, we lower bound
$I(\widetilde{X}_0^{n-1};Z^q|\Theta)$ and upper bound 
$I(\widetilde{X}_0^{n-1};Y_0^{n-1},Z^q|\Theta)$. Vice-versa to get the 
lower bound.

\subsection{Lower bounding $I(\widetilde{X}_0^{n-1};Z^q|\Theta)$}
The first term is easily lower bounded for $\Delta$ small enough since
\begin{eqnarray} 
I(\widetilde{X}_0^{n-1};Z^q|\Theta) 
& = & H(Z^q|\Theta) - H(Z^q|\widetilde{X}_1^{n},\Theta) \nonumber\\
& = & H(Z^q|\Theta) \nonumber\\
& \geq & H(Z^q|\Theta,W_0^{n-2}) \nonumber\\
& \geq & \lfloor \log_2 \lambda^{n-1} \rfloor \nonumber\\
& = &    \lfloor (n-1) \log_2 \lambda \rfloor.
\label{eqn:lowerboundonendpoint}
\end{eqnarray}
This holds since conditioned on the final dither $\Theta$, the
quantized endpoint is a discrete random variable that is a
deterministic function of $\widetilde{X}_{n-1}$ and conditioning
reduces entropy. But $Z^q$ conditioned on the driving noise
$W_0^{n-2}$ is just the $\Delta$-precision quantization of
$\lambda^{n-1}$ times a uniform random variable of width $\Delta$ and
hence has discrete entropy $\geq \log_2 \lambda^{n-1}$. 

\subsection{Upper bounding $I(\widetilde{X}_0^{n-1};Z^q|\Theta)$}
To upper-bound this term, Lemma~\ref{lem:momentboundentropyquantized} can be
used to see 
\begin{eqnarray} 
I(\widetilde{X}_0^{n-1};Z^q|\Theta) 
& = & H(Z^q|\Theta) \nonumber\\
& < & 7 + \frac{\log_2 K'}{\eta} + 2\log_2 \frac{\log_2 K'}{\eta}
+ \log_2\frac{1}{\Delta} + 2 \log_2 \log_2\frac{1}{\Delta} 
+ \frac{5 + \ln2}{\eta \ln 2}
\end{eqnarray}
where $K'$ is an upper-bound to $E[|Z|^\eta]$. Such an upper-bound is readily
available since 
\begin{eqnarray*}
E[|Z|^\eta] & = & 
E[|\widetilde{X}_n|^\eta] \\
& \leq & 
E[(\frac{\Delta}{2} + |\sum_{i=0}{n-1} \lambda^{n-1-i} W_i|)^\eta] \\
& = & 
E[(\frac{\Delta}{2} + \lambda^{n-1} |\sum_{i=0}{n-1} \lambda^{-i}
W_i|)^\eta] \\
& = & 
\lambda^{\eta(n-1)}E[(\frac{\Delta}{2 \lambda^{n -1}} +
|\sum_{i=0}{n-1} \lambda^{-i} W_i|)^\eta] \\ 
& \leq &
\lambda^{\eta(n-1)}E[(2 \max(\frac{\Delta}{2 \lambda^{ n
    -1}},|\sum_{i=0}{n-1} \lambda^{-i} W_i|)^\eta] \\ 
& < &
\lambda^{\eta(n-1)}(\frac{\Delta^\eta}{\lambda^{\eta(n -1)}} + 2^{\eta}K).
\end{eqnarray*}
Using this for $K'$ and taking logs shows
\begin{eqnarray*}
\frac{\log_2 K'}{\eta}
& = & 
\frac{\log_2(\lambda^{\eta(n-1)}(\frac{\Delta^\eta}{\lambda^{\eta(n
      -1)}} + 2^{\eta}K))}{\eta} \\
& = &
1 + (n-1)\log_2 \lambda + \frac{\log_2 (\frac{\Delta^\eta}{\lambda^{\eta(n
      -1)}} + 2^{\eta}K)}{\eta}
\end{eqnarray*}
Substituting this in gives the desired bound
\begin{eqnarray} 
& & I(\widetilde{X}_0^{n-1};Z^q|\Theta)  \nonumber \\
& < &
8 + (n-1)\log_2 \lambda + 2\log_2 (1+ (n-1)\log_2 \lambda +
\frac{\log_2 (\frac{\Delta^\eta}{\lambda^{\eta(n -1)}} +
  2^{\eta}K)}{\eta}) 
+ \log_2\frac{1}{\Delta} + 2 \log_2 \log_2\frac{1}{\Delta} 
+ \frac{5 + \ln2}{\eta \ln 2}. \label{eqn:upperboundendpoint}
\end{eqnarray}

There is only a single $O(n)$ term above, and it is $(n-1)\log_2
\lambda$. Everything else is $o(n)$. 

\subsection{Lower bounding
  $I(\widetilde{X}_0^{n-1};Y_0^{n-1},Z^q|\Theta)$}
We need to establish 
\begin{equation} \label{eqn:historyconditionalboundlower}
I(\widetilde{X}_0^{n-1};Y_0^{n-1},Z^q|\Theta) \geq
n(R_\infty^X(d) - \epsilon_2).
\end{equation}
This is immediately obvious from
\begin{eqnarray*}
I(\widetilde{X}_0^{n-1};Y_0^{n-1},Z^q|\Theta) 
& = & 
I(\widetilde{X}_0^{n-1};Y_0^{n-1}|\Theta) +
H(Z^q|\Theta,Y_0^{n-1})
- H(Z^q|\Theta,Y_0^{n-1},\widetilde{X}_0^{n-1}) \\
& = & 
I(\widetilde{X}_0^{n-1};Y_0^{n-1}|\Theta) + H(Z^q|\Theta,Y_0^{n-1}) \\
& \geq & 
n(R_\infty^X(d) - \epsilon_2).
\end{eqnarray*}
The first equality is just expanding the mutual information and
recognizing the fact that $Z^q$ is discrete once conditioned on the
dither $\Theta$ and so $H$ is the regular discrete entropy here. Let
$Q_{(\Delta,\Theta)}$ denote the dithered scalar quantizer used to
generate the encoded checkpoints, just appropriately translated so it
can apply to the $\widetilde{X}$ giving $Z^q =
Q_{(\Delta,\Theta)}(\widetilde{X}_{n-1})$. The next equality is a
consequence of this deterministic relationship. Finally, the discrete
entropy is always positive and can be dropped to give a lower bound. 

\subsection{Upper bounding $I(\widetilde{X}_0^{n-1};Y_0^{n-1},Z^q|\Theta)$}
The second term of (\ref{eqn:conditionalbreakdown}) is upper bounded
in a way similar to the first term. We need to establish 
\begin{equation} \label{eqn:historyconditionalbound}
I(\widetilde{X}_0^{n-1};Y_0^{n-1},Z^q|\Theta) \leq
n(R_\infty^X(d)+\epsilon_2) + o(n).
\end{equation}
Expand the mutual information as before
\begin{eqnarray*}
I(\widetilde{X}_0^{n-1};Y_0^{n-1},Z^q|\Theta) 
& = & 
I(\widetilde{X}_0^{n-1};Y_0^{n-1}|\Theta) +
H(Z^q|\Theta,Y_0^{n-1}) \\
& \leq & 
I(\widetilde{X}_0^{n-1};Y_0^{n-1}|\Theta) +
H(Z^q|\Theta,Y_{n-1}) \\
& = & 
n(R_\infty^X(d)+\epsilon_2) +
H(Z^q-Q_{(\Delta,\Theta)}(Y_{n-1})|\Theta,Y_{n-1}) \\
& \leq & 
n(R_\infty^X(d)+\epsilon_2) + \log_2 3 +
H(Q_{(\Delta,\Theta)}(\widetilde{X}_{n-1} - Y_{n-1})|\Theta).
\end{eqnarray*}
The first inequality comes from dropping conditioning. After that,
the quantizer $Q_{(\Delta,\Theta)}$ can be applied to $Y_{n-1}$ so that 
$Z^q-Q_{(\Delta,\Theta)}(Y_{n-1}) = S \Delta$ where $S$ is an  
integer-valued random variable representing how many steps up or down
the $\Delta$-quantization ladder are needed to get from
$Q_{(\Delta,\Theta)}(Y_{n-1})$ to $Z^q$. The difference of two
quantized numbers differs by at most 1 quantization bin from the
quantization of the difference. This slack of up to 1 bin in either 
direction can be encoded using $\log_2 3$ bits.

At this point, Lemma~\ref{lem:momentboundentropyquantized} applies
using the trivial upper bound $n(d+\epsilon_3)$ for the $\eta$-th
moment of $\widetilde{X}_{n-1} - Y_{n-1}$, since the worst case is for
the entire distortion to fall on the last component of the vector.
\begin{eqnarray*}
 & & H(Q_{(\Delta,\Theta)}(\widetilde{X}_{n-1} - Y_{n-1})|\Theta) \\
& < & 
7 + \frac{\log_2 n(d+\epsilon_3)}{\eta} + 2\log_2 \frac{\log_2
  n(d + \epsilon_2)}{\eta}
+ \log_2\frac{1}{\Delta} + 2 \log_2 \log_2\frac{1}{\Delta} 
+ \frac{5 + \ln2}{\eta \ln 2} 
\end{eqnarray*}

The $\log_2 n$ term is certainly $o(n)$. The only other
term that might raise concern is $\log_2 \frac{1}{\Delta}$, but that
is $o(n)$ since (\ref{eqn:checkpointRconstraint}) tells us that we are 
already required to choose $n$ much larger than that to have $R_1$
close to $\log_2 \lambda$ in the first stream. The order of limits is
to always let $n$ go to infinity before $\Delta$ goes to zero.

\subsection{Putting pieces together}
With (\ref{eqn:historyconditionalbound}) established, it can be applied
along with (\ref{eqn:lowerboundonendpoint}) to 
(\ref{eqn:conditionalbreakdown}) and gives
\begin{equation} \label{eqn:conditionalinfo}
I(\widetilde{X}_0^{n-1};Y_0^{n-1}|Z^q,\Theta) \leq 
n(R_\infty^X(d) - \log_2 \lambda +\epsilon_2) + o(n).
\end{equation}
Taking $n$ to $\infty$ and dividing through by $n$ establishes the
desired result on the upper bound. 

Similarly putting together (\ref{eqn:upperboundendpoint}) and
(\ref{eqn:historyconditionalboundlower}) gives
\begin{equation} \label{eqn:conditionalinfoupper}
I(\widetilde{X}_0^{n-1};Y_0^{n-1}|Z^q,\Theta) \geq 
n(R_\infty^X(d) - \log_2 \lambda - \epsilon_2) - o(n).
\end{equation}
Taking $n$ to $\infty$ and dividing through by $n$ establishes the
desired result on the lower bound. 

But $\epsilon_2$ was arbitrary and this establishes the desired
result. \hfill $\Box$

\section{Proof of
  Theorem~\ref{thm:anytime_codes_exist}} \label{app:randomanytimecodesexist}
 Interpret the random ensemble of infinite
tree codes as a single code with both encoder and decoder having
access to the common-randomness used to generate the
code-tree. Populate the tree with iid~channel inputs drawn from the
distribution that achieves $E_r(R)$ for block codes. Theorem~7 in
\cite{ForneyML} tells us that the code achieves anytime reliability
$\alpha = E_r(R)$ since the analysis uses the same infinite ensemble
for all $i$ and delays.

Alternatively, this can be seen from first principles for ML decoding
by observing that any false path $\widetilde{B}_1^i$ can be divided into a
true prefix $B_1^{j-1}$ and a false suffix $\widetilde{B}_j^i$. The iid
nature of the channel inputs on the code tree tells us that the true
code-suffix corresponding to the received channel outputs from time
$\frac{j}{R}$ to $t$ is independent of any false code-suffix. Since
there are $\leq 2^{R(t-\frac{j}{R})}$ such false code-suffixes
(ignoring integer effects) at depth $j$, Gallager's random
block-coding analysis from \cite{Gallager} applies since all that it
requires is pairwise independence between true and false codewords.
\begin{eqnarray*}
&  & {\cal P}(\widehat{B}_j(t) \neq B_j |B_1^{j-1} \mbox{ already
known}) \\
& \leq &
{\cal P}(\mbox{error on random code with }2^{R(t-\frac{j}{R})}\mbox{
  words and block length }t - \lceil \frac{j}{R} \rceil ) \\
& \leq &
2^{-(t-\lceil \frac{j}{R} \rceil)E_r(R)} \\
& \leq &
2^{-(t-\frac{j}{R} - 1)E_r(R)}
\end{eqnarray*}
The probability of error on $B_1^i$ can be bounded by the union bound
over $j=1 \ldots i$. 
\begin{eqnarray*}
{\cal P}(\widehat{B}_1^i(t) \neq B_1^i) 
& \leq &
\sum_{j=1}^{i} {\cal P}(\widehat{B}_j(t) \neq B_j |B_1^{j-1} \mbox{ already known}) \\
& \leq &
\sum_{j=1}^{i} 2^{-(t-\frac{j}{R} - 1)E_r(R)} \\
& < &
\sum_{j=0}^{\infty} 2^{-(t-\frac{i}{R} - j -1)E_r(R)} \\
& = & K 2^{-(t-\frac{i}{R})E_r(R)} 
\end{eqnarray*}
The exponent for the probability of error is dominated by the
shortest codeword length in the union bound, and this corresponds to
$t - \frac{i}{R}$. \hfill $\Box$

\section{Proof of
  Property~\ref{gap_property}} \label{app:propertygapbound}

\begin{eqnarray*}
|X'_{kn} - \bar{X}'_{kn}|
& \geq & \lambda^{n(k - j)} \beta (|M_j - \bar{M}_j| - \left|\sum_{i=j+1}^{\infty}
\lambda^{-n(i-j)} (M_i - \bar{M}_i)\right|) \\
& \geq & \lambda^{n(k - j)} \beta (|M_j - \bar{M}_j| - 2 \lambda^{-n} \sum_{i=0}^{\infty}
\lambda^{-ni}) \\
& \geq & \lambda^{n(k - j)} \beta (2^{1-nR} - 2 \frac{\lambda^{-n}}{1-\lambda^{-n}}) \\
& = & \lambda^{n(k - j)} 2 \beta (2^{-nR} - \frac{1}{\lambda^n-1})
\end{eqnarray*}
which is positive as long as $2^{-nR} > \frac{1}{\lambda^n-1}$ or $nR <
\log_2(\lambda^n - 1)$. We can thus use $K = 2 \beta (2^{-nR} -
\frac{1}{\lambda^n-1}) = \frac{\delta}{\lambda} (2^{n(\log_2 \lambda - R)} -
\frac{\lambda^n}{\lambda^n - 1})$ and the property is proved. \hfill $\Box$

\section*{Acknowledgments}
The authors would like to give special thanks to Mukul Agarwal for the
discussions regarding this paper and the resulting contributions in
\cite{OurDirectConverse} where one of the key results required is
proved. We also thank Nigel Newton, Nicola Elia, and Sekhar Tatikonda
for several constructive discussions about this general subject over a
long period of time which have influenced this work in important ways.

\bibliographystyle{IEEEtran}
\bibliography{IEEEabrv,./MyMainBibliography}

\end{document}